\documentclass[fleqn,
    usenatbib,
]{mnras}

\usepackage{newtxtext,newtxmath}

\usepackage[T1]{fontenc}

\DeclareRobustCommand{\VAN}[3]{#2}
\let\VANthebibliography\thebibliography
\def\thebibliography{\DeclareRobustCommand{\VAN}[3]{##3}\VANthebibliography}

\usepackage{graphicx}	%
\usepackage{amsmath}	%

\usepackage{mathtools}
\usepackage[normalem]{ulem}
\usepackage{tensor, comment}
\usepackage{enumitem}
\usepackage{color}
\usepackage{subcaption}
\usepackage{multicol}
\usepackage{multirow}
\usepackage{duckuments}
\usepackage{multirow, tabularx}
\usepackage{tikz}
\usetikzlibrary{arrows.meta}

\setlist[itemize]{
    leftmargin=1.6em,
    labelsep=.6em,
    labelwidth=1em,
}

\newcommand{\e}[1]{\times 10^{#1}}
\newcommand{\pdev}[2]{\frac{\partial #1}{\partial #2}}

\newcommand{\inner}[1]{#1_{\rm in}}
\renewcommand{\outer}[1]{#1_{\rm out}}
\newcommand{\kms}{{\rm \, km \, s^{-1}}}

\newcommand{\vkick}{V_{\rm kick}}

\newcommand{\athenak}[1]{\texttt{AthenaK}}

\graphicspath{{./}{Figures/}{VolumeRendering/}}

\title[Kicked hydrodynamic disks]{Tilt, warp, and dissipation flare in hydrodynamic disks around kicked compact objects}

\author[Y. Kim]{
Yoonsoo Kim$^{1,2}$\thanks{E-mail: yoonsoo@princeton.edu}
\\
$^{1}$Princeton Center for Theoretical Science,
Princeton University, Princeton, NJ 08544, USA\\
$^{2}$Princeton Gravity Initiative, Princeton
University, Princeton, NJ 08544, USA
}

\date{Accepted XXX. Received YYY; in original form ZZZ}

\pubyear{\the\year{}}

\begin{document}
\label{firstpage}
\pagerange{\pageref{firstpage}--\pageref{lastpage}}
\maketitle

\begin{abstract}
    We study the dynamical response of a hydrodynamic disk to an impulsive kick
    imparted to a central massive object, focusing on the global properties of
    the material that remains gravitationally bound. We construct a simple
    analytic model and show that the bound material can acquire a significant
    tilt---or even a retrograde rotation---if the kick is directed close to the
    disk midplane. We perform 3D numerical simulations and show that (1) the
    bound material relaxes to a torus-like circularized flow of which total mass
    and angular momentum is in fair agreement with the analytic estimate, and
    (2) in spite of nonlinear hydrodynamic effects, a highly tilted or
    retrograde disk can be sufficiently formed. The simulations additionally
    reveal a transient warp in the kicked disk, suggesting a kick as a formation
    mechanism of tilted or warped disks around compact objects.
    About 10\% of the kinetic energy the bound material acquired from the kick
    is converted to the internal energy, while the exact conversion efficiency
    depends on the kick angle and the initial surface mass density profile. The
    luminosity of this kick-driven dissipation flare can reach $\sim
    10^{43}\,{\rm erg\,s^{-1}}$ for strong kicks ($\vkick \sim 1000\kms$).
    We discuss the implications of our results for recoiling supermassive black
    holes, recoiled stellar-mass black holes in the disks of active galactic
    nuclei, and natal kicks on compact objects.
\end{abstract}


\section{Introduction}

Under certain astrophysical conditions, a massive object can acquire linear
momentum via an internal process or interaction with the surrounding
environment, commonly referred to as a kick or recoil. If the kicked object is
surrounded by a mass distribution (e.g. gas disks, clouds, or stars), the
response of the environment to the kick can result in prompt electromagnetic
emissions or unique dynamical signatures.

A prominent example is the gravitational radiation recoil
\citep{Peres:1962zz,Bekenstein:1973zz,Fitchett:1983qzq} onto the remnant black
hole (BH) in the mergers of binary black holes (BBHs). Numerical relativity
simulations have shown that the recoil speed spans from $\vkick \lesssim
200\kms$ for non-spinning BH binaries \citep{Baker:2006vn,Gonzalez:2006md} to
$\vkick \approx \mathcal{O} (1000) \kms$ for rapidly spinning binaries
\citep{Campanelli:2007cga,Campanelli:2007ew,Gonzalez:2007hi,Lousto:2011kp}. In
the merger of binary supermassive black holes (SMBHs)
\citep{Begelman:1980vb,Milosavljevic:2001vi}, a recoiled SMBH can generate
electromagnetic counterparts in various forms owing to the matter-rich
environment in the galactic nuclear region \cite[see
e.g.][]{Bogdanovic:2021aav}, including electromagnetic flares from the perturbed
circumbinary disk
\cite[e.g][]{Shields:2008va,Lippai:2008fx,Schnittman:2008ez,Megevand:2009yx,Corrales:2009nv,Rossi:2009nk,Zanotti:2010xs,Ponce:2011kv,Bode:2011pqa,Meliani:2016rwn,deSimone:2025cqg,Kim:2026qad}
and characteristic dynamical signatures in dense stellar clusters around the
SMBH
\citep{Merritt:2008kg,OLeary:2008tpn,OLeary:2011ykg,Akiba:2021,Akiba:2023ami,Bright:2024soo,Akiba:2025}.

Accretion disks of active galactic nuclei (AGNs) may harbor a population of
stellar remnants and provide a site for stellar-mass BBH mergers
\citep{McKernan:2020lgr,Tagawa:2019osr,McKernan:2023xio,Ford:2025bbk}. The
(magneto-) hydrodynamic interactions between the recoiled remnant BH and the AGN
disk have been suggested to produce observable electromagnetic counterparts
\citep{McKernan:2019hqs,Graham:2020gwr,Kimura:2021xxu,Tagawa:2023uqa,Chen:2023xrm,Kim:2024zjb,McPike:2026ugj,He:2026rle,Tagawa:2026lqy}.

Most previous studies on kicked hydrodynamic disks have mainly focused on
tightly bound disks for which the recoil is perturbative. In other words, the
size of the disk is assumed to be much smaller than the radius $GM/\vkick^2$
where $M$ is the mass of the recoiled object and $\vkick$ is the kick speed.
\cite{Rossi:2009nk} and \cite{Ponce:2011kv} have relaxed this assumption and
tracked the disruption of the disk into gravitationally bound and unbound
material. A notable result from those studies is that the angular momentum
vector of the bound material acquires a finite tilt angle relative to the
original rotation axis of the disk, although the phenomenology of this
kick-driven tilt has not been explored in further detail.

In this paper, we systematically examine the global properties of kicked
hydrodynamic disks, focusing on how the mass and orientation of the kicked disks
vary with the kick geometry and the initial surface mass density profile. We
first analytically estimate the total bound mass and angular momentum of a
kicked mass distribution in Sec.~\ref{sec:analysis-infinite-disk}, and discuss
their implications for hydrodynamic (gaseous) disks in
Sec.~\ref{sec:tilt-and-size}. We present the results from numerical simulations
and assess the impact of hydrodynamic effects in
Sec.~\ref{sec:numerical-experiment}. Discussions and applications in
astrophysical systems are presented in Sec.~\ref{sec:discussions}, then we
conclude in Sec.~\ref{sec:conclusion}.

\section{Analysis}
\label{sec:analysis-infinite-disk}

\subsection{Description of the problem}

\begin{figure}
    \centering
    \begin{tikzpicture}[>=Latex,
        axis/.style={->, thick},
        kick/.style={blue!90!black},
        orbit/.style={color={rgb,1: red,1; green,0.95; blue,0.7}},
    ]
    \def\az{23} \def\el{16}
    \settowidth{\dimen0}{\includegraphics{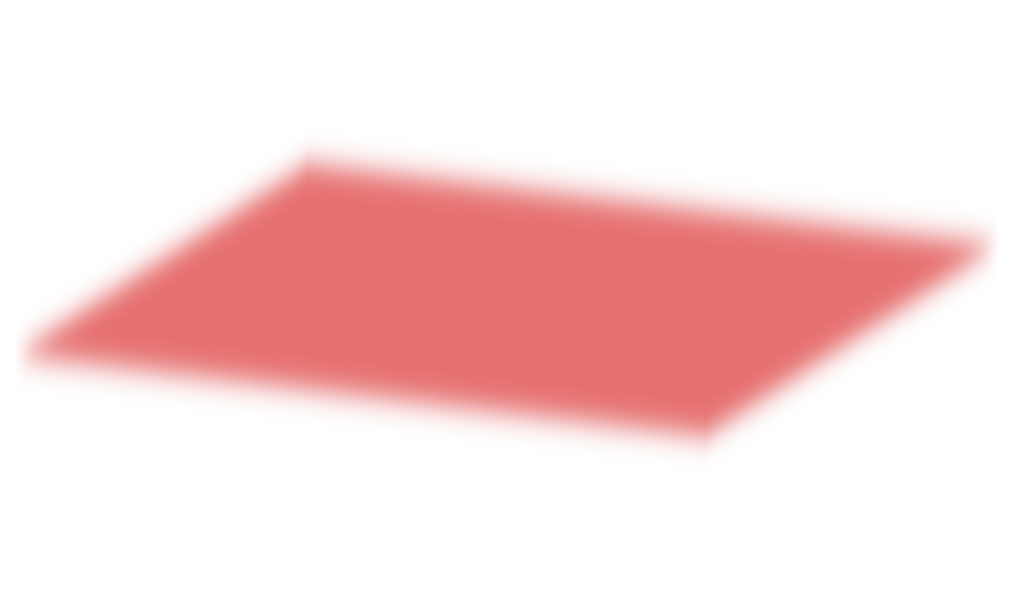}}
    \pgfmathsetmacro{\imgw}{\the\dimen0/1in}
    \pgfmathsetmacro{\un}{\linewidth/\imgw/1cm}
    \pgfmathsetmacro{\imgscale}{\un/2.54}
    \pgfmathsetmacro{\Xx}{\un*cos(\az)}
    \pgfmathsetmacro{\Xy}{-\un*sin(\az)*sin(\el)}
    \pgfmathsetmacro{\Yx}{\un*sin(\az)}
    \pgfmathsetmacro{\Yy}{\un*cos(\az)*sin(\el)}
    \pgfmathsetmacro{\Zy}{\un*cos(\el)}
    \tikzset{x={(\Xx cm,\Xy cm)}, y={(\Yx cm,\Yy cm)}, z={(0cm,\Zy cm)}}

    \def\alp{60}   %
    \def\phid{40}  %
    \def\rr{1.55}   %
    \def\vk{2.68}     %

    \node[inner sep=0] at (0,0)
        {\includegraphics[scale=\imgscale]{disk_parallelogram.png}};

    \draw[black!70, thin] (-2.1, 1.1, 0) -- ++ (-0.5cm, 0.8cm)
        node[above left, black] {$\Sigma(r) \propto r^p$};

    \draw[orbit, semithick]
        (\rr,0,0) \foreach \t in {5,10,...,360} {-- ({\rr*cos(\t)},{\rr*sin(\t)},0)};
    \draw[orbit, ->, very thick]
        plot[variable=\t, domain=180:255, samples=20]
        ({\rr*cos(\t)},{\rr*sin(\t)},0);
    \node[orbit, left] at ({\rr*cos(215)-0.06},{\rr*sin(215)},0) {$v_K(r)$};

    \draw[orbit, thick, ->] (0,0,0) -- ({\rr*cos(\phid)},{\rr*sin(\phid)},0)
        node[pos=0.6, above, sloped, inner sep=2pt] {$r$};
    \draw[orbit]
        plot[variable=\t, domain=0:\phid, samples=20]
        ({0.6*cos(\t)},{0.6*sin(\t)},0); \node[orbit] at
        ({0.95*cos(\phid/2)},{0.95*sin(\phid/2)},0) {$\phi$};

    \coordinate (O) at (0,0,0);
    \draw[axis] (O) -- (3.1, 0, 0)  node[right] {$x$};
    \draw[axis] (O) -- (0, 4.1, 0)  node[above right] {$y$};
    \draw[axis] (O) -- (0, 0, 3)  node[above] {$z$};

    \draw[kick, dashed]
        ({\vk*cos(\alp)},0,{\vk*sin(\alp)}) -- ({\vk*cos(\alp)},0,0);
    \draw[kick, dashed]
        ({\vk*cos(\alp)},0,{\vk*sin(\alp)}) -- (0,0,{\vk*sin(\alp)});
    \draw[kick, very thick, ->] (O) --
        ({\vk*cos(\alp)}, 0, {\vk*sin(\alp)})
        node[above right] {$\mathbf{V}_{\rm kick}$};
    \draw[kick, thick]
        plot[variable=\t, domain=0:\alp, samples=15]
            ({0.8*cos(\t)},0,{0.8*sin(\t)});
        \node[kick] at
            ({1*cos(0.82*\alp)},0,{1*sin(0.82*\alp)}) {$\alpha$};

    \fill (O) circle (2pt); \node[left, inner sep=5pt] at (O) {$M$};

    \end{tikzpicture}
    \vspace{-4em}
    \caption{Schematic of the problem. A massive object with the mass $M$ is
        located at the coordinate origin. A non self-gravitating, geometrically
        thin mass distribution on the $xy$ plane with the surface mass density
        $\Sigma(r)$ is initially rotating in circular orbits with the Keplerian
        velocity $v_K(r) = \sqrt{GM/r}$. At $t=0$, the central object is kicked
        with the speed $\vkick$ and the angle $\alpha$ from the midplane, i.e.
        $\mathbf{V}_{\rm kick} = \vkick(\cos \alpha, 0, \sin \alpha)$.}
    \label{fig:problem-schematic}
\end{figure}

Fig.~\ref{fig:problem-schematic} illustrates the idealized problem considered in
this work. A central massive object with the mass $M$ is located at the
coordinate origin $r=0$. We consider a disk on the $z=0$ plane initially
rotating in circular orbits with the Keplerian velocity $v_K(r) = \sqrt{GM/r}$.
We assume that the gravitational force in this system is dominated by the
central object $M$, and neglect the self-gravity of the disk.

At $t=0$, the central object is impulsively kicked with the velocity
$\mathbf{V}_{\rm kick} = \vkick(\cos\alpha \,\mathbf{\hat{x}} + \sin\alpha
\,\mathbf{\hat{z}})$, where $\vkick$ is the kick speed and $\alpha$ is the kick
angle measured from the disk midplane ($z=0$). In the rest frame of the kicked
object, the instantaneous velocity of a particle at the coordinate $\mathbf{r} =
(r \cos \phi, r \sin \phi, 0)$ is
\begin{equation} \label{eq:velocity-vector}
    \mathbf{v} = (-\vkick \cos\alpha -v_K(r) \sin \phi,\,\, v_K(r) \cos \phi,\,\, -\vkick  \sin \alpha) \, .
\end{equation}

The gravitational mass of the kicked object may change concurrently with the
recoil. For example, in the merger of a BH binary, gravitational wave emission
carries away net mass-energy and leads to the mass loss in the merger remnant of
several percent \cite[e.g.][]{Campanelli:2007cga,Healy:2014yta}. In such cases,
the gravitational potential at $t \geq 0$ should be modified to
\begin{equation}
    \Phi(r) = -(1-\epsilon) \frac{GM}{r}
\end{equation}
where $\epsilon$ is the fraction of the mass loss.

The mass of the kicked object $M$ and the kick speed $\vkick$ set the
characteristic length and time scales
\begin{align}
    \label{eq:scale-radius}
    r_0 & \equiv \frac{GM}{\vkick^2} \, , \\[1ex]
    \label{eq:scale-time}
    t_0 & \equiv \frac{GM}{\vkick^3} \, ,    
\end{align}
namely the radius at which the Keplerian orbital speed equals the kick speed and
the orbital time scale at that radius.

\subsection{Gravitationally bound region}
\label{sec:gravitationally-bound-region}

\begin{figure}
    \centering
    \includegraphics[width=0.96\linewidth]{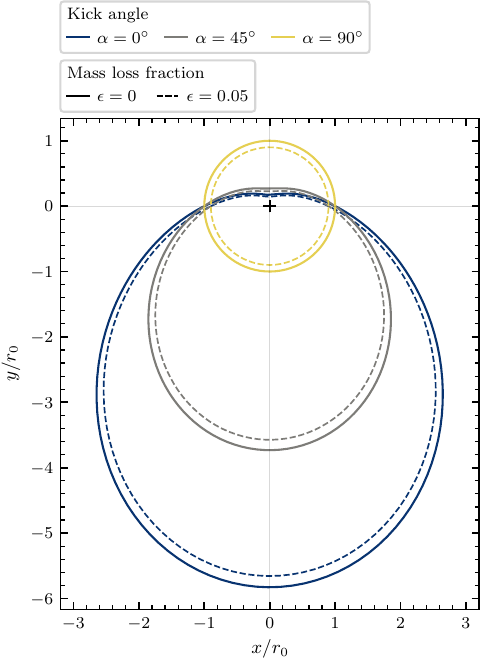}
    \caption{The boundary $r_b(\phi)$ between the gravitationally bound and
         unbound region for different kick angles and the mass loss fraction of
         the central object. $\alpha=0$ corresponds to the in-plane kick $\vkick
         \mathbf{\hat{x}}$ (pointing right in this figure) and $\alpha=90^\circ$
         corresponds to $\vkick \mathbf{\hat{z}}$ (out of the page).}
    \label{fig:bound-radius}
\end{figure}

Using Eq.~\eqref{eq:velocity-vector}, the specific orbital energy of a particle
in the rest frame of the kicked object is
\begin{equation}
\begin{split}
    \mathcal{E} & = v^2/2 + \Phi(r) \\[1ex]
    & = \frac{\vkick^2}{2} + \vkick v_K(r) \cos\alpha\sin\phi - (1-2\epsilon) \frac{v_K^2(r)}{2}
    \, .
\end{split}
\end{equation}
The material with $\mathcal{E}<0$ remains gravitationally bound to the kicked
object. This condition corresponds to
\begin{equation} \label{eq:bound-velocity}
    \frac{\vkick}{v_K(r)} < \sqrt{1 - 2\epsilon +\cos^2\alpha\sin^2\phi} \, - \cos\alpha\sin\phi \, ,
\end{equation}
or equivalently
\begin{equation} \label{eq:bound-radius}
    r < r_b(\phi) \equiv r_0 \left( \sqrt{1-2\epsilon+\cos^2\alpha\sin^2\phi} \, - \cos\alpha\sin\phi \right)^2 .
\end{equation}

In Fig.~\ref{fig:bound-radius}, we plot the curve $r = r_b(\phi)$ for different
kick angles $\alpha$ and the mass loss fraction $\epsilon$. The gravitationally
bound region of the disk appears as an ellipsoid-like shape, of which areal
extent and asymmetry are greater for smaller kick angles. For $\epsilon=0$, the
perpendicular kick ($\alpha=90^\circ$) gives a circular bound region $r_b(\phi)
= r_0$, and the in-plane kick ($\alpha=0$) gives a lopsided bound region
spanning $[(\sqrt{2}-1)^2, (\sqrt{2}+1)^2] r_0$. A small mass loss fraction
$\epsilon$ shrinks the overall bound region a bit, but generally has less impact
on the shape and extent of $r_b(\phi)$.

For convenience, hereafter we denote the factor on the right hand side of
Eq~\eqref{eq:bound-radius} as
\begin{equation} \label{eq:rb-over-r0}
    I \equiv \frac{r_b(\phi)}{r_0}
        = \left( \sqrt{1-2\epsilon+\cos^2\alpha\sin^2\phi}
            \, - \cos\alpha\sin\phi \right)^2 \, .
\end{equation}

\subsection{Total bound mass}
\label{sec:total-mass}

\begin{figure}
    \centering
    \includegraphics[width=0.97\linewidth]{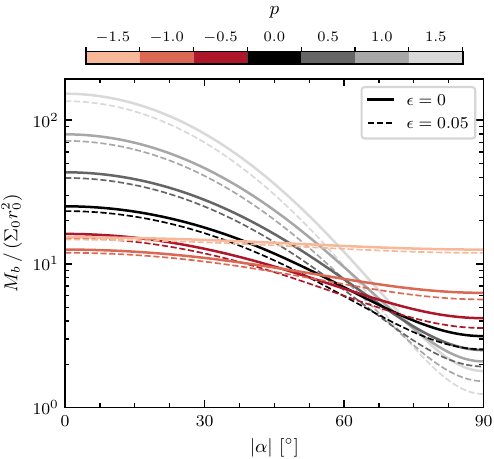}
    \caption{Total mass of the gravitationally bound material over different
        kick angles ($\alpha$) and the disk surface mass density profiles
        ($\Sigma \propto r^p$).}
    \label{fig:bound-mass}
\end{figure}

We assume the (vertically integrated) surface mass density of the disk before
the kick follows a power law
\begin{equation} \label{eq:surface-density}
    \Sigma(r) = \Sigma_0 \, \left(\frac{r}{r_0}\right)^{p} \, ,
\end{equation}
with the power exponent $p > -2$, for simplicity.\footnote{The standard
alpha-disk model \citep{Shakura:1972te}, for example, exhibits the radial
exponent ranging from $p=3/2$ to $p=-3/4$ \cite[see e.g.][]{Abramowicz:2011xu}.}
The total mass of the gravitationally bound material is
\begin{equation} \label{eq:integral-bound-mass}
\begin{split}
    M_b &= \int_0^{2\pi} d\phi \int_0^{r_b(\phi)} \Sigma(r) \,r \, dr \\[1ex]
    &= \frac{\Sigma_0 r_0^2}{p+2} \int_0^{2\pi} d\phi \, I^{p+2} \, .
\end{split}
\end{equation}

Fig.~\ref{fig:bound-mass} plots the total bound mass $\bar{M}_b\equiv
M_b/(\Sigma_0 r_0^2)$ for different kick angles and surface mass density
exponents. The variation of the total bound mass is larger for higher values of
$p$, since the integral Eq.~\eqref{eq:integral-bound-mass} picks up the mass
more rapidly from the radial expansion of the curve $r_b(\phi)$ for smaller kick
angles (see Fig.~\ref{fig:bound-radius}).

\subsection{Total angular momentum of the bound material}
\label{sec:total-angular-momentum}

\begin{figure*}
    \centering
    \includegraphics[width=0.98\linewidth]{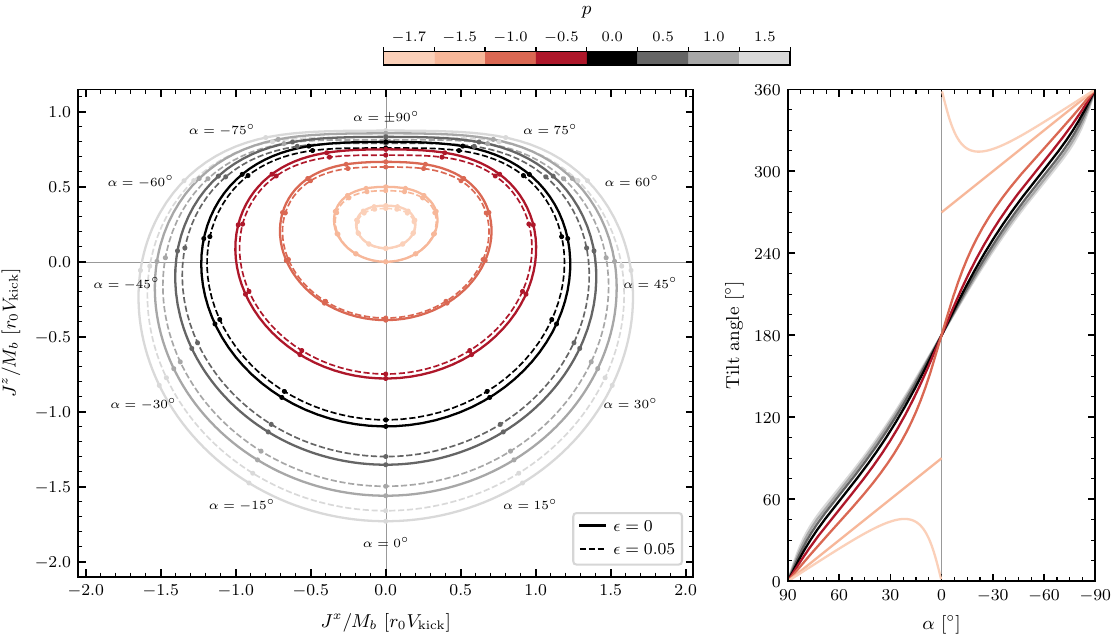}
    \caption{{\em Left:} The total angular momentum of the gravitationally bound
        material normalized with the total bound mass. The squared distance from
        the origin in this plot corresponds to the circularization radius in the
        units of $r_0 = GM/\vkick^2$; see the discussions in
        Sec.~\ref{sec:tilt-and-size}.
        {\em Right:} Vertical tilt angle of the total bound angular momentum
        vector $J^i$ as a function of the kick angle. We only show the case
        without mass loss ($\epsilon=0$) for brevity.}
    \label{fig:normalized-angular-momentum-vector}
\end{figure*}

Following the kick, the specific angular momentum of a particle at $\mathbf{r} =
(r\cos\phi, r\sin\phi, 0)$ is
\begin{equation} \label{eq:specific-angular-momentum}
\begin{split}
    \mathbf{j} & = \mathbf{r} \times \mathbf{v}
        = \big( 0 ,\, 0 ,\, r v_K(r) \big) \\
        & \hspace{5em} + r \vkick \big( -\sin\alpha\sin\phi ,\, \sin\alpha\cos\phi ,\, \cos\alpha\sin\phi \big) .
\end{split}
\end{equation}
Eq.~\eqref{eq:specific-angular-momentum} shows that, on top of the original
Keplerian angular momentum $rv_K(r)\,\hat{\mathbf{z}}$, misaligned components
proportional to the kick speed $\vkick$ are excited if the kick is directed
off-plane ($\alpha \neq 0$).

The total angular momentum contained in the gravitationally bound material is
\begin{equation} \label{eq:total-angular-momentum}
    \mathbf{J} = \int_0^{2\pi} d\phi \int_0^{r_b(\phi)} \Sigma(r) \, \mathbf{j} \,r dr \, .
\end{equation}
Evaluating each spatial component in Eq.~\eqref{eq:total-angular-momentum}, we
get
\begin{subequations} \label{eq:total-angular-momentum-components}
\begin{align} 
\label{eq:integral-jx}
    \bar{J}^x & = \frac{-\sin\alpha}{p+3}
        \int_0^{2\pi} I^{p+3} \sin\phi \,d\phi \, , \\[1ex]
\label{eq:integral-jy}
    \bar{J}^y & =  \frac{\sin\alpha}{p+3}
        \int_0^{2\pi} I^{p+3} \cos\phi \,d\phi \, , \\[1ex]
\label{eq:integral-jz}
    \bar{J}^z & = \int_0^{2\pi} \left[
        \frac{2}{2p + 5} I^{p + 5/2} + \frac{\cos\alpha}{p+3} I^{p+3} \sin\phi
    \right] d\phi \, ,
\end{align}
\end{subequations}
where $\bar{J}^i \equiv J^i / \Sigma_0 r_0^3 \vkick$.
\begin{itemize}[itemsep=1ex]
\item The integral in Eq.~\eqref{eq:integral-jx} is negative since $I(\phi)$ has
    a maximum at $\phi = 3\pi/2$. Therefore $J^x$ has the same sign as the kick
    angle $\alpha$.
\item The integral in Eq.~\eqref{eq:integral-jy} is zero by symmetry, i.e.
    $J^y=0$. The total bound angular momentum vector $\mathbf{J}$ lies within
    the $x$-$z$ plane, being coplanar with the kick vector and the original
    rotation axis of the disk.
\item The first term in the square bracket on the right hand side of
    Eq.~\eqref{eq:integral-jz} is the Keplerian (orbital) angular momentum
    contained within the bound region $r < r_b(\phi)$. The second term sums up
    the effect from the in-plane component of the kick, and the extra $\sin
    \phi$ factor gives a negative net contribution to the total integral. This
    opens up a possibility of yielding $J^z < 0$ depending on the exponent $p$
    and the kick angle $\alpha$.
\end{itemize}

The first panel in Fig.~\ref{fig:normalized-angular-momentum-vector} shows
numerically computed values of the total bound angular momentum $J^i$ normalized
with the total bound mass $M_b$. We denote the kick angle $\alpha$ at 15$^\circ$
intervals along each constant-$p$ curve. A perpendicular kick $\alpha=90^\circ$
yields the maximum value of $J^z/M_b$, whereas an in-plane kick $\alpha=0$
yields the minimum. For the surface mass density profiles with $p > -3/2$, it is
possible to have $J^z < 0$ if the absolute value of the kick angle $\alpha$ is
below a certain threshold, namely when the kick is directed close enough to the
disk midplane. For the surface mass density profiles more steeply decreasing
than $p = -3/2$, we see $J^z$ always remains positive.

\section{Relaxation into a tilted disk}
\label{sec:tilt-and-size}

\begin{figure*}
    \centering
    \vspace{1em}
    \includegraphics[width=0.98\linewidth]{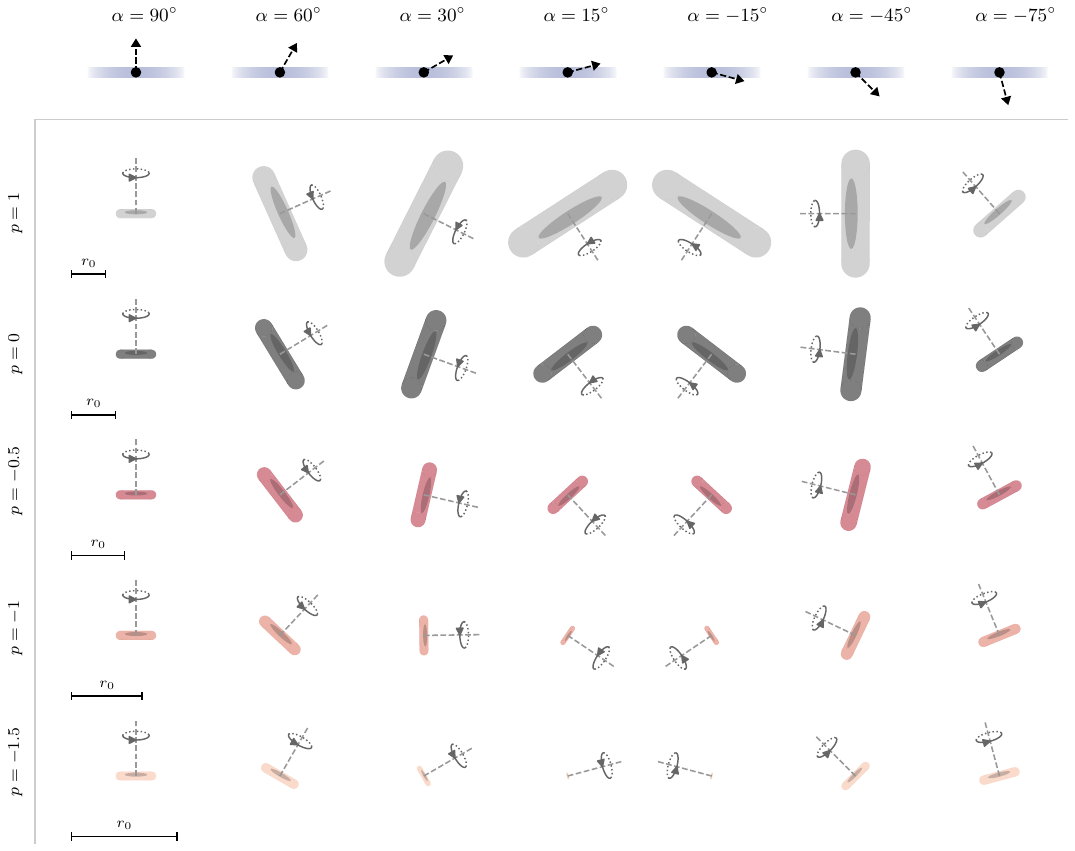}
    \vspace{1em}
    \caption{A combined illustration of the tilt angle and the circularization
        radius of the gravitationally bound gas. Shown are the cases without
        mass loss (i.e. $\epsilon=0$). See Sec.~\ref{sec:tilt-and-size} for the
        detailed discussions. The dashed lines indicate the rotation axis, and
        the size of torus is proportional to the circularization radius
        Eq.~\eqref{eq:circularization-radius-scale} rescaled in the units of
        $r_0 = GM/\vkick^2$ for each row.}
    \label{fig:disk-panels}
\end{figure*}

The total angular momentum $J^i$ of the gravitationally bound material
calculated in the previous section requires some care in its interpretation, for
we have not yet made any particular assumptions on the nature of the material
comprising the disk.

If the disk consists of collisionless or non-interacting particles, individual
particles will be excited into independent orbits with different eccentricities,
semimajor axes, and orbital inclinations
(Eq.~\ref{eq:specific-angular-momentum}; see \citealt{Akiba:2023ami} for
detailed analysis on the orbital parameters). The kick on the central object
excites a `swarm' of orbits with a broad dispersion in their orbital parameters.
If the particles evolve independently without any dynamical coupling, the total
angular momentum does not fully characterize their global dynamics, since their
orbital motions might lack a coherent feature.\footnote{In the presence of
gravitational interactions between the particles, however, nontrivial signatures
can develop in their phase space over time. See
\citet{Akiba:2021,Akiba:2023ami,Bright:2024soo,Akiba:2025} for N-body
simulations and comprehensive discussions on the kicked stellar clusters.}

If there exists a physical mechanism that can dissipate the kinetic energy of
the particles while conserving the global angular momentum, the gravitationally
bound material will be able to relax into a coherently rotating configuration
with its rotation axis aligned with the vector $J^i$ given in
Eq~\eqref{eq:total-angular-momentum-components}. In hydrodynamic flows, kinetic
energy can be dissipated and converted into fluid internal energy (heat) via
self-collision while respecting the angular momentum conservation. This suggests
that the gravitationally bound gaseous material can relax into a single
disk-like structure after several times the dynamical time scale $GM/\vkick^3$.

Based on this interpretation, we revisit the results shown in the first panel of
Fig.~\ref{fig:normalized-angular-momentum-vector} for hydrodynamic disks. For a
given surface mass density profile and a kick angle $\alpha$, we can read off
the values of $(\bar{J}^x,\bar{J}^z)/\bar{M}_b$ from the curves. We see that a
certain region in the parameter space permits $J^z \sim 0$ or $J^z < 0$,
indicating that the relaxed configuration of the bound material is heavily
tilted or rotating in retrograde. The final tilt angle relative to the original
rotation axis is plotted in the second panel of
Fig.~\ref{fig:normalized-angular-momentum-vector}. The kick angle $\alpha$ plays
a salient role in determining the overall tilt; the surface mass density
exponent $p$ and the mass loss fraction $\epsilon$ give extra degrees of
variation, but their effects are somewhat secondary. If the initial surface mass
density profile decreases more steeply than $p = -1.5$, the possible range of
the final tilt angle is limited to a finite range. Otherwise, loosely speaking,
the tilt angle is roughly twice the angle between the original rotation axis and
the kick.

The circularization radius of the bound material is
\begin{equation} \label{eq:circularization-radius}
    r_{\rm circ} \sim \frac{(J/M_b)^2}{GM} \, ,
\end{equation}
and, with $GM = r_0 \vkick^2$, we get
\begin{equation} \label{eq:circularization-radius-scale}
    \frac{r_{\rm circ}}{r_0} \sim \left(\frac{J}{M_b r_0 \vkick}\right)^2
    = \left(\frac{\bar{J}}{\bar{M}_b}\right)^2 \, ,
\end{equation}
which is the squared distance from the origin to a given data point in the first
panel of Fig.~\ref{fig:normalized-angular-momentum-vector}. We see that the
circularization radius does not vary significantly across the parameter space
and remains fairly close to $r_{\rm circ} \sim r_0$. The actual radial extent of
the circularized disk will depend on various factors such as its history of
nonlinear hydrodynamic evolution and the final density/rotation profiles.
Fig.~\ref{fig:disk-panels} illustrates the interpretations above in a pictorial
form.

Several caveats need to be noted. In case of the in-plane kick ($\alpha \simeq
0$), gravitationally bound part of the disk can immediately collide with the
outer, gravitationally unbound part of the disk. A gain or loss in the bound
material from this collision was not taken into account in our analysis.
Therefore, our estimates on the mass, size, and tilt angle of the kicked disks
are applicable only down to a finite threshold angle $\alpha_{\rm th} > 0$,
above which the gravitationally tightly bound inner region ($r\lesssim r_0$) can
geometrically avoid the collision with the outer region ($r \gtrsim r_0$). A
reasonable estimate would be several multiples of the disk aspect ratio (i.e.
$\alpha_{\rm th} > H/r_0$). Another aspect in hydrodynamic disks is the
kick-driven shockwaves
\cite[e.g.][]{Megevand:2009yx,Corrales:2009nv,Zanotti:2010xs,Meliani:2016rwn}. A
thinner disk generally requires a higher orbital Mach number, which makes the
disk more vulnerable to shocks and the consequent deviations from a simple
analytic picture we developed in Sec.~\ref{sec:analysis-infinite-disk}.

In brief, the closer the kick is directed to the midplane and the thinner the
disk is, the more deviation from the analytic estimate is induced due to
nonlinear hydrodynamic effects. Whether a heavily tilted or retrograde disk can
be formed from low kick angles, analyzed in
Sec.~\ref{sec:analysis-infinite-disk} and naively argued in this section, needs
to be verified with numerical simulations.

\section{Numerical experiment}
\label{sec:numerical-experiment}

\subsection{Initial data}

We set up a vertically stratified hydrodynamic disk as initial data. In terms of
the cylindrical coordinates $R \equiv (x^2 + y^2)^{1/2}$ and $z$, the fluid mass
density is initialized as
\begin{equation}
    \rho(R, z) = \frac{\Sigma(R)}{(2\pi)^{1/2} H}
        \exp \left(-\frac{z^2}{2H^2}\right) \, .
\end{equation}
$H$ is the disk scale height and $\Sigma (R) = \int \rho dz$ is the surface mass
density with the form
\begin{equation} \label{eq:id-surface-density}
    \Sigma(R) = \Sigma_0 \left(\frac{R}{r_0}\right)^{p}
        \inner{f} (R) \outer{f} (R) \, ,
\end{equation}
where
\begin{equation}\label{eq:disk-truncation-inner}
    \inner{f} (R) = \left[1 + \exp\left(-\frac{R - \inner{R}}{\inner{a}}\right)\right]^{-1}
\end{equation}
and
\begin{equation} \label{eq:disk-truncation-outer}
    \outer{f} (R) = \left[1 + \exp\left(\frac{R - \outer{R}}{\outer{a}}\right)\right]^{-1}
\end{equation}
are the sigmoid functions truncating the mass density at $R=\inner{r}$ and
$R=\outer{r}$ over the smoothening lengths $\inner{a}$ and $\outer{a}$
\citep{Duffell:2024fwy}. The inner truncation $f_{\rm in}$ is applied to reduce
spurious numerical behavior from the steeply diverging gravitational force
around the coordinate origin, and the outer truncation $f_{\rm out}$ is applied
to reduce contaminations from the domain boundary. The initial thermal profile
of the disk is assumed to be locally isothermal
\begin{equation} \label{eq:sound-speed-profile}
    c_s^2(R, z) = \frac{GM}{R} \left(\frac{H}{R}\right)^2 \, .
\end{equation}
The initial rotation profile $\Omega(R,z) = v_\phi/R$ is computed from the
stationary equilibrium in the thin-disk approximation
\begin{equation}
    \Omega^2 R = \frac{\partial \Phi}{\partial R} + \frac{1}{\rho}\frac{\partial P}{\partial R} \, ,
\end{equation}
with
\begin{equation}
\begin{split}
    \frac{1}{\rho} \pdev{P}{R}
    &= \pdev{c_s^2}{R} + c_s^2 \pdev{\ln \rho}{R} \, ,
\end{split}
\end{equation}
and keeping only the first correction term in $z^2/R^2$.
The scale height of the disk is set to a constant value $H/r_0 = 0.05$. We use
$(\inner{R}, \inner{a}) = (0.1, 0.02)r_0$ and $(\outer{R}, \outer{a}) = (30,
0.5)r_0$.
Once the initialization is complete, a uniform velocity offset $(-\vkick \cos
\alpha, 0, -\vkick \sin \alpha)$ is added everywhere to emulate the kick.

\subsection{Time evolution}

Simulations are performed in the rest frame of the kicked object. We use the
astrophysical fluid dynamics code AthenaK \citep{Stone:2026} to solve the ideal
Newtonian hydrodynamics equations
\begin{subequations} \label{eq:hydro-evolution}
\begin{align}
    \label{eq:hydro-evolution-mass}
    \frac{\partial \rho}{\partial t} + \nabla \cdot (\rho \mathbf{v}) &= 0, \\[1ex]
    \label{eq:hydro-evolution-momentum}
    \frac{\partial (\rho \mathbf{v})}{\partial t} + \nabla \cdot (\rho \mathbf{v} \mathbf{v} + P \mathbf{I}) &= -\rho \nabla \Phi, \\[1ex]
    \label{eq:hydro-evolution-energy}
    \frac{\partial E}{\partial t} + \nabla \cdot \left[ (E + P) \mathbf{v} \right] &= -\rho \mathbf{v} \cdot \nabla \Phi,
\end{align}
\end{subequations}
where $\rho$ is mass density, $\mathbf{v}$ is velocity, $e$ is the internal
energy density, $P$ is the gas pressure, $\mathbf{I}$ is the identity tensor, $E
\equiv \rho v^2/2 + e$ is the total energy density, and $\Phi = -GM/r$ is the
background gravitational potential of the kicked object. The gravitational mass
loss in the kicked object is not included. We adopt the ideal gas equation of
state $P = (\Gamma - 1) e$ with the adiabatic index $\Gamma = 5/3$. To
regularize the numerical behavior of $\Phi$ at the origin, we use the softening
kernel \citep{Springel:2009aa}:
\begin{equation} \label{eq:softening-kernel}
    \Phi = 
    - \frac{GM}{r} \times \begin{dcases}
        \frac{14}{5}u - \frac{16}{3}u^3 + \frac{48}{5} u^5 - \frac{32}{5}u^6
            & \text{for} \,\,\, r < r_{s}/2   \\[2ex]
        \begin{aligned}[c]
            & -\frac{1}{15} + \frac{16}{5}u - \frac{32}{3}u^3 \\
            & \quad\,\, + 16u^4 - \frac{48}{5}u^5 + \frac{32}{15}u^6
        \end{aligned}
            & \text{for} \,\,\, r_{s}/2 \leq r < r_{s}  \\[2ex]
        1 & \text{for} \,\,\, r \geq r_{s} \, ,
    \end{dcases}
\end{equation}
where $u\equiv r/r_s$ and $r = \sqrt{R^2 + z^2} = \sqrt{x^2 + y^2 + z^2}$ is the
spherical radius. We use the softening length $r_s = 0.05 r_0$, and do not apply
any gas removal schemes (i.e. sink particle prescription).

The computational domain has the size $[-40r_0, 40r_0]^3$ with the base grid
resolution $320^3$. Four levels of nested static mesh refinement are applied
toward the coordinate origin. The finest grid level covers $[-2.5r_0, 2.5r_0]^3$
with the resolution of 64 grid points per $r_0$. Time integration is performed
with a second order Runge-Kutta stepper using the piecewise parabolic
reconstruction \citep{Colella:1982ee} and HLLC Riemann fluxes
\citep{Harten1983,Einfeldt1991}. The density atmosphere is set to $\rho_{\rm
floor} = 10^{-8}$. No-inflow (diode) boundary condition is imposed at the
external boundary.

\subsection{Results}
\label{sec:results}

\begin{figure*}
\centering
\subcaptionbox{$\alpha = 90^\circ$\label{fig:3d-p00-alphas-a90}}{
\includegraphics[width=0.328\linewidth]{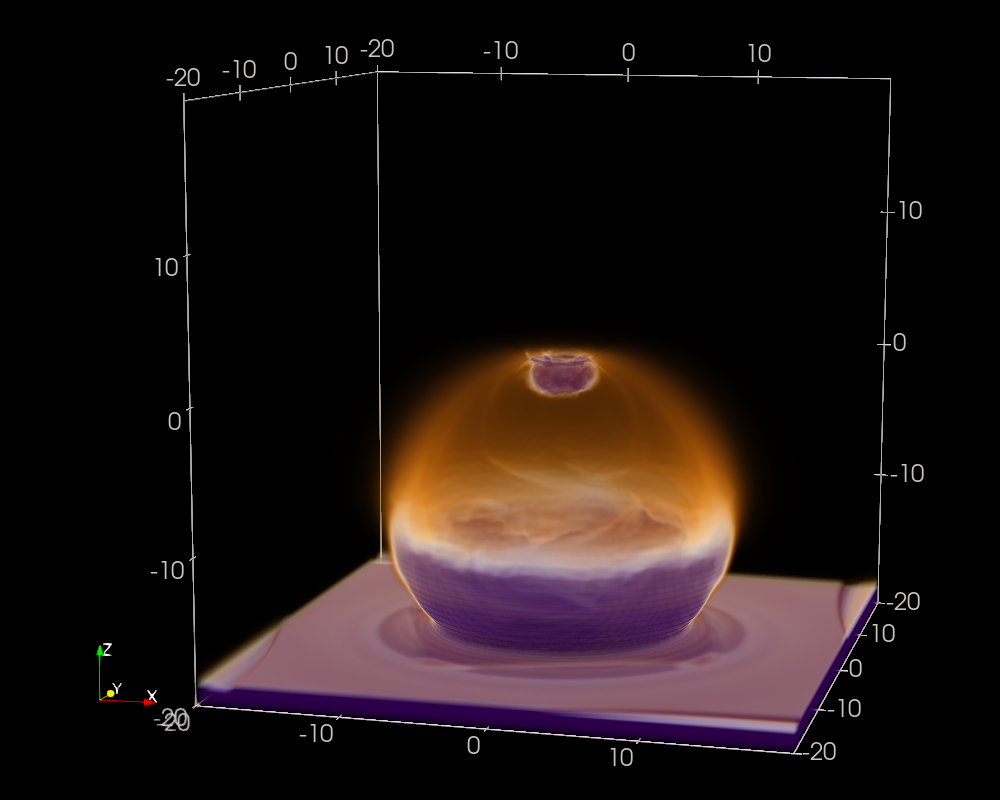}
\includegraphics[width=0.328\linewidth]{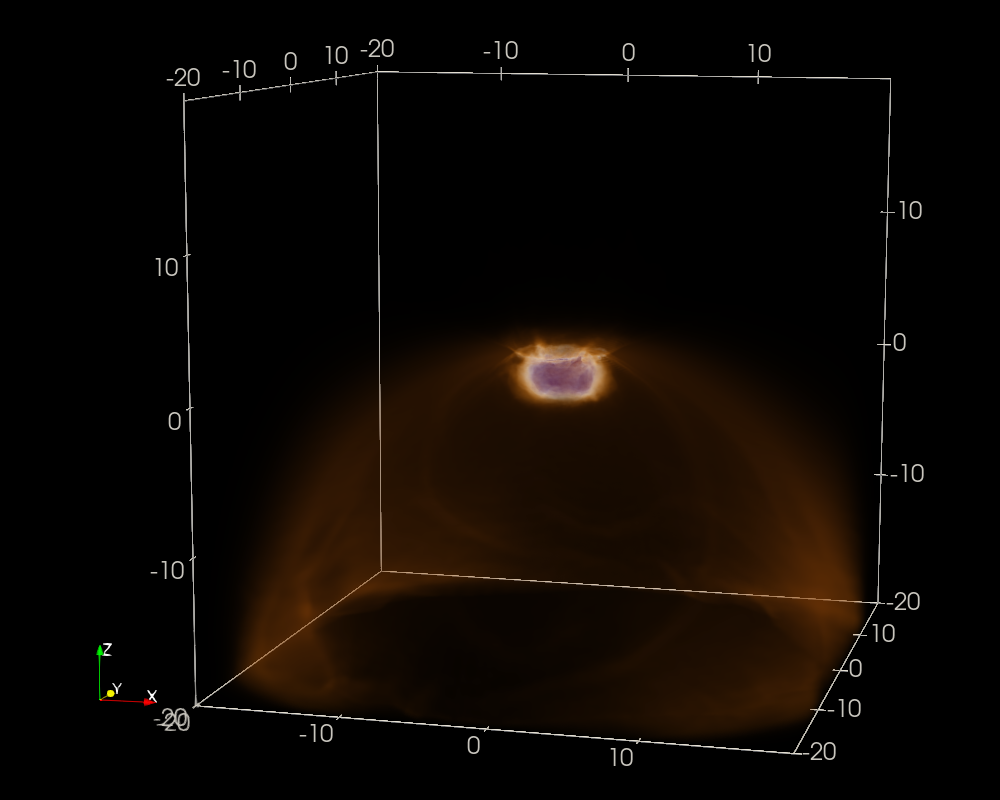}
\includegraphics[width=0.328\linewidth]{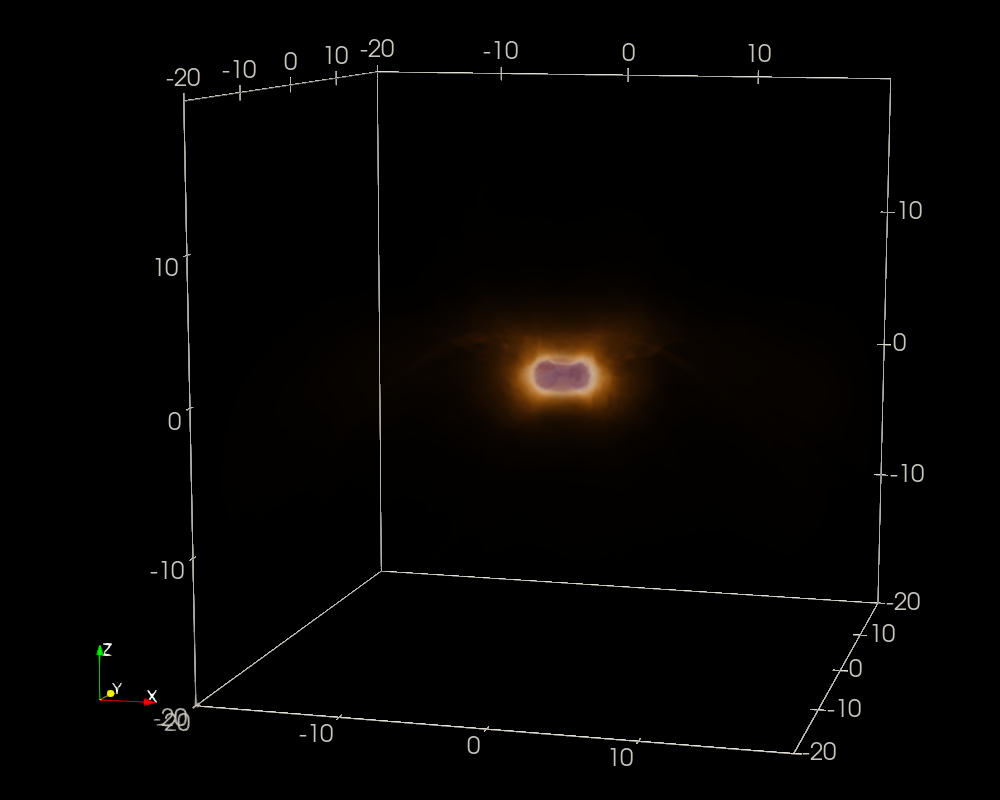}
}\\[1em]
\subcaptionbox{$\alpha = 45^\circ$\label{fig:3d-p00-alphas-a45}}{
\includegraphics[width=0.328\linewidth]{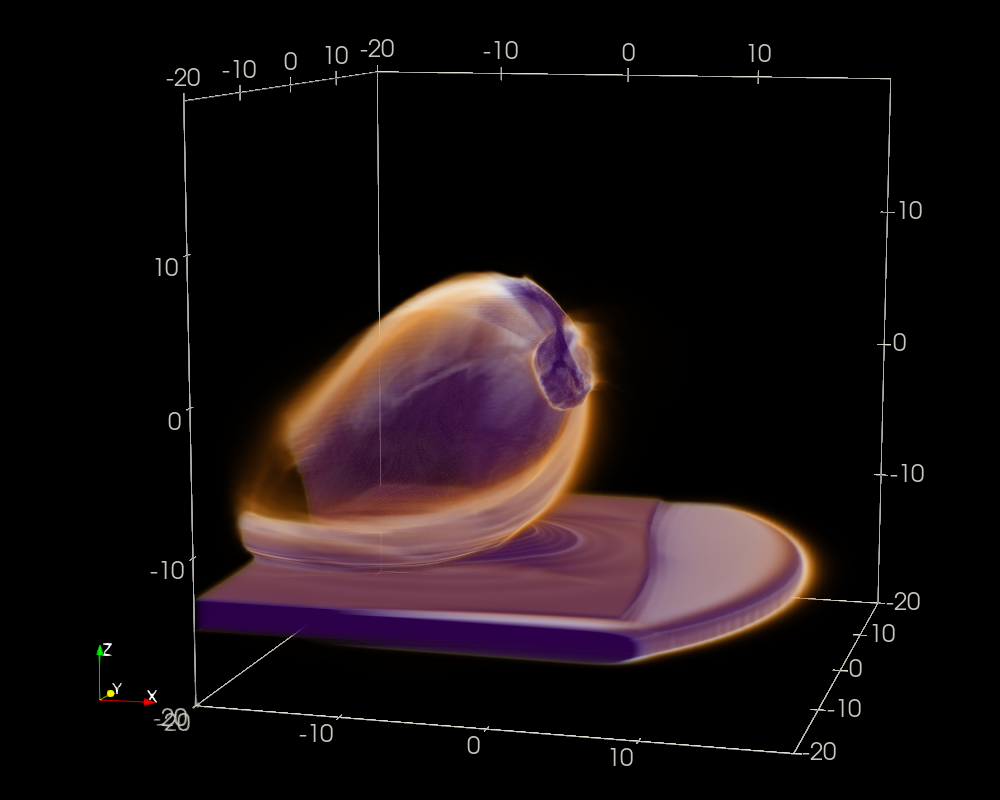}
\includegraphics[width=0.328\linewidth]{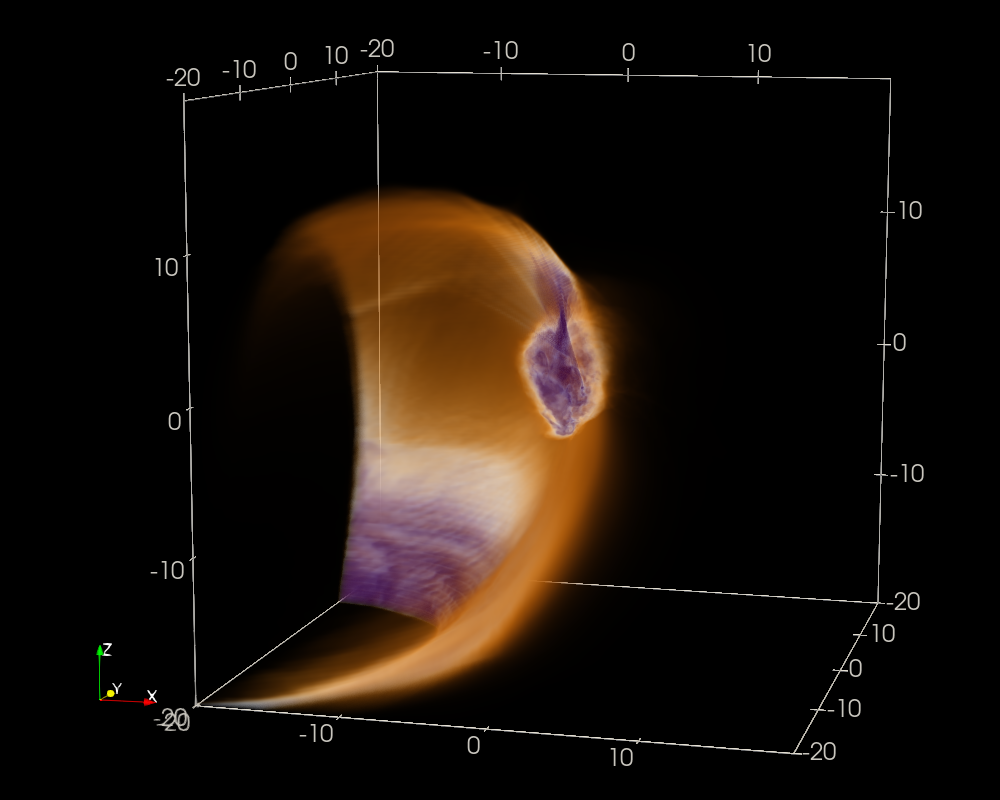}
\includegraphics[width=0.328\linewidth]{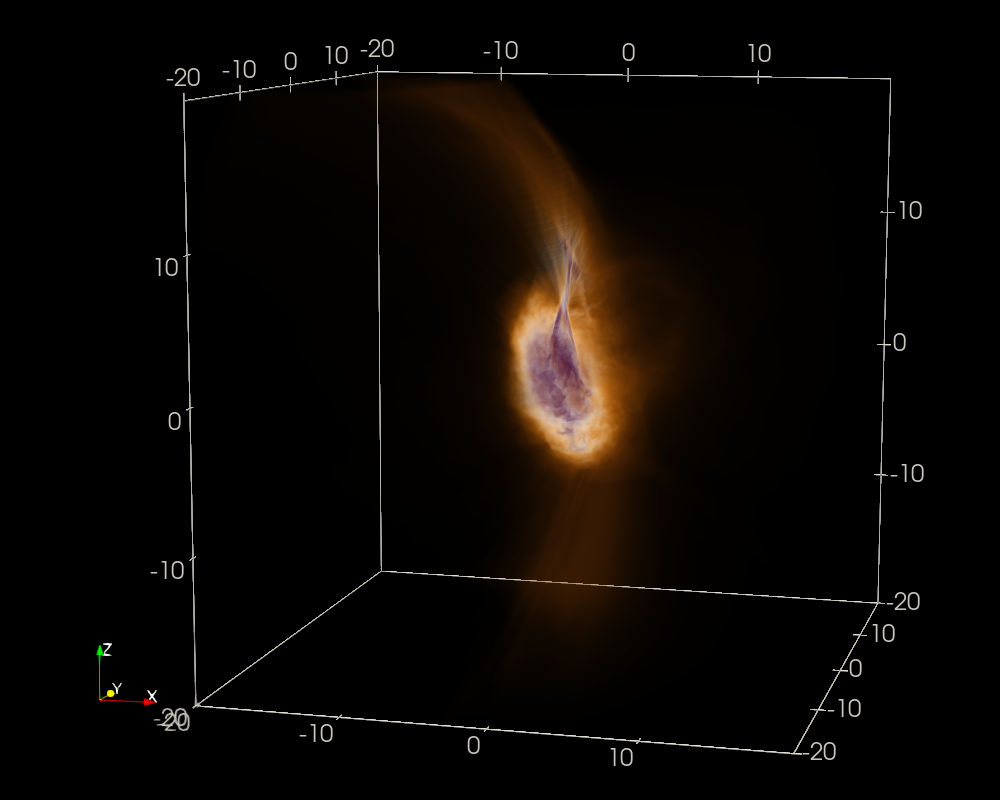}
}\\[1em]
\subcaptionbox{$\alpha = 15^\circ$\label{fig:3d-p00-alphas-a15}}{
\includegraphics[width=0.328\linewidth]{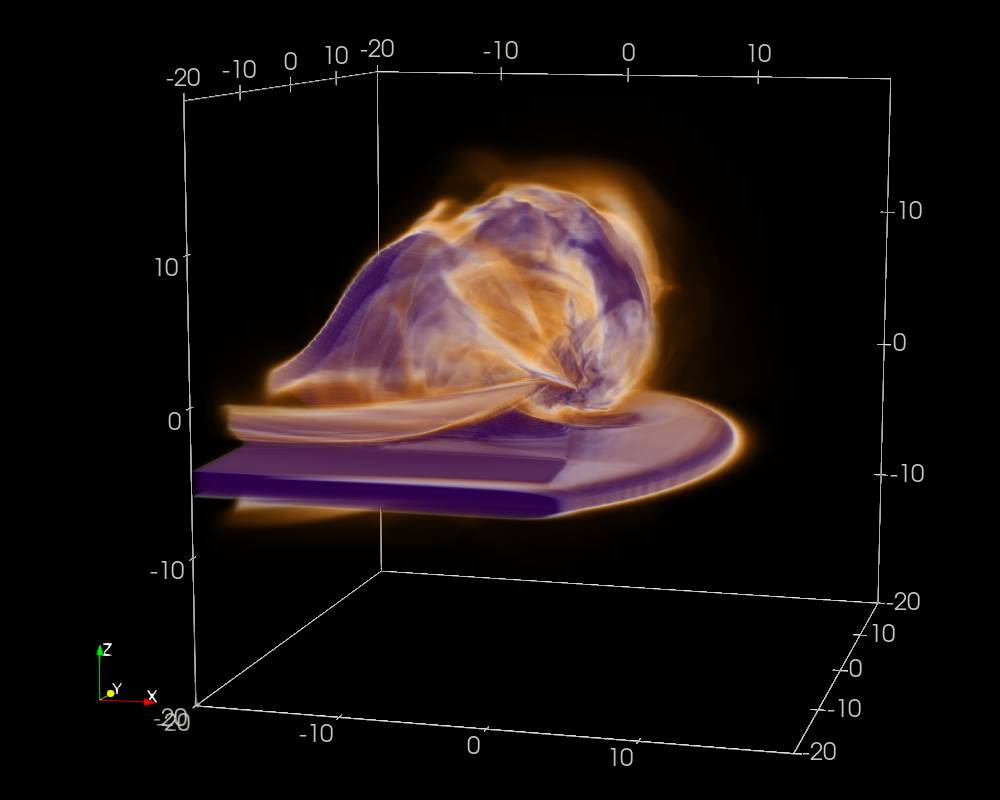}
\includegraphics[width=0.328\linewidth]{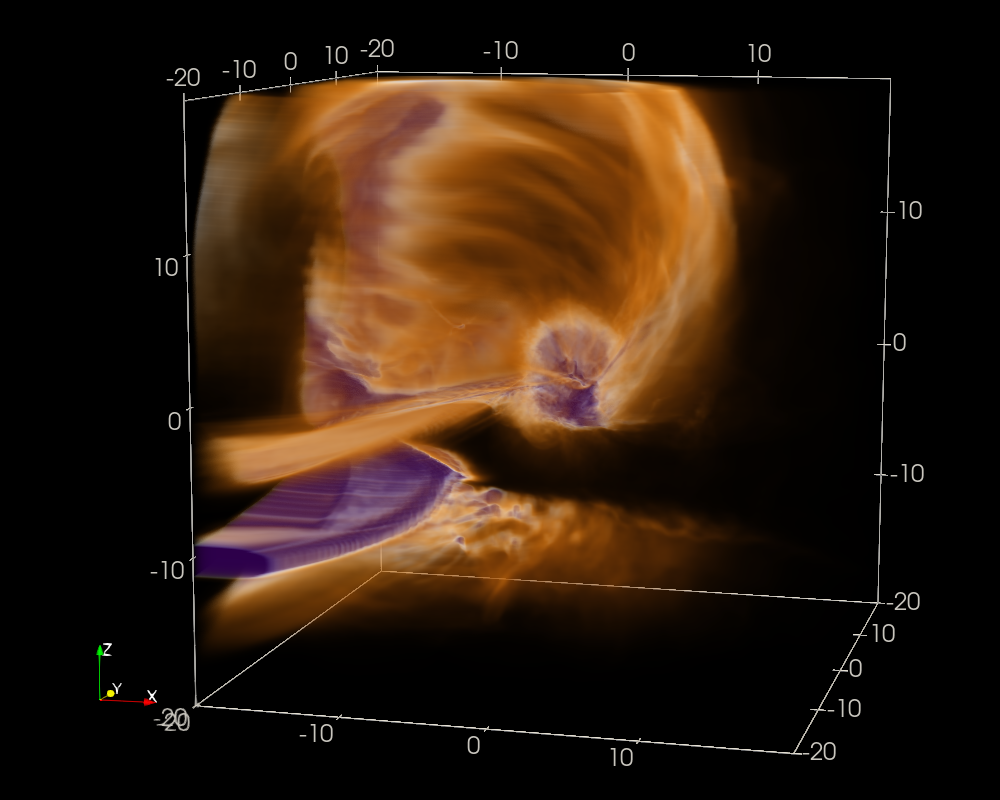}
\includegraphics[width=0.328\linewidth]{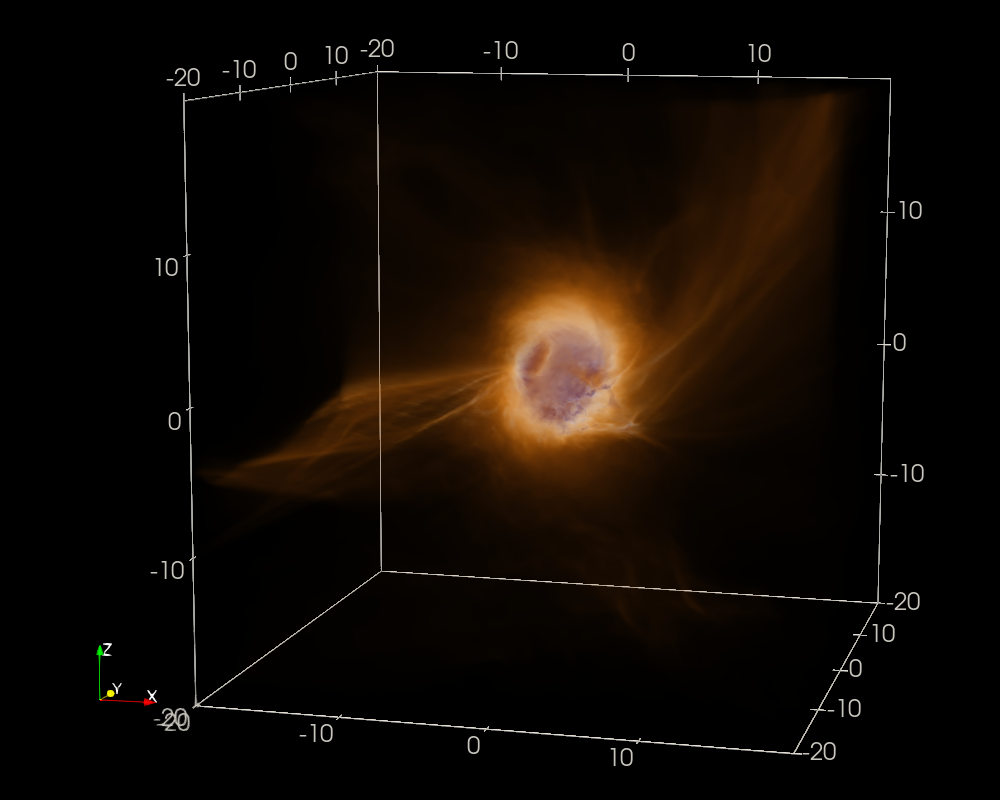}
}
\caption{Volume rendering of the gas mass density from the simulations of
    constant-surface mass density ($p = 0$) disks with the kick angles
    $\alpha=90^\circ$ (top), $45^\circ$ (middle), and $15^\circ$ (bottom). Each
    column shows the simulation time $\bar{t} = 20$ (left), $\bar{t} = 40$
    (center), and $\bar{t} = 100$ (right). Grid axes are in the units of $r_0 =
    GM/\vkick^2$.}
\label{fig:3d-p00-alphas}
\end{figure*}

Simulations are carried out in dimensionless code units $r_0 = \vkick = \Sigma_0
= 1$, which also sets $GM = 1$. Throughout this section, we quote the radius and
simulation times in the units of $\bar{r}^i \equiv r^i / r_0$ and $\bar{t}
\equiv t / (r_0 \vkick^{-1}) = t/(GM/\vkick^3)$. Other physical quantities with
the dimension of mass, energy, and angular momentum can be recovered from the
code units as
\begin{align}
    \bar{m} & \equiv m / \Sigma_0 r_0^2 \,, \\
    \bar{E} & \equiv E / \Sigma_0 r_0^2 \vkick^2 \,, \\
    \bar{J} & \equiv J / \Sigma_0 r_0^3 \vkick \, .
\end{align}

We take the constant-surface mass density profile ($p=0$) as a baseline and
perform simulations for six different kick angles: $\alpha = 15^\circ$,
$22.5^\circ$, $30^\circ$, $45^\circ$, $60^\circ$, and $90^\circ$. The exact
in-plane kick ($\alpha=0^\circ$) was avoided since it leads to a continued
collision between gravitationally bound and unbound material, which makes the
final outcome highly dependent on the arbitrary choice of the outer truncation
radius $\outer{r}$. In order to study the effect of the non-constant surface
mass density, we perform additional simulations with the $p=1$ and $p=-1$
profiles for three kick angles $\alpha=30^\circ$, $60^\circ$, and $90^\circ$.
All simulations are performed up to $\bar{t}_f=100$. 

We discuss the results from $p=0$ simulations in detail in
Sec.~\ref{sec:results-evolution-overview}--\ref{sec:warp}, and summarize the
$p=\pm1$ results in Sec.~\ref{sec:results-ppm1}.

\subsubsection{Evolution overview}
\label{sec:results-evolution-overview}

In Fig.~\ref{fig:3d-p00-alphas}, we show three-dimensional views of the gas mass
density at $\bar{t} = 20$, 40, and 100 for selected kick angles from $p=0$
simulations. We outline the hydrodynamic evolution as below.
\begin{itemize}[itemsep=1ex]
\item The inner region $r \lesssim r_0$ undergoes a rapid reorientation up to
    $\bar{t}\simeq 10$, followed by a more gradual but continued increase of the
    tilt. The inner region relaxes close to its final state by $\bar{t} \simeq
    30$, namely after several Keplerian orbital periods at $r=r_0$.
\item The marginally bound region in the vicinity of $r = r_b(\phi)$ expands
    into a long diffuse ejecta that exhibits a characteristic "$\Omega$"-shaped
    morphology
    \cite[e.g.][]{Lippai:2008fx,Shields:2008va,Rossi:2009nk,Akiba:2023ami,Kim:2026qad}.
    The inner part of the ejecta falls back and forms a diffuse torus
    surrounding the tightly bound inner disk, while the outer part of the ejecta
    continues to expand and leaves the computational domain with the unbound
    material.
    The fallback ejecta often forms a filamentary streamer connected to the
    inner disk; see the panels in the right column of
    Fig.~\ref{fig:3d-p00-alphas}.
\item The gravitationally unbound, outer region $r> r_b(\phi)$ is decoupled and
    drifts away. A cylindrical shock is launched onto this unbound portion of
    the disk, with a higher degree of asymmetry for lower kick angles (e.g. see
    the first panels of Fig.~\ref{fig:3d-p00-alphas-a90} and
    Fig.~\ref{fig:3d-p00-alphas-a45}). As a result, the unbound disk advects
    away with an expanding central cavity on it.
\end{itemize}

The shock heating and the dissipation of gas kinetic energy lead to a
geometrically thick morphology of the bound gas, especially given that our
simulations are performed adiabatically using the ideal gas equation of state.
As previously noted, kicks directed close to the midplane can make the inner
bound region and the outer unbound region immediately collide with each other,
leading to a more violent launching of the weakly bound ejecta and stripping of
the bound mass. Although the smallest kick angle $\alpha=15^\circ$ ($=
0.26\,\text{rad}$) explored is several times larger than the disk scale height
$H/r_0 = 0.05$, we still see copious shock heating and mass stripping.

\subsubsection{Mass, energy dissipation, and angular momentum}

\begin{figure}
    \centering
    \includegraphics[width=0.98\linewidth]{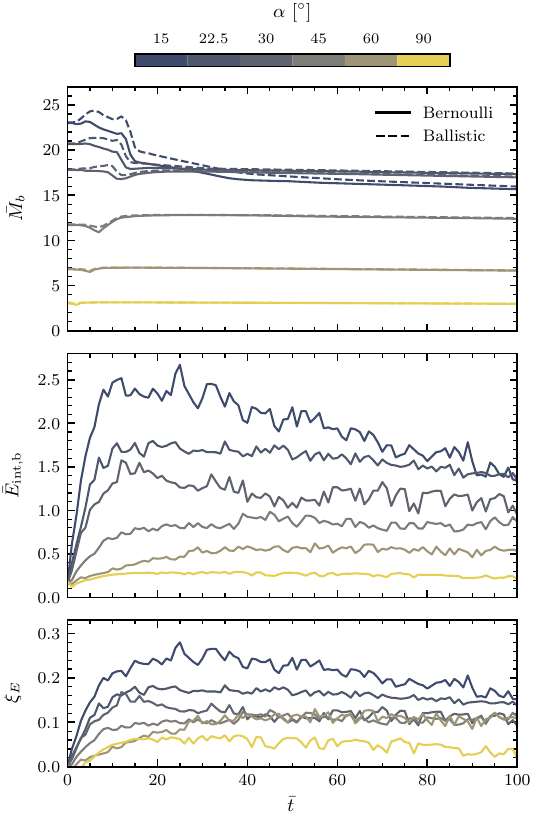}
    \caption{Mass (top), internal energy (middle) and the energy dissipation
        efficiency (bottom, see Eq.~\ref{eq:energy-dissipation-efficiency} for
        definition) of the bound material in the $p=0$ simulations. The
        Bernoulli criterion (Eq.~\ref{eq:bernoulli-criterion}) is used to filter
        out gravitationally bound fluid elements in the simulation volume data.
        In the top panel, we also show the total bound mass computed with the
        ballistic boundness criterion $v^2/2 + \Phi < 0$.}
    \label{fig:p00-mass-eint-bound}
\end{figure}

\begin{figure}
    \centering
    \includegraphics[width=0.98\linewidth]{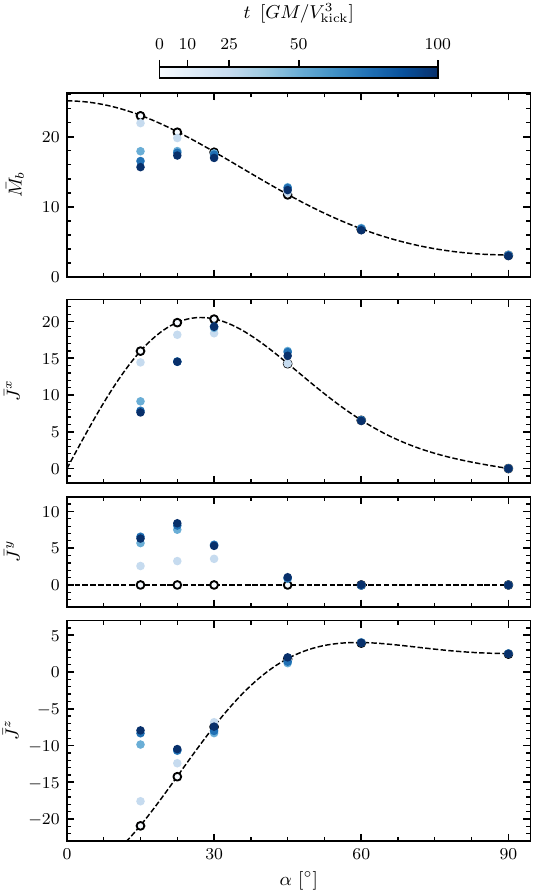}
    \caption{The total mass and angular momentum of the bound gas from the
        simulations (markers) and the analytic estimates (dashed curves) for the
        $p=0$ disk. Simulation data are shown for the times $\bar{t} \in \{0,
        10, 25, 50, 100\}$.}
    \label{fig:comparison-p00}
\end{figure}

\begin{figure*}
\centering
    \subcaptionbox{$\alpha=90^\circ$ \label{fig:p00-a90}}
    {\includegraphics[width=0.33\linewidth]{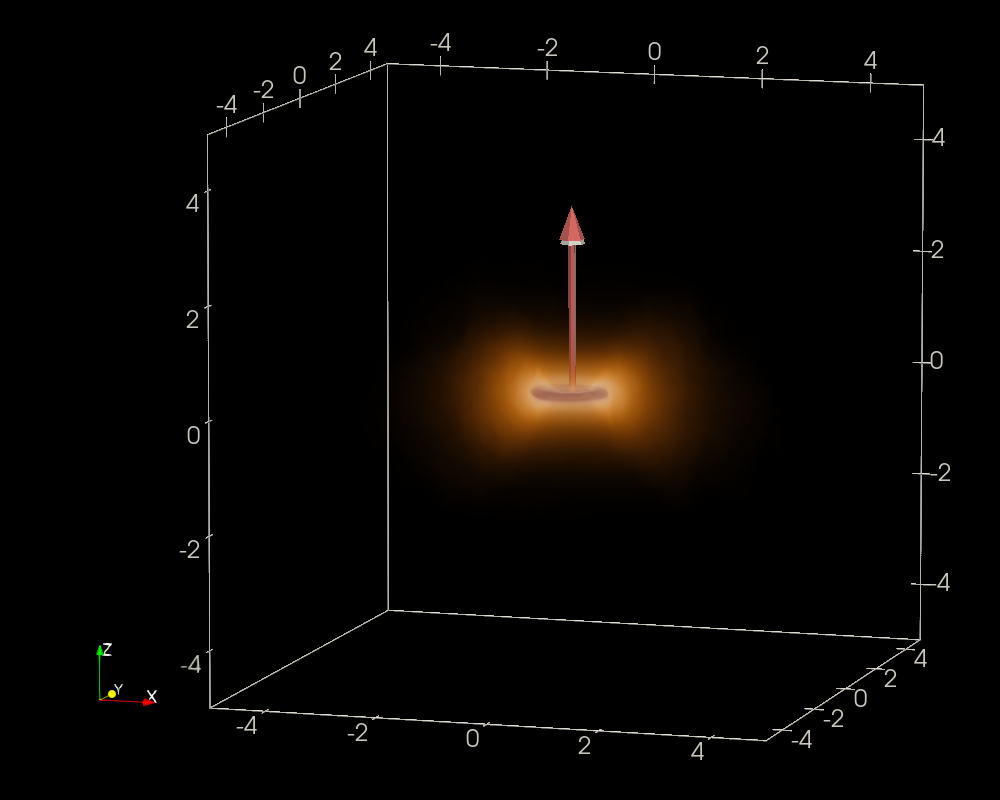}}
    \subcaptionbox{$\alpha=60^\circ$ \label{fig:p00-a60}}
    {\includegraphics[width=0.33\linewidth]{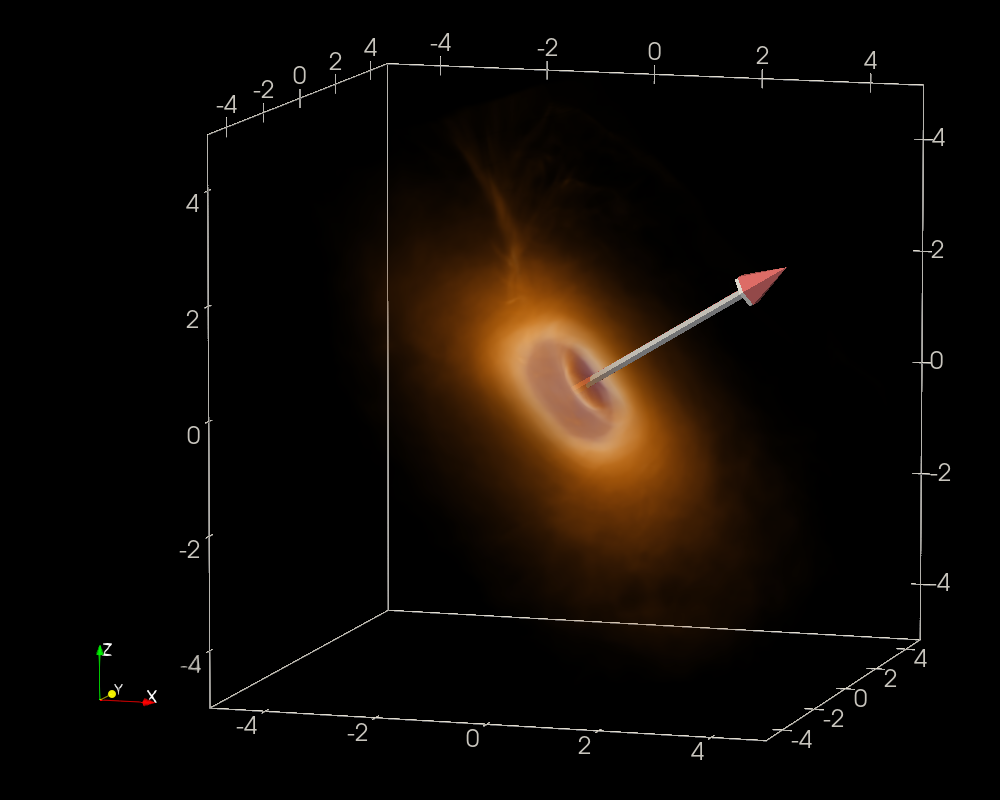}}
    \subcaptionbox{$\alpha=45^\circ$ \label{fig:p00-a45}}
    {\includegraphics[width=0.33\linewidth]{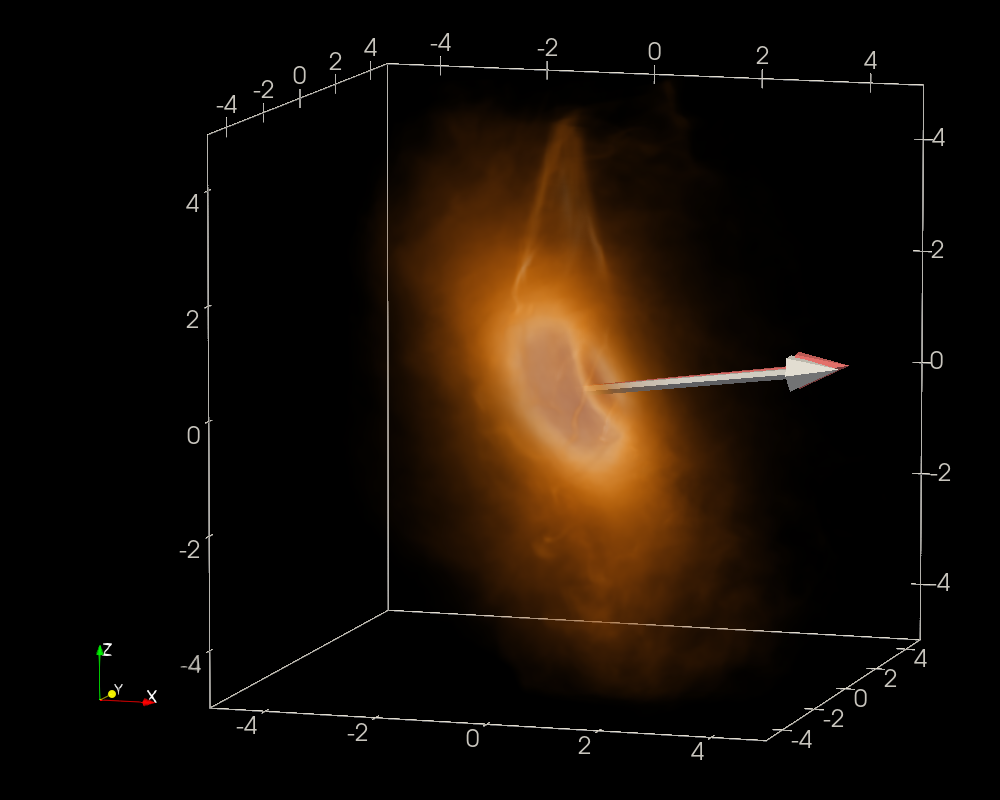}} \\[1em]
    \subcaptionbox{$\alpha=30^\circ$ \label{fig:p00-a30}}
    {\includegraphics[width=0.33\linewidth]{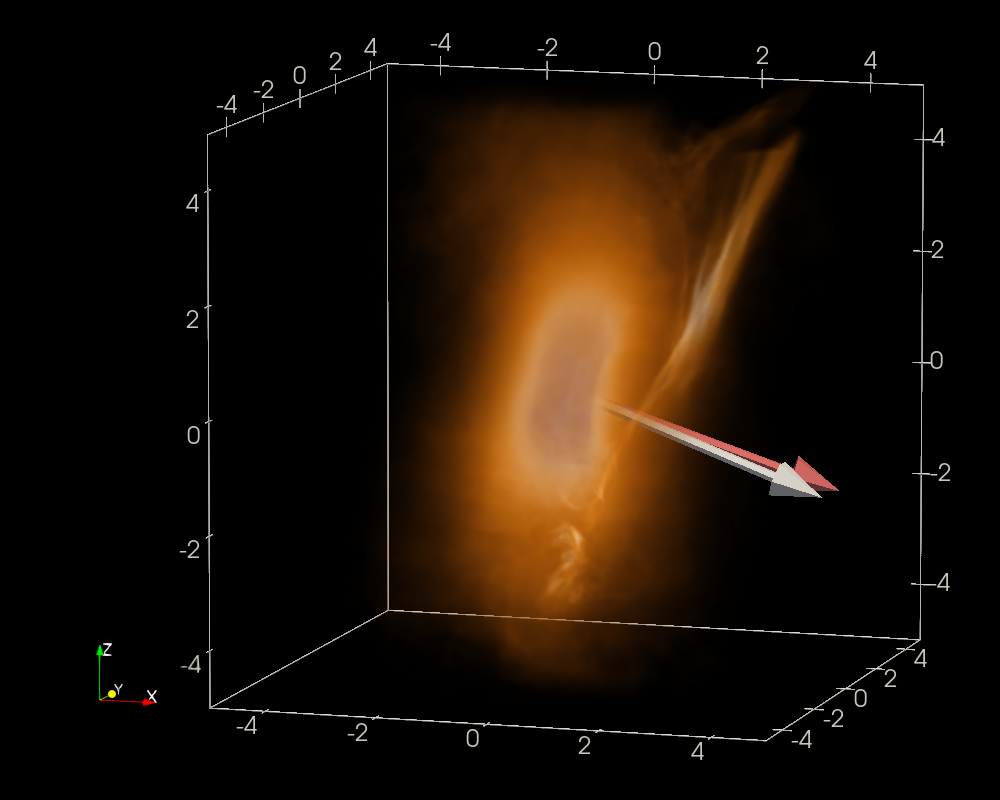}}
    \subcaptionbox{$\alpha=22.5^\circ$ \label{fig:p00-a22.5}}
    {\includegraphics[width=0.33\linewidth]{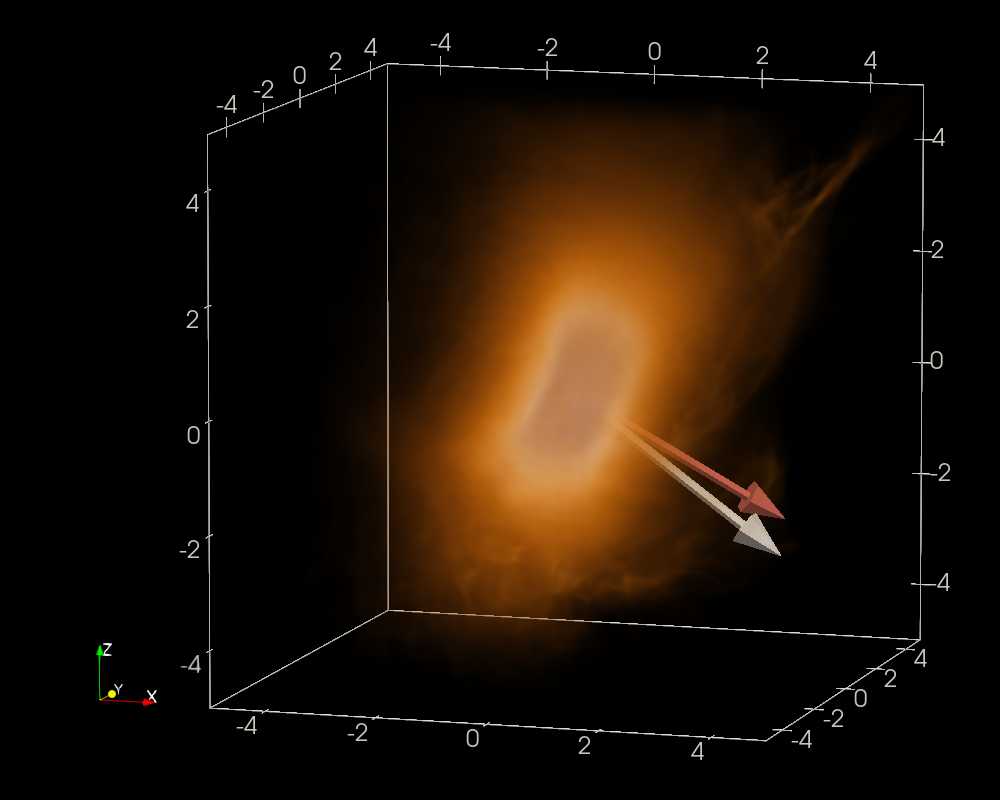}}
    \subcaptionbox{$\alpha=15^\circ$ \label{fig:p00-a15}}
    {\includegraphics[width=0.33\linewidth]{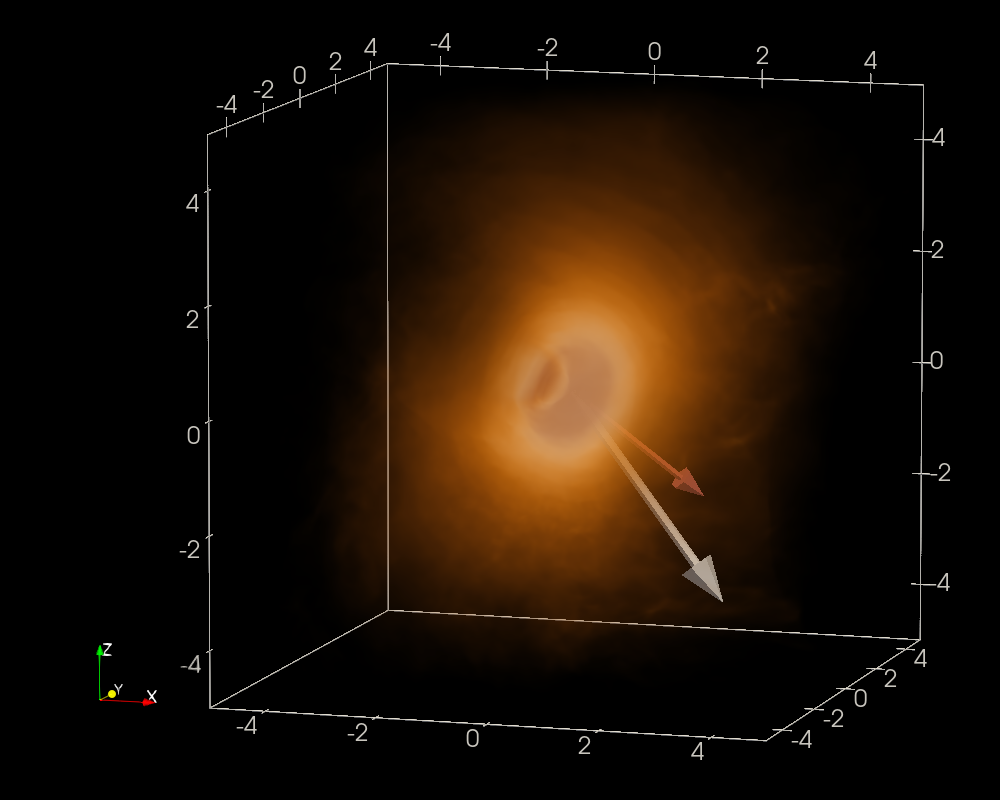}} \\[1em]
    \caption{Volume rendering of the mass density at $\bar{t}_f=100$ in the
        $p=0$ simulations. Colormap range and opacity are set different from
        Fig.~\ref{fig:3d-p00-alphas} to better expose the inner torus-like
        structure. Arrows in each panel denote the angular momentum of the bound
        material normalized with its mass (i.e. $\bar{J}^i/\bar{M}_b$) from the
        analytic estimate (white, cf. Fig.~\ref{fig:disk-panels}) and from the
        simulations (red).}
    \label{fig:3d-p00-final-tilt}
\end{figure*}

Fig.~\ref{fig:p00-mass-eint-bound} shows the time variation of the total mass
and total internal energy of the bound material, providing a quantitative
picture of the hydrodynamic evolution described above. For masking the
gravitationally bound fluid elements in the simulation volume data, we adopt the
Bernoulli criterion
\begin{equation} \label{eq:bernoulli-criterion}
    \frac{1}{2}v^2 + h + \Phi < 0 \, ,
\end{equation}
where $h = (e+P)/\rho$ is the specific enthalpy. In the upper panel of
Fig.~\ref{fig:p00-mass-eint-bound}, we included the total bound mass estimated
with the ballistic boundness criterion $v^2/2 + \Phi < 0$ for comparison. While
the Bernoulli and ballistic criteria show a good agreement overall, we see a
slight discrepancy in $\bar{t} \lesssim 20$, which is due to the active
dissipation of the gas kinetic energy as the disk rearranges itself into a new
orientation. This dissipative heating manifests as a heightened contribution of
the enthalpy term $h$ in the Bernoulli criterion
Eq.~\eqref{eq:bernoulli-criterion}, increasing the discrepancy between the two
criteria. It is more pronounced for lower kick angles which are subject to more
severe self-collision and shock heating.
The middle panel of Fig.~\ref{fig:p00-mass-eint-bound} shows the total internal
energy of the bound material. Lower kick angles show stronger heating and
confirm this interpretation. The total internal energy rapidly rises until
$\bar{t} \sim 10$, then slowly declines over time as the bound material relaxes
into a new equilibrium.

The dissipation of the kinetic energy in the bound material can source
electromagnetic flares in the post-kick phase. The total dissipated energy
$E_{\rm flare}$ and the time scale of the flare $t_{\rm flare}$ can be
parametrized as \citep{Shields:2008va}
\begin{align}
    \label{eq:shields-parametrization-energy}
    E_{\rm flare} &= \xi_E \left(\frac{1}{2} M_b \vkick^2 \right) \, , \\[1ex]
    \label{eq:shields-parametrization-time}
    t_{\rm flare} &= \xi_t \left(\frac{2\pi r_0}{\vkick}\right)
        = \xi_t \left(2\pi \frac{GM}{\vkick^3}\right) \, .
\end{align}
We estimate the energy dissipation efficiency $\xi_E$ from our simulations as
\begin{equation} \label{eq:energy-dissipation-efficiency}
    \xi_E = \frac{\bar{E}_{\mathrm{int, b}} (t) - \bar{E}_{\mathrm{int, b}} (0)}{\bar{M}_b/2} \, \, .
\end{equation}
Strictly speaking, Eq.~\eqref{eq:energy-dissipation-efficiency} includes not
only the dissipative heating but also the contribution from adiabatic
compression and expansion of the bound material. We do not make a distinction
between these two contributions and only monitor the total variation in $E_{\rm
int, b}$.

The bottom panel of Fig.~\ref{fig:p00-mass-eint-bound} shows the result. Our
simulations show the dissipation time scale $\xi_t \approx 2$. The dissipation
efficiency $\xi_E$ shows a larger dependence on the kick angle, ranging from
$\xi_E \approx 0.05$ ($\alpha=90^\circ$) to $\xi_E \approx 0.25$
($\alpha=15^\circ$).
A previous study approximating the kicked disk as a collection of collisionless
particles obtained $\xi_E \approx 1$ and $\xi_t \approx 2$ for $\alpha=45^\circ$
kick \citep{Shields:2008va}. Our hydrodynamic simulation with the same kick
angle shows about an order of magnitude smaller value of the efficiency and a
similar time scale of dissipation.

In Fig.~\ref{fig:comparison-p00}, we show the total mass and angular momentum of
the bound material computed with the Bernoulli criterion. Data points are
color-coded from light (earlier) to dark (later) tones to indicate their change
in time. The dashed black curves are the analytic estimates from
Eq.~\ref{eq:integral-bound-mass} and
Eq.~\ref{eq:total-angular-momentum-components}. For all kick angles, the total
bound mass and angular momentum at $\bar{t}=0$ are very close to the analytic
estimate, which is not surprising since the initial rotation profile of the disk
is nearly Keplerian with only small corrections.

We see that the final values of the total bound mass and angular momentum
generally agree well with the analytic estimates for $\alpha \geq 45^\circ$,
where the deviations due to hydrodynamic effects manifest as the kick is
directed closer to the midplane. The difference in the total bound mass at the
end of the simulation amounts to less than about 5\% for $\alpha\geq 45^\circ$,
while reaching $\approx$30\% for $\alpha=15^\circ$. The loss in the total
angular momentum $J^i$ results from a combination of (i) the stripping of the
bound mass itself and (ii) the torque exerted on the bound material in the early
phase of the evolution during which the bound and unbound regions of the disk
collide. Notably, the final state picks up nonzero $y$-component of the angular
momentum $J^y > 0$.
Following the notation in Fig.~\ref{fig:problem-schematic}, this could be
qualitatively understood as follows. The $x>0$ side of the bound region is
subject to the grazing collision with the unbound region neighboring on the same
side, and experiences a net force directed opposite to the kick vector
$\mathbf{V}_{\rm kick}$. This results in acquiring the angular momentum $(r_0
\mathbf{\hat{x}}) \times (-\vkick \cos \alpha \mathbf{\hat{x}} - \vkick \sin
\alpha \mathbf{\hat{z}}) = r_0 \vkick \sin \alpha \, \mathbf{\hat{y}}$,
yielding a positive $y$ component. The $x<0$ side of the bound region, in
contrast, smoothly moves away from the unbound region that it is facing and does
not experience such torque.

In Fig.~\ref{fig:3d-p00-final-tilt}, we show the close-up 3D view of the
gravitationally bound material at $\bar{t}_f = 100$. The mass-normalized angular
momentum vector $\bar{J}^i/\bar{M}_b$ is shown with the arrows. The white arrow
shows the analytic prediction (cf.
Fig.~\ref{fig:normalized-angular-momentum-vector} and Fig.\ref{fig:disk-panels})
and the red arrow shows the simulation results. In spite of the nonlinear
hydrodynamic effects discussed so far, our simulations show that a large tilt or
retrograde rotation ($J^z<0$) can still occur for low kick angles.

\subsubsection{Transient warp}
\label{sec:warp}

\begin{figure}
    \centering
    {\includegraphics[width=\linewidth]{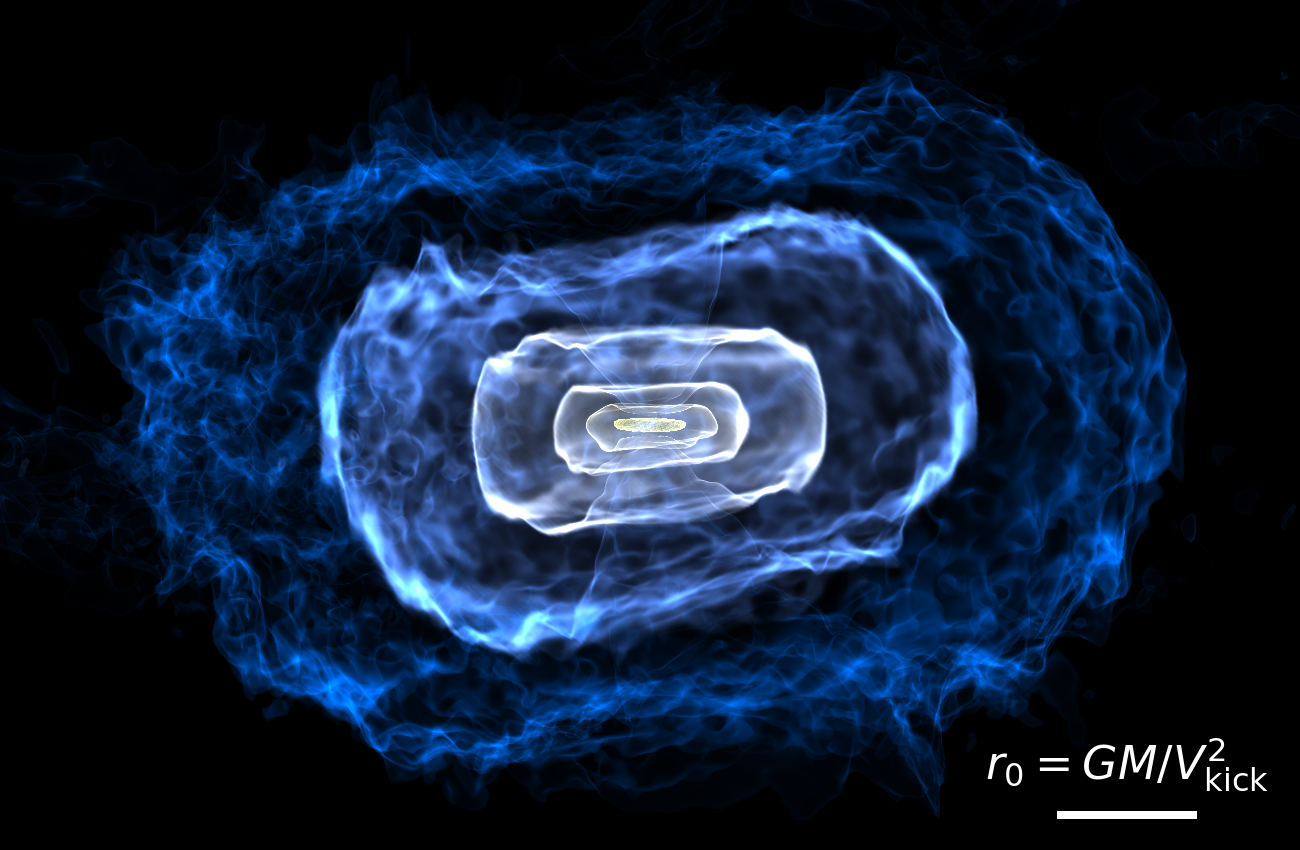}}
    \caption{The edge-on anatomy of the bound material at the end of the
        simulation ($\bar{t}_f = 100$) with $p=0$ and $\alpha = 15^\circ$.
        Iso-density surfaces (color contours) reveal a radial variation in the
        disk orientation.}
    \label{fig:3d-warp}
\end{figure}

\begin{figure*}
    \centering
    \includegraphics[width=0.98\linewidth]{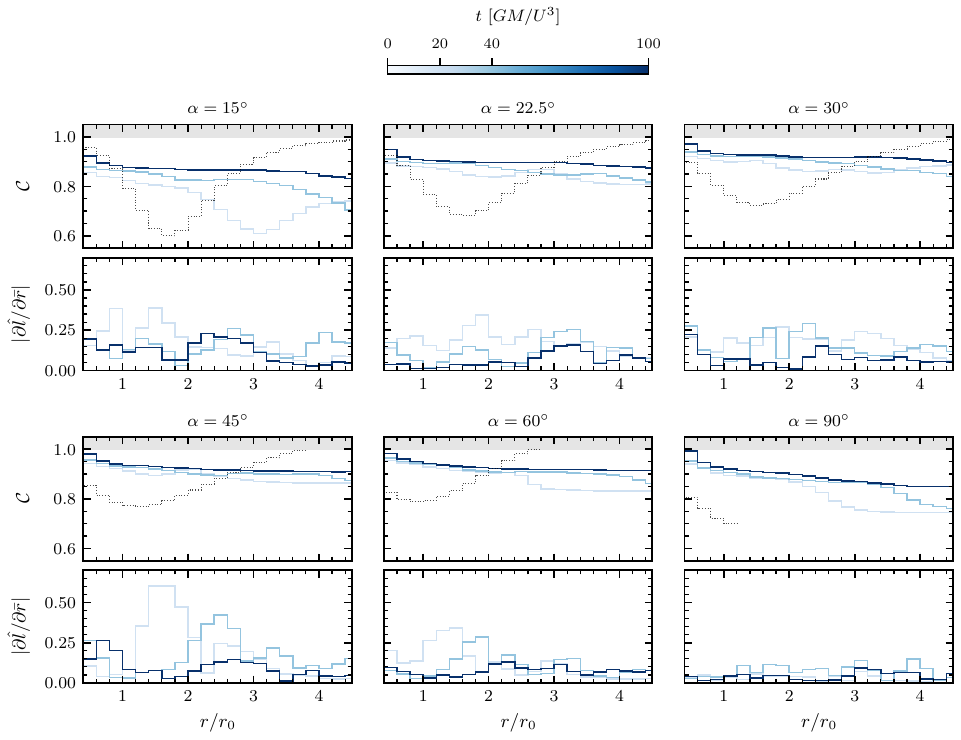}
    \caption{The rotation coherence parameter $\mathcal{C}(r)$
        (Eq.~\ref{eq:coherence-definition}) and the tilt gradient $|\partial
        \hat{l}/\partial \bar{r}|$ of the gravitationally bound material within
        $0.5 \leq r/r_0 \leq 4.5$ at $\bar{t} = 20$, 40, 100 in the $p=0$
        simulations. The coherence parameter $\mathcal{C}$ at $\bar{t}=0$ is
        shown with the gray dashed lines in each top subpanel.}
    \label{fig:coherence-and-warp-p00}
\end{figure*}

The scaling $\vkick/v_K(r) = (r/r_0)^{1/2}$ implies the outer part of the disk
`feels' the recoil more strongly. Such a differential influence over radius
suggests the radius-dependent tilt, namely warp can develop in obliquely kicked
disks.

Apparent mismatch between the orientation of the iso-density surface and the
total bound angular momentum vector, for example in Fig.~\ref{fig:p00-a45} (see
also Fig.~\ref{fig:pm10-a60}), indicates that a misalignment of the angular
momentum between the inner and outer radii of the bound material is still
present at the end of the simulations. Fig.~\ref{fig:3d-warp} shows an edge-on
view of the mass density from the $p=0$, $\alpha=15^\circ$ simulation at
$\bar{t}_f = 100$. The camera is aligned with the innermost part of the disk
($r<r_0$ in the figure). We observe that the outer region ($r>r_0$) shows a
progressive tilt relative to the innermost part.

The warp in a disk is characterized as a radius-dependent tilt of the disk
midplane. A quantitative measure of the warp commonly adopted in the accretion
disk literature is the dimensionless warp amplitude
$\psi = r |\partial \hat{l} / \partial r| = |\partial \hat{l} / \partial \ln r|$
\cite[e.g.][]{Ogilvie:1998js,Lodato:2007hf}, where $\hat{l}(r)$ is the unit
angular momentum vector of the disk annulus at radius $r$. We instead simply use
the magnitude of the tilt gradient
\begin{equation} \label{eq:tilt-gradient}
    \bigg| \pdev{\hat{l}}{r} \bigg|
        = r_0^{-1} \bigg| \pdev{\hat{l}}{\bar{r}} \bigg|
        = \frac{\psi}{r}
\end{equation}
to gauge the variation in the tilt angle over radius, since we are only
interested in examining the presence of warp near $r\sim r_0$.
We divide the simulation domain into multiple spherical shells, and compute the
unit vector $\hat{l}(r)$ from the total angular momentum of the bound material
contained in each shell.
Additionally, we compute
\begin{equation} \label{eq:coherence-definition}
    \mathcal{C} \equiv \frac{\left\lvert\int d\mathbf{J}\right\rvert}{\int \left\lvert d\mathbf{J} \right\rvert} \, ,
\end{equation}
for each spherical shell, which measures a directional coherence of the angular
momentum. If the motion of the fluid within a shell is well aligned to a single
rotation axis, Eq.~\eqref{eq:coherence-definition} yields $\mathcal{C} \approx
1$, and if the motion is highly dispersed or randomized, it gives $\mathcal{C}
\approx 0$. For instance, a separately performed simulation evolving the initial
condition without any kick maintains $\mathcal{C}(r) \gtrsim 0.99$ throughout
$\bar{t}_f = 100$.

Interpretation of the quantity $|\partial \hat{l}(r) / \partial r|$, as well as
the dimensionless warp amplitude $\psi(r)$, as a measure of warp in a rotating
fluid body is valid only if the coherence measure $\mathcal{C}(r)$ is
sufficiently high. In other words, the motion of the fluid contained within a
given spherical shell should be coherent enough that its rotation axis and
midplane could be identified.

Fig.~\ref{fig:coherence-and-warp-p00} plots the rotation coherence measure
$\mathcal{C}(r)$ and the tilt gradient $|\partial \hat{l} / \partial \bar{r}|$
from $p=0$ simulations. In each top subpanel, we also show $\mathcal{C}(r)$ at
$\bar{t}=0$ with the dashed lines. Initially, the misalignment between the kick
vector and the orbital velocity at each point of the disk induces a wide
dispersion in the (instantaneous) orbital tilt angle \citep{Akiba:2023ami}
leading to an overall drop in $\mathcal{C}(r)$. This effect is most pronounced
where $v_K(r) \approx \vkick$, appearing as a clear dip in the $\mathcal{C}(r)$
curves near $r\approx r_0$. As hydrodynamic evolution continues, we see that
relaxation of the bound material increases the coherence to progressively larger
radii over time.
The $\alpha=90^\circ$ case should ideally exhibit zero tilt gradient by the
symmetry, while nonzero gradient $|\partial \hat{l} / \partial \bar{r}| \lesssim
0.1$ was observed; this is likely due to numerical noise from the break of
bitwise cylindrical symmetry in the nested Cartesian AMR grid. In this regard,
we consider $|\partial \hat{l} / \partial \bar{r}| \simeq 0.1$ as a noise floor.

Several kick angles and simulation times, still, exhibit the tilt gradient above
this noise threshold. For instance, $\alpha=15^\circ$ simulation shows a warp
developed around $r = 2.5 r_0$. In the $\alpha=45^\circ$ and $\alpha=60^\circ$
simulations, we observe a warp moving radially outward with its amplitude
decreasing over time. Such a short-lasting warp in an obliquely kicked disk has
also been briefly reported in the smoothed-particle hydrodynamics simulation of
\cite{Rossi:2009nk}. For a large warp with $\psi > 0.1$ (which roughly
corresponds to $|\partial \hat{l}/\partial \bar{r}| > 0.1$ in our figure),
several previous studies have shown that periodic scale height variations can
lead to shock dissipation
\cite[e.g.][]{Kaaz:2024zhy,Fairbairn:2021,Fragile:2008sv}. We therefore expect a
fraction of the total dissipation ($\Delta E_{\rm int, b}$) is associated with
the relaxation of the warp, although disentangling its contribution from that of
radial circularization is nontrivial.

The warp introduced in the bound material in our simulation is an outcome of a
kick, namely a linear momentum imparted to the central gravitating object. This
suggests an additional formation mechanism of warped disks on top of having an
internal source of misaligned torque (e.g. Lense--Thirring precession in black
hole accretion disks or magnetospheric torques from an accreting star). However,
the physical lifetime of this kick-driven warp has not been very well resolved
in our simulations and needs further investigations.

\subsubsection{Non-constant initial surface mass density}
\label{sec:results-ppm1}

Here we briefly summarize the simulation results from $p=\pm1$ initial surface
mass density profiles. All corresponding figures are compiled in
Appendix~\ref{sec:non-const-p-results}.

The overall trend is similar to the $p=0$ simulations, showing a good agreement
with the analytic estimates for high kick angles and larger deviations for lower
angles. We see a higher $p$ value generally leads to the higher coherence
measure $\mathcal{C}(r)$. This could be understood as the radial hydrodynamic
mixing is weaker for lower $p$ values, since the lower mass density (and
consequently lower ram-pressure) makes it harder for the gas flows in the outer
regions to penetrate into a more tightly bound inner region. For
$\alpha=60^\circ$ kick angle, we could observe a transient warp for both $p$
indices.

The dissipation efficiency $\xi_E$ is fairly well clustered around $\xi_E
\approx 0.08$ for $p=1$, showing weaker variation on the kick angle compared to
the $p=0$ simulations. In contrast, $p=-1$ simulations show a more dramatic
variation over the kick angle, reaching $\xi_E \sim 0.3$ for $\alpha=30^\circ$
but dropping to $\xi_E \approx -0.2$ for $\alpha=90^\circ$. A negative value of
$\xi_E$ indicates that the cool-down of the bound material from its adiabatic
expansion dominates over the dissipation of the kinetic energy. This is due to
the initial thermal profile (Eq.~\ref{eq:sound-speed-profile}), combined with
the steep $\Sigma \propto r^{-1}$ surface mass density profile, depositing a
large portion of the total internal energy into the innermost region ($r \ll
r_0$), which remains relatively undisturbed from the kick but spreads out to a
new equilibrium.

\section{Discussions}
\label{sec:discussions}

\begin{figure*}
    \centering
    \includegraphics[width=0.98\linewidth]{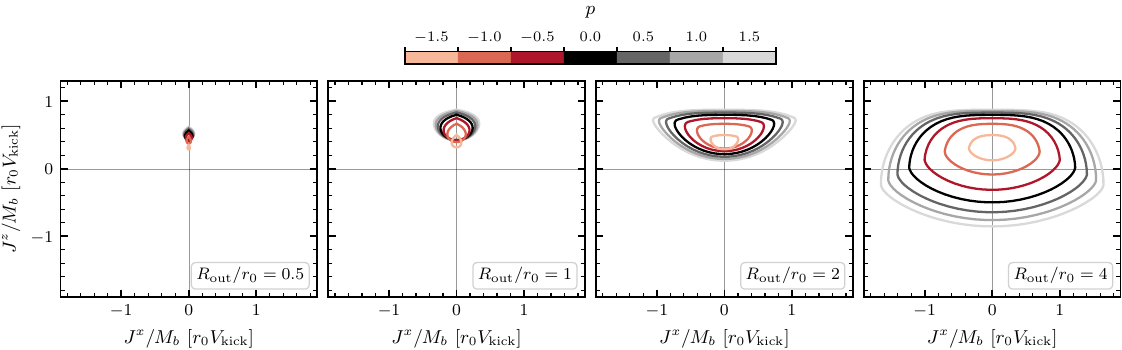}\\[1ex]
    \caption{Total angular momentum of the bound material normalized with its
        mass (same as the left panel in
        Fig.~\ref{fig:normalized-angular-momentum-vector}) for the disks with a
        finite radial size $R_{\rm out}$.}
    \label{fig:angularmomentum-truncated}
\end{figure*}

\subsection{Effect of the finite initial disk size}
\label{sec:finite-size-effect}

Depending on a specific astrophysical scenario under consideration, the size of
the disk $\outer{R}$ before the recoil can be comparable to or smaller than $r_0
= GM/\vkick^2$. The analysis in Sec.~\ref{sec:analysis-infinite-disk} then could
be repeated with a finite upper limit $\outer{R}$ in the integrals. For example,
the total bound mass can now be computed as
\begin{equation}
\begin{split}
    M_b & = \int_0^{2\pi} d\phi \int_0^{\min(r_b(\phi), \outer{R})} \Sigma(r) r dr \\[1ex]
        & = \frac{\Sigma_0 r_0^2}{p+2} \int_0^{2\pi} d\phi
            \left[ \frac{\min(r_b(\phi), \outer{R})}{r_0} \right]^{p+2} \, ,
\end{split}
\end{equation}
and the total bound angular momentum Eq.~ \eqref{eq:total-angular-momentum} and
\eqref{eq:total-angular-momentum-components} can be modified in the same manner.

If $\outer{R}/r_0 < (\sqrt{2}-1)^2 \approx 0.2$, the whole disk remains
gravitationally bound regardless of the recoil angle and the total angular
momentum does not change. It is the asymmetric \emph{loss} of the
gravitationally unbound material that introduces a tilt to the bound gas.
On the other hand, if $\outer{R}/r_0 > (\sqrt{2}+1)^2 \approx 6$, the analysis
made in Sec.~\ref{sec:analysis-infinite-disk} is applicable without
modifications.

Fig.~\ref{fig:angularmomentum-truncated} shows the mass-normalized angular
momentum of the bound material $J^i/M_b$ for different finite disk sizes
$\outer{R}$, in the same manner as the first panel of
Fig.~\ref{fig:normalized-angular-momentum-vector}. We see that the intermediate
regime $0.2 \lesssim \outer{R}/r_0 \lesssim 6$ shows a rapid expansion in the
possible range of the final tilt angle as $\outer{R}/r_0$ increases.

\subsection{Luminosity of electromagnetic flare}
\label{sec:recoil-flare-luminosity}

Hydrodynamic relaxation and fallback of the bound material can source
electromagnetic transients following the kick. From
Eqs.~\eqref{eq:shields-parametrization-energy} and
\eqref{eq:shields-parametrization-time}, the luminosity of this recoil-driven
flare (or a `kick afterglow', e.g. \citealt{Lippai:2008fx}) can be estimated as
\begin{equation} \label{eq:recoil-flare-luminosity-form-1}
    L_{\rm flare} = \frac{E_{\rm flare}}{t_{\rm flare}}
        = \frac{\xi_E}{4\pi\xi_t} \left(\frac{M_b}{M}\right) \frac{\vkick^5}{G} \, ,
\end{equation}
assuming a radiatively efficient environment. In terms of the characteristic
surface mass density of the original disk $\Sigma_0$, we can approximate $M_b
\approx \Sigma_0 r_0^2 = \Sigma_0 (GM/\vkick^2)^2$ and write
Eq.~\eqref{eq:recoil-flare-luminosity-form-1} into an alternative form
\begin{equation} \label{eq:recoil-flare-luminosity-form-2}
    L_{\rm flare}
        = \frac{\xi_E}{4\pi\xi_t} \Sigma_0 GM \vkick \, .
\end{equation}
In order to drive a bright electromagnetic emission, it is preferred to have (1)
a high surface density (equivalently, higher bound mass $M_b$), (2) a low kick
angle to increase the dissipation efficiency $\xi_E$, and (3) a high kick
velocity. As previously seen in Sec.~\ref{sec:results}, the energy dissipation
efficiency $\xi_E$ and the time scale $\xi_t$ vary with the kick angle and the
surface mass density profile.
For the purpose of an order-of-magnitude estimate, we simply take $\xi_E \sim
0.1$ and $\xi_t \sim 2$ as a flat estimate and set the fudge factor $\xi_E/(4\pi
\xi_t) \sim 10^{-2.5}$, which is adopted for the discussions below.

\subsection{Applications}
\label{sec:applications}

The applicability of the results presented in this work to a specific
astrophysical system depends on the following aspects:
\begin{itemize}[itemsep=1ex, labelsep=1.5ex]
\item The key parameter is the ratio between the characteristic size of the disk
    and $r_0 = GM/\vkick^2$:
    \begin{equation} \label{eq:system-size-ratio}
        \frac{\outer{R}}{r_0}
            = \left(\frac{GM}{\outer{R} \vkick^2}\right)^{-1} \, ,
    \end{equation}
    as previously shown in Sec.~\ref{sec:finite-size-effect} and
    Fig.~\ref{fig:angularmomentum-truncated}. The feasibility of a high tilt
    angle and the efficient dissipation of the kinetic energy ($\xi_E \sim 0.1$)
    within the bound material critically depend on whether the size of the
    original disk $\outer{R}$ is at least comparable to or a few times larger
    than $r_0 = GM/\vkick^2$.
\item If there is an external gravitational field, the radius of gravitational
    influence of the kicked object $R_{\rm ROI}$ sets an upper bound on the
    possible size of the kicked disk,
    i.e. the factor $\outer{R}$ in Eq.~\eqref{eq:system-size-ratio} needs to be
    replaced with $\min[R_{\rm out}, R_{\rm ROI}]$.
\item The rotation profile of the disk before the kick should be reasonably
    close to the Keplerian profile $\Omega \propto R^{-3/2}$.
\end{itemize}
Keeping these caveats in mind, we consider several astrophysical systems and
discuss the implications.

\subsubsection{SMBH mergers}

For the remnants of binary SMBH mergers, the characteristic radius and time
scales of the environmental response to the kick are
\begin{equation}
\begin{split}
    r_0 & = 10
        \left(\frac{M_{\rm SMBH}}{10^8 M_\odot}\right)
        \left(\frac{\vkick}{200 \kms}\right)^{-2} \, {\rm pc} \\[1ex]
        & \hspace{1em} \approx 2 \times 10^6
        \left(\frac{\vkick}{200 \kms}\right)^{-2}
        r_g \, ,
\end{split}
\end{equation}
and
\begin{equation} \label{eq:timescale-smbh}
    t_0 = \frac{G M_{\rm SMBH}}{\vkick^3}
    \approx 5\e{4} \left(\frac{M_{\rm SMBH}}{10^8 M_\odot}\right)
        \left(\frac{\vkick}{200 \kms}\right)^{-3} {\rm yr} \, ,
\end{equation}
where $r_g \equiv GM_{\rm SMBH}/c^2$.
The background galactic potential sets the SMBH region of influence $R_{\rm ROI}
\approx 2\e{6} \, (\sigma / 200 \kms)^{-2}\,r_g$,
where $\sigma$ is the stellar velocity dispersion in the nuclear region.
Also, outer part of the AGN accretion disk beyond $10^3$\textendash$10^4 r_g$ is
prone to gravitational instability and fragmentation
\citep{Goodman:2002gv,Sirko:2002ex,Jiang:2010ba}, which sets the finite radial
extent of the disk. Therefore the gravitational recoil should be strong, i.e.
$\vkick \gtrsim 1000\kms$ to have $R_{\rm out}/r_0 \gtrsim 10^0$. Otherwise,
dynamical response of the galactic nuclear region to the recoiling SMBH remains
in a perturbative regime.

A strong kick with $\vkick \gtrsim 1000\kms$ requires specific spin
configurations in the BH binary \citep{Campanelli:2007cga,Lousto:2011kp}, and is
preferentially directed perpendicular to the binary orbital plane. Such fast
kicks misaligned with the axis of the circumbinary disk therefore may be
uncommon (although the circumbinary disk may still be misaligned with the BBH),
but their possibility cannot be ruled out.\footnote{See, e.g.
\cite{Lousto:2009ka,Lousto:2012su} for quantitative estimates.}

Below we describe several possible outcomes of a strong SMBH kick reorienting
the AGN disk, which can introduce unusual observational features. Assuming
$\vkick \approx 1000\kms$, the characteristic timescale $t_0$
(Eq.~\ref{eq:timescale-smbh}) ranges from several decades for low-mass SMBHs
($M_{\rm SMBH} \approx 10^{6.5}\text{--}10^{7}M_\odot$) to thousands of years
for very massive ones ($M_{\rm SMBH} \approx 10^{8.5}\text{--}10^9M_\odot$).
\begin{enumerate}[leftmargin=1.5em, labelsep=1ex, labelwidth=1em, itemsep=1ex]
\item {\em Dissipation flare from the kicked disk}: The luminosity can reach
    \begin{align}
        L_{\rm flare}
            & \approx 10^{43} \left(\frac{M_b/M}{10^{-2}}\right)
            \left(\frac{\vkick}{1000\kms}\right)^5 \, {\rm erg\,s^{-1}} \, ,
    \end{align}
    in the soft X-ray band \cite[e.g.][]{Lippai:2008fx,Shields:2008va} or
    reprocessed into infrared afterglow \citep{Schnittman:2008ez}, lasting
    hundreds to thousands of years for a $10^8 M_\odot$ SMBH. If the total mass
    of the gravitationally bound material is comparable to the SMBH mass, the
    dissipation luminosity can reach $L_{\rm flare} \sim 10^{45} \, {\rm
    erg\,s^{-1}}$.
\item {\em Change in the AGN illumination/obscuration}: The accretion disk
    ejected with the SMBH is not only displaced from the original merger site,
    but also geometrically misaligned with the outer dusty torus (or any other
    environmental features that were aligned before the merger). Depending on
    the viewing angle, this `disengaged' state of the accretion disk might make
    the AGN appear distinct from ordinary ones, owing to its altered
    illumination and obscuration geometry. A quantitative model of these
    systems, from future studies involving radiative transfer calculations,
    could help identify such anomalies in the survey data. \\[1ex]
    We note that the time scale $t_0 = GM_{\rm SMBH}/\vkick^3$ over which such
    transition occurs is typically much longer than the variability time scales
    of months to several years seen in the changing-look AGNs
    \cite[e.g.][]{Shappee:2014,LaMassa:2015,MacLeod:2016}, unless for a low-mass
    SMBH kicked with an extremal velocity (e.g. $M_{\rm SMBH} = 10^{6.5}M_\odot$
    with $\vkick=1000\kms$ yields $t_0 \approx 13$ yr.).
\item {\em Warp-driven AGN variability}: If the remnant SMBH is spinning,
    Lense--Thirring torque can cause precession and warp in the inner part of
    the disk, leading to warp-driven variability in the accretion luminosity.
    General relativistic magnetohydrodynamics (GRMHD) simulations of tilted and
    warped thin accretion disks have shown that these effects can drive rapid
    variability resembling that observed in the changing-look AGNs
    \cite[e.g.][]{Kaaz:2024zhy,Kaaz:2025udg}. Since our simulations in the
    current work are Newtonian and adiabatic, an additional study incorporating
    cooling and relativistic effects would be required to validate this picture.
\item {\em Reorientation of jets}: GRMHD simulations of accreting BBHs in the
    late inspiral show relativistic outflows or common jets from the binary are
    being launched mostly perpendicular to the circumbinary disk plane
    \cite[e.g.][]{Most:2024qus,Manikantan:2025eqa,Ennoggi:2025ijc,Kalinani:2026mqe}.
    Jets from a (single) spinning BH are launched in alignment with the BH spin
    \citep{Blandford:1977ds,Tchekhovskoy:2011zx} close to the horizon, but the
    subsequent outward propagation of the jets can be steered to align with the
    rotation axis of the outer disk
    \citep{McKinney:2012wd,Liska:2017alm,Liska:2018ayk,Kim:2026qad}
    or exhibit intermittent behavior if highly misaligned
    \citep{Chatterjee:2023ber}. An extreme kick tilting the accretion disk from
    its original orientation as a circumbinary disk can consequently reorient
    the relativistic outflows at large distances.
    \\[1ex]
    The spin-flip of the remnant in binary BH mergers \citep{Campanelli:2006fy}
    is often invoked as a possible formation mechanism of the reoriented lobes
    in radio galaxies
    \cite[e.g.][]{Merritt:2002hc,Dennett-Thorpe:2001pdn,Liu:2003fx,Roberts:2015wja}.
    Our results suggest an additional mechanism for reoriented radio lobes,
    namely from the reorientation of the AGN disk, without invoking the
    reorientation of the SMBH spin itself. A key difference between the two jet
    reorientation mechanisms is the time scale; the spin-flip occurs in several
    light-crossing times of the SMBH horizon ($\sim GM_{\rm SMBH}/c^3$) whereas
    the recoil-driven tilt of the AGN disk takes place over a much longer time
    scale.
\end{enumerate}

An important caveat regarding (iii) and (iv) is that the GRMHD simulations of
tilted disks commonly \emph{begin} with a globally tilted torus.\footnote{See
\cite{Kim:2026qad} for the development of a tilted accretion disk following the
SMBH recoil, beginning from an initially aligned disk midplane.} The kick-driven
tilt, in contrast, develops gradually over the time scale $t_0$, during which it
can be affected by the Bardeen--Petterson alignment
\citep{1975ApJ...195L..65B,Nelson:2000mw,Lodato:2006kv} and nonlinear warp
dynamics \citep{Lodato:2007hf,Nixon:2012fm}. The realistic evolution of the
radial tilt profile and the resulting jet reorientation depend on the complex
interplay between these effects. We leave this investigation for future work.

\subsubsection{Stellar-mass BBH mergers in AGN disks}

The accretion disks of AGNs may host mergers of stellar-mass BH binaries
\cite[e.g.][]{McKernan:2020lgr,Ford:2025bbk}. The interactions between a kicked
remnant BH and the surrounding gas in the AGN disk can drive post-merger
electromagnetic counterparts
\citep{McKernan:2019hqs,Chen:2023xrm,McPike:2026ugj}. The characteristic length
and time scales associated with the kick are
\begin{align}
    r_0 & \approx 2 \, \left(\frac{m_{\rm BH}}{100 M_\odot}\right)
        \left(\frac{\vkick}{200\kms}\right)^{-2} \, {\rm AU} \, , \\[1ex]
    t_0 & \approx 20 \, \left(\frac{m_{\rm BH}}{100 M_\odot}\right)
        \left(\frac{\vkick}{200\kms}\right)^{-3} \, {\rm days} \, .
\end{align}
Although the size of an AGN disk is typically much larger than $r_0 \sim 2 {\rm
AU}$, gravitational potential of the supermassive black hole sets the Hill
sphere around the remnant BH with the radius
\begin{equation}
    R_{\rm Hill} = a_{\rm BH} \left(\frac{m_{\rm BH}}{3M_{\rm SMBH}}\right)^{1/3}
    = a_{\rm BH} \left(\frac{q}{3}\right)^{1/3} \, ,
\end{equation}
where $a_{\rm BH}$ is the semimajor axis of the remnant BH around the SMBH and
$q = m_{\rm BH} / M_{\rm SMBH}$ is the mass ratio.

The effect of the finite scale height $H_{\rm disk}$ of the AGN disk, meanwhile,
should be taken into consideration as well. If $H_{\rm disk} / r_0 > 1$, the
thin-geometry assumption that we made in Sec.~\ref{sec:analysis-infinite-disk}
is no longer valid and the accretion flow around the remnant BH will more
closely resemble spherical Bondi-type flows.

From the discussions above, we see that our results are best applicable when
$H_{\rm disk} < r_0 < R_{\rm Hill}$. The first part of the inequality $H_{\rm
disk} < r_0$ corresponds to
\begin{equation} \label{eq:agn-inequality-1}
        \left(\frac{H_{\rm disk}/a_{\rm BH}}{10^{-2}}\right)
        \left(\frac{a_{\rm BH}}{10^3 r_g}\right)
        \left(\frac{q}{10^{-6}}\right)^{-1}
        \left(\frac{\vkick}{200 \kms}\right)^{2}
        \lesssim 0.2 \, ,
\end{equation}
and the second part $r_0 < R_{\rm Hill}$ is\footnote{Our argument relies on the
assumption that the gas flow around the merging BH binary is close to the
Keplerian profile up to several times the recoil radius $r_0$. Hydrodynamic
simulations of an accreting binary embedded in a shearing flow show the
development of the circumbinary disk, but a substantial fraction of the Hill
sphere can still be occupied by horseshoe flows from the surrounding AGN disk
\cite[e.g.][]{Dittmann:2023sha,Dempsey:2022iws,Li:2022pnc,Li:2022eup}. For the
velocity profile surrounding the remnant BH to be approximately Keplerian, it is
favorable to have a high ratio of $R_{\rm Hill}/r_0$, to ensure the inspiralling
BBH and its circumbinary disk are placed deep inside the Hill sphere.}
\begin{equation} \label{eq:agn-inequality-2}
        \left(\frac{q}{10^{-6}}\right)^{-2/3}
        \left(\frac{a_{\rm BH}}{10^3 r_g}\right)
        \left(\frac{\vkick}{200 \kms}\right)^{2} \gtrsim 0.3 \, ,
\end{equation}
where $r_g = GM_{\rm SMBH}/c^2$ is the gravitational radius of the SMBH. While
the conditions Eqs.~\eqref{eq:agn-inequality-1} and \eqref{eq:agn-inequality-2}
exclude certain regimes, a fraction of parameter space still remains viable, for
which a kicked BH acquires a tilted disk of bound material after tens to a
hundred days following the merger. If relativistic jets are launched from the
remnant BH \citep{Yuan:2021jjt,Tagawa:2023uqa,Chen:2023xrm,Kim:2024zjb}, their
directions could be affected by the tilt angle of the kicked disk as discussed
before. The characteristic dissipative luminosity is
\begin{equation}
    L_{\rm flare} \sim
        10^{39}
        \left(\frac{M_b}{1M_\odot}\right)
        \left(\frac{m_{\rm BH}}{100M_\odot}\right)^{-1}
        \left(\frac{\vkick}{200 \kms}\right)^{5}
        \, {\rm erg\,s^{-1}} \, .
\end{equation}

\cite{Graham:2020gwr} reported a candidate electromagnetic counterpart to a BBH
merger GW190521 \citep{LIGOScientific:2020ufj} with the bolometric luminosity $L
\sim 10^{45}{\rm erg\,s^{-1}}$, claimed to originate from the AGN disk
interacting with the kicked remnant BH. Based solely on the kick-driven
dissipation in the bound material assuming $M_b \sim 10^{-2} m_{\rm BH} \approx
1 M_\odot$, either the kick should be extreme ($\vkick >
1000\kms$),\footnote{See \cite{LIGOScientific:2020ufj} for the posterior
distributions of the BH kick magnitude in this event.} or the surface mass
density around the merging BBH needs to be unreasonably high ($\Sigma_0 \sim
10^{12} {\rm g \, cm^{-2}}$) to match the observed luminosity, making it
unlikely to be the primary source. Additional mechanisms invoking a prolonged
interaction between the kicked BH and the AGN disk, such as the
Bondi-Hoyle-Lyttleton accretion onto the BH (adopted in \citealt{Graham:2020gwr}
to account for this issue) or the ram-pressure stripping of the Hill sphere
\citep{McKernan:2019hqs}, can provide an additional energy budget for the
post-kick flare.

\subsubsection{Natal kicks on stellar evolution remnants}

Newly born neutron stars and black holes can acquire natal kicks from asymmetric
mass ejection and neutrino emissions during the supernova explosion
\cite[e.g.][]{Wongwathanarat:2012zp,Chan:2020lnd,Coleman:2022lwr,Janka:2024xbp}.
Recent numerical studies have reported that the natal kick can reach several
hundred to a thousand km\,s\textsuperscript{-1} depending on the progenitor and
explosion process \citep{Burrows:2023ffl,Popov:2025hva,Li:2026pxl}. Whether the
fallback material can form a disk around the remnant depends on the rotation
profile of the progenitor and the angular momentum transport in highly
asymmetric 3D explosion dynamics. Postponing these complications to future work,
here we only quote the relevant scales
\begin{align}
    r_0 & \approx 0.8 \, \left(\frac{M}{M_\odot}\right)
        \left(\frac{\vkick}{500\kms}\right)^{-2} \, R_\odot \, , \\[1ex]
    \label{eq:recoil-flare-timescale-natal}
    t_0 & \approx 20 \, \left(\frac{M}{M_\odot}\right)
        \left(\frac{\vkick}{500\kms}\right)^{-3} \, {\rm minute} \, , \\[1ex]
    \label{eq:recoil-flare-luminosity-natal}
    L_{\rm flare} & \approx 10^{41} \left(\frac{M_b/M}{10^{-2}}\right)
        \left(\frac{\vkick}{500\kms}\right)^5 \, {\rm erg\,s^{-1}} \, .
\end{align}

If a circularized disk of debris is formed and its size is comparable to $\sim 1
R_\odot$, the kicked natal compact object can acquire a tilted disk around it
within several hours. If the natal compact object has a spin, the kick can
naturally introduce a spin-disk misalignment to the system. The dissipation
luminosity $L_{\rm flare}$ inherits a large uncertainty in the bound mass $M_b$,
where we estimate a fiducial value in \eqref{eq:recoil-flare-luminosity-natal}.
The distinguishing signature of this kick-driven flare, if observable, would be
the relatively short time scale \eqref{eq:recoil-flare-timescale-natal}.

\section{Summary}
\label{sec:conclusion}

In this paper, we have investigated the dynamical response of a hydrodynamic
disk to a kicked central object. Based on a simple analytic calculation, we show
that the kicked gaseous disk can feature a large tilt or retrograde rotation
depending on the kick geometry. We perform numerical simulations to verify this
claim, and have assessed the impact of the nonlinear hydrodynamic effects.

After a highly dynamic initial phase during which the disk undergoes an active
rearrangement and dissipation of its orbital kinetic energy, we see the bound
gas gradually relaxes into a coherently rotating structure after a few tens of
$GM/\vkick^3$. Kicks directed closer to the midplane, as expected, exhibit
stronger shock heating and self-collision of the gas flow. The net increase in
the internal energy of the bound material reaches up to a few tens of percent of
the characteristic kinetic energy $M_b \vkick^2/2$, although its efficiency
varies with the kick angle and the surface mass density profile. We also observe
a transient warp induced in the kicked disks, while a more careful analysis with
a higher grid resolution would be needed to properly resolve its lifetime and
detailed evolution.

Our results suggest a kick as a possible formation mechanism of tilted or warped
astrophysical disks, in particular with no need to assume the presence of an
internal source of misaligned torque. Since it is the non-axisymmetric
gravitational unbinding of the disk that introduces a net tilt to the bound
material, the feasibility of the kick-driven tilt heavily depends on the initial
size of the disk compared to the bound radius scale $r_0 = GM/\vkick^2$. Several
possible applications in astrophysics have been discussed with caveats.

While we have considered a vertically thin Keplerian disk as an exploratory
setup, a more sophisticated analysis with the problem-specific variations (e.g.
non-Keplerian rotation profile or geometrically thick disks) should be possible.
In future work, we plan to carry out targeted studies on specific astrophysical
systems, incorporating cooling effects, finite disk truncation radii, and
gravitational mass loss of the kicked object.

\section*{Acknowledgements}

The author is grateful to Adam Burrows, Yixian Chen, Alexander Dittmann, Chris
Hamilton, Jeremy Goodman, Nick Kaaz, Chang-Goo Kim, Barry McKernan, Eliot
Quataert, and Nicholas Rui for insightful discussions and suggestions, and
especially to Alexander Dittmann for helpful advice on simulations and
presentation of the data.
Y.K. acknowledges support by a joint postdoctoral fellowship at the Princeton
Center for Theoretical Science and the Princeton Gravity Initiative.

This work used the Anvil cluster at Purdue University
\citep{McCartney2014,10.1145/3491418.3530766} and the Delta system at the
National Center for Supercomputing Applications (award OAC 2005572) through
allocation PHY260163 from the Advanced Cyberinfrastructure Coordination
Ecosystem: Services \& Support (ACCESS) program, which is supported by National
Science Foundation grants \#2138259, \#2138286, \#2138307, \#2137603, and
\#2138296.
The simulations presented in this article were partially performed on
computational resources managed and supported by Princeton Research Computing, a
consortium of groups including the Princeton Institute for Computational Science
and Engineering (PICSciE) and Research Computing at Princeton University.

Figures in this article were produced using matplotlib
\citep{Hunter:2007,the_matplotlib_development_team_2026_22700776}, numpy
\citep{harris2020array}, scipy \citep{2020SciPy-NMeth}, Paraview
\citep{paraview:2005,paraview:2015}, yt \citep{2011ApJS..192....9T}, with the
assistance of Anthropic Claude under the supervision of the author.

\section*{Data Availability}

Data underlying this article will be shared upon a reasonable request to the
author.

\bibliographystyle{mnras}
\bibliography{references}

@article{Rossi:2009nk,
  archiveprefix = {arXiv},
  author        = {Rossi, Elena M. and Lodato, G. and Armitage, P. J. and Pringle, J. E. and King, A. R.},
  doi           = {10.1111/j.1365-2966.2009.15802.x},
  eprint        = {0910.0002},
  journal       = {Mon. Not. Roy. Astron. Soc.},
  pages         = {2021},
  primaryclass  = {astro-ph.HE},
  title         = {{Black hole mergers: the first light}},
  volume        = {401},
  year          = {2010}
}

@article{Akiba:2021,
  author        = {{Akiba}, Tatsuya and {Madigan}, Ann-Marie},
  title         = {{On the Formation of an Eccentric Nuclear Disk following the Gravitational Recoil Kick of a Supermassive Black Hole}},
  journal       = {Astrophys. J. Lett.},
  year          = 2021,
  month         = nov,
  volume        = {921},
  number        = {1},
  eid           = {L12},
  pages         = {L12},
  doi           = {10.3847/2041-8213/ac30d9},
  archiveprefix = {arXiv},
  eprint        = {2110.10163},
  primaryclass  = {astro-ph.GA},
  adsurl        = {https://ui.adsabs.harvard.edu/abs/2021ApJ...921L..12A}
}

@article{Akiba:2023ami,
  author        = {Akiba, Tatsuya and Madigan, Ann-Marie},
  title         = {{Anisotropic Star Clusters around Recoiling Supermassive Black Holes}},
  eprint        = {2305.03054},
  archiveprefix = {arXiv},
  primaryclass  = {astro-ph.GA},
  doi           = {10.3847/1538-4357/acffb6},
  journal       = {Astrophys. J.},
  volume        = {958},
  number        = {2},
  pages         = {137},
  year          = {2023}
}

@article{Bright:2024soo,
  author        = {Bright, Jane C. and Akiba, Tatsuya and Madigan, Ann-Marie},
  title         = {{Counter-rotation and slow precession in aligned eccentric nuclear discs due to gravitational wave recoil kicks}},
  eprint        = {2408.13353},
  archiveprefix = {arXiv},
  primaryclass  = {astro-ph.GA},
  doi           = {10.1093/mnrasl/slae079},
  journal       = {Mon. Not. Roy. Astron. Soc.},
  volume        = {534},
  number        = {1},
  pages         = {L42--L47},
  year          = {2024}
}

@article{Akiba:2025,
  author        = {{Akiba}, Tatsuya and {Naoz}, Smadar and {Madigan}, Ann-Marie},
  title         = {{On the Formation of S stars from a Recent Massive Black Hole Merger in the Galactic Center}},
  journal       = {Astrophys. J. Lett.},
  year          = 2025,
  month         = jul,
  volume        = {987},
  number        = {2},
  eid           = {L27},
  pages         = {L27},
  doi           = {10.3847/2041-8213/addc5d},
  archiveprefix = {arXiv},
  eprint        = {2410.19881},
  primaryclass  = {astro-ph.GA},
  adsurl        = {https://ui.adsabs.harvard.edu/abs/2025ApJ...987L..27A}
}

@article{Kim:2026qad,
  author        = {Kim, Yoonsoo and Most, Elias R. and Wang, Hai-Yang},
  title         = {{Recoil geometry determines electromagnetic counterparts from supermassive black hole merger remnants}},
  eprint        = {2603.26641},
  archiveprefix = {arXiv},
  primaryclass  = {astro-ph.HE},
  month         = {3},
  year          = {2026},
  journal       = {arXiv e-prints},
  doi           = {10.48550/arXiv.2603.26641}
}

@article{Stone:2026,
  doi       = {10.3847/1538-4365/ae3717},
  url       = {https://doi.org/10.3847/1538-4365/ae3717},
  year      = {2026},
  month     = {feb},
  publisher = {The American Astronomical Society},
  volume    = {283},
  number    = {1},
  pages     = {27},
  author    = {Stone, James M. and Mullen, Patrick D. and Fielding, Drummond and Grete, Philipp and Guo, Minghao and Kempski, Philipp and Most, Elias R. and White, Christopher J. and Wong, George N.},
  title     = {{AthenaK: A Performance-portable Version of the Athena++ Adaptive Mesh Refinement Framework}},
  journal   = {The Astrophysical Journal Supplement Series}
}

@article{Duffell:2024fwy,
  author        = {Duffell, Paul C. and others},
  title         = {{The Santa Barbara Binary{\ensuremath{-}}disk Code Comparison}},
  eprint        = {2402.13039},
  archiveprefix = {arXiv},
  primaryclass  = {astro-ph.SR},
  doi           = {10.3847/1538-4357/ad5a7e},
  journal       = {Astrophys. J.},
  volume        = {970},
  number        = {2},
  pages         = {156},
  year          = {2024}
}

@article{Springel:2009aa,
  author        = {Springel, Volker},
  title         = {{E pur si muove: Galiliean-invariant cosmological hydrodynamical simulations on a moving mesh}},
  eprint        = {0901.4107},
  archiveprefix = {arXiv},
  primaryclass  = {astro-ph.CO},
  doi           = {10.1111/j.1365-2966.2009.15715.x},
  journal       = {Mon. Not. Roy. Astron. Soc.},
  volume        = {401},
  pages         = {791},
  year          = {2010}
}

@article{Graham:2020gwr,
  author        = {Graham, M. J. and others},
  title         = {{Candidate Electromagnetic Counterpart to the Binary Black Hole Merger Gravitational Wave Event S190521g}},
  eprint        = {2006.14122},
  archiveprefix = {arXiv},
  primaryclass  = {astro-ph.HE},
  doi           = {10.1103/PhysRevLett.124.251102},
  journal       = {Phys. Rev. Lett.},
  volume        = {124},
  number        = {25},
  pages         = {251102},
  year          = {2020}
}

@article{Peres:1962zz,
  author  = {Peres, Asher},
  title   = {{Classical Radiation Recoil}},
  doi     = {10.1103/PhysRev.128.2471},
  journal = {Phys. Rev.},
  volume  = {128},
  pages   = {2471--2475},
  year    = {1962}
}

@article{Bekenstein:1973zz,
  author  = {Bekenstein, Jacob D.},
  title   = {{Gravitational-Radiation Recoil and Runaway Black Holes}},
  doi     = {10.1086/152255},
  journal = {Astrophys. J.},
  volume  = {183},
  pages   = {657--664},
  year    = {1973}
}

@article{Fitchett:1983qzq,
  author  = {Fitchett, M. J.},
  title   = {{The influence of gravitational wave momentum losses on the centre of mass motion of a Newtonian binary system}},
  doi     = {10.1093/mnras/203.4.1049},
  journal = {Mon. Not. Roy. Astron. Soc.},
  volume  = {203},
  number  = {4},
  pages   = {1049--1062},
  year    = {1983}
}

@article{Baker:2006vn,
  author        = {Baker, John G. and Centrella, Joan and Choi, Dae-Il and Koppitz, Michael and van Meter, James R. and Miller, M. Coleman},
  title         = {{Getting a kick out of numerical relativity}},
  eprint        = {astro-ph/0603204},
  archiveprefix = {arXiv},
  doi           = {10.1086/510448},
  journal       = {Astrophys. J. Lett.},
  volume        = {653},
  pages         = {L93--L96},
  year          = {2006}
}

@article{Campanelli:2007ew,
  archiveprefix = {arXiv},
  author        = {Campanelli, Manuela and Lousto, Carlos O. and Zlochower, Yosef and Merritt, David},
  doi           = {10.1086/516712},
  eprint        = {gr-qc/0701164},
  journal       = {Astrophys. J.},
  pages         = {L5--L8},
  title         = {{Large merger recoils and spin flips from generic black-hole binaries}},
  volume        = {659},
  year          = {2007}
}

@article{Campanelli:2007cga,
  author        = {Campanelli, Manuela and Lousto, Carlos O. and Zlochower, Yosef and Merritt, David},
  title         = {{Maximum gravitational recoil}},
  eprint        = {gr-qc/0702133},
  archiveprefix = {arXiv},
  doi           = {10.1103/PhysRevLett.98.231102},
  journal       = {Phys. Rev. Lett.},
  volume        = {98},
  pages         = {231102},
  year          = {2007}
}

@article{Campanelli:2006fy,
  author        = {Campanelli, Manuela and Lousto, Carlos O. and Zlochower, Yosef and Krishnan, Badri and Merritt, David},
  title         = {{Spin Flips and Precession in Black-Hole-Binary Mergers}},
  eprint        = {gr-qc/0612076},
  archiveprefix = {arXiv},
  doi           = {10.1103/PhysRevD.75.064030},
  journal       = {Phys. Rev. D},
  volume        = {75},
  pages         = {064030},
  year          = {2007}
}

@article{Gonzalez:2006md,
  archiveprefix = {arXiv},
  author        = {Gonzalez, Jose A. and Sperhake, Ulrich and Bruegmann, Bernd and Hannam, Mark and Husa, Sascha},
  doi           = {10.1103/PhysRevLett.98.091101},
  eprint        = {gr-qc/0610154},
  journal       = {Phys. Rev. Lett.},
  pages         = {091101},
  title         = {{Maximum Kick from Nonspinning Black-Hole Binary Inspiral}},
  volume        = {98},
  year          = {2007}
}

@article{Gonzalez:2007hi,
  author        = {Gonzalez, J. A. and Hannam, M. D. and Sperhake, U. and Bruegmann, Bernd and Husa, S.},
  title         = {{Supermassive recoil velocities for binary black-hole mergers with antialigned spins}},
  eprint        = {gr-qc/0702052},
  archiveprefix = {arXiv},
  doi           = {10.1103/PhysRevLett.98.231101},
  journal       = {Phys. Rev. Lett.},
  volume        = {98},
  pages         = {231101},
  year          = {2007}
}

@article{Lousto:2011kp,
  author        = {Lousto, Carlos O. and Zlochower, Yosef},
  title         = {{Hangup Kicks: Still Larger Recoils by Partial Spin/Orbit Alignment of Black-Hole Binaries}},
  eprint        = {1108.2009},
  archiveprefix = {arXiv},
  primaryclass  = {gr-qc},
  doi           = {10.1103/PhysRevLett.107.231102},
  journal       = {Phys. Rev. Lett.},
  volume        = {107},
  pages         = {231102},
  year          = {2011}
}

@article{Lousto:2012su,
  author        = {Lousto, Carlos O. and Zlochower, Yosef and Dotti, Massimo and Volonteri, Marta},
  title         = {{Gravitational Recoil From Accretion-Aligned Black-Hole Binaries}},
  eprint        = {1201.1923},
  archiveprefix = {arXiv},
  primaryclass  = {gr-qc},
  doi           = {10.1103/PhysRevD.85.084015},
  journal       = {Phys. Rev. D},
  volume        = {85},
  pages         = {084015},
  year          = {2012}
}

@article{Megevand:2009yx,
  archiveprefix = {arXiv},
  author        = {Megevand, Miguel and Anderson, Matthew and Frank, Juhan and Hirschmann, Eric W. and Lehner, Luis and Liebling, Steven L. and Motl, Patrick M. and Neilsen, David},
  doi           = {10.1103/PhysRevD.80.024012},
  eprint        = {0905.3390},
  journal       = {Phys. Rev. D},
  pages         = {024012},
  primaryclass  = {astro-ph.HE},
  title         = {{Perturbed disks get shocked. Binary black hole merger effects on accretion disks}},
  volume        = {80},
  year          = {2009}
}

@article{Corrales:2009nv,
  author        = {Corrales, Lia R. and Haiman, Zoltan and MacFadyen, Andrew},
  title         = {{Hydrodynamical Response of a Circumbinary Gas Disk to Black Hole Recoil and Mass Loss}},
  eprint        = {0910.0014},
  archiveprefix = {arXiv},
  primaryclass  = {astro-ph.HE},
  doi           = {10.1111/j.1365-2966.2010.16324.x},
  journal       = {Mon. Not. Roy. Astron. Soc.},
  volume        = {404},
  pages         = {947},
  year          = {2010}
}

@article{Meliani:2016rwn,
  author        = {Meliani, Zakaria and Mizuno, Yosuke and Olivares, Hector and Porth, Oliver and Rezzolla, Luciano and Younsi, Ziri},
  title         = {{Simulations of recoiling black holes: adaptive mesh refinement and radiative transfer}},
  eprint        = {1606.08192},
  archiveprefix = {arXiv},
  primaryclass  = {astro-ph.HE},
  doi           = {10.1051/0004-6361/201629191},
  journal       = {Astron. Astrophys.},
  volume        = {598},
  pages         = {A38},
  year          = {2017}
}

@article{Zanotti:2010xs,
  archiveprefix = {arXiv},
  author        = {Zanotti, Olindo and Rezzolla, Luciano and Del Zanna, Luca and Palenzuela, Carlos},
  doi           = {10.1051/0004-6361/201014969},
  eprint        = {1002.4185},
  journal       = {Astron. Astrophys.},
  pages         = {A8},
  primaryclass  = {astro-ph.HE},
  title         = {{EM counterparts of recoiling black holes: general relativistic simulations of non-Keplerian discs}},
  volume        = {523},
  year          = {2010}
}

@article{Ponce:2011kv,
  author        = {Ponce, Marcelo and Faber, Joshua A. and Lombardi, Jr, James C.},
  title         = {{Accretion disks around kicked black holes: Post-kick Dynamics}},
  eprint        = {1107.1711},
  archiveprefix = {arXiv},
  primaryclass  = {astro-ph.CO},
  doi           = {10.1088/0004-637X/745/1/71},
  journal       = {Astrophys. J.},
  volume        = {745},
  pages         = {71},
  year          = {2012}
}

@phdthesis{Bode:2011pqa,
  author = {Bode, Jason Nathaniel},
  title  = {{Black Hole Mergers and Their Electromagnetic Counterparts}},
  doi    = {10.7907/YVVG-R137},
  school = {Caltech},
  year   = {2011}
}

@article{Lippai:2008fx,
  archiveprefix = {arXiv},
  author        = {Lippai, Zoltan and Frei, Zsolt and Haiman, Zoltan},
  doi           = {10.1086/587034},
  eprint        = {0801.0739},
  journal       = {Astrophys. J. Lett.},
  pages         = {L5},
  primaryclass  = {astro-ph},
  title         = {{Prompt Shocks in the Gas Disk Around a Recoiling Supermassive Black Hole Binary}},
  volume        = {676},
  year          = {2008}
}

@article{Shields:2008va,
  archiveprefix = {arXiv},
  author        = {Shields, G. A. and Bonning, E. W.},
  doi           = {10.1086/589427},
  eprint        = {0802.3873},
  journal       = {Astrophys. J.},
  pages         = {758},
  primaryclass  = {astro-ph},
  title         = {{Powerful flares from recoiling black holes in quasars}},
  volume        = {682},
  year          = {2008}
}

@article{Schnittman:2008ez,
  author        = {Schnittman, Jeremy D. and Krolik, Julian H.},
  title         = {{The Infrared Afterglow of Supermassive Black Hole Mergers}},
  eprint        = {0802.3556},
  archiveprefix = {arXiv},
  primaryclass  = {astro-ph},
  doi           = {10.1086/590363},
  journal       = {Astrophys. J.},
  volume        = {684},
  pages         = {835--844},
  year          = {2008}
}

@article{McPike:2026ugj,
  author        = {{McPike}, Emily and {Perna}, Rosalba and {Ford}, K.~E. Saavik and {McKernan}, Barry and {Cook}, Harrison E. and {Delfavero}, Vera and {McCarthy}, Miranda and {Nathaniel}, Kaila and {Postiglione}, Jake and {Posner}, Nicolas and {Pritmani}, Varun and {Ray}, Shawn and {O'Shaughnessy}, Richard},
  title         = {{McFACTS. IV. Electromagnetic Counterparts to AGN-disk-embedded Binary Black Hole Mergers}},
  journal       = {Astrophys. J.},
  year          = 2026,
  month         = aug,
  volume        = {1007},
  number        = {2},
  eid           = {105},
  pages         = {105},
  doi           = {10.3847/1538-4357/ae8396},
  archiveprefix = {arXiv},
  eprint        = {2602.04135},
  primaryclass  = {astro-ph.HE},
  adsurl        = {https://ui.adsabs.harvard.edu/abs/2026ApJ..1007..105M}
}

@article{Colella:1982ee,
  author       = {Colella, Phillip and Woodward, Paul R.},
  title        = {{The Piecewise Parabolic Method (PPM) for Gas Dynamical Simulations}},
  reportnumber = {LBL-14661},
  doi          = {10.1016/0021-9991(84)90143-8},
  journal      = {J. Comput. Phys.},
  volume       = {54},
  pages        = {174--201},
  year         = {1984}
}

@article{Harten1983,
  author  = {Harten, Amiram and Lax, Peter D. and Leer, Bram van},
  doi     = {10.1137/1025002},
  journal = {SIAM Review},
  number  = {1},
  pages   = {35-61},
  title   = {{On Upstream Differencing and Godunov-Type Schemes for Hyperbolic Conservation Laws}},
  url     = {https://doi.org/10.1137/1025002},
  volume  = {25},
  year    = {1983}
}

@article{Einfeldt1991,
  author  = {B Einfeldt and C.D Munz and P.L Roe and B Sjögreen},
  doi     = {https://doi.org/10.1016/0021-9991(91)90211-3},
  issn    = {0021-9991},
  journal = {Journal of Computational Physics},
  number  = {2},
  pages   = {273-295},
  title   = {{On Godunov-type methods near low densities}},
  url     = {https://www.sciencedirect.com/science/article/pii/0021999191902113},
  volume  = {92},
  year    = {1991}
}

@article{Kaaz:2024zhy,
  author        = {Kaaz, Nicholas and Lithwick, Yoram and Liska, Matthew and Tchekhovskoy, Alexander},
  title         = {{Extreme Scale Height Variations and Nozzle Shocks in Warped Disks}},
  eprint        = {2410.00961},
  archiveprefix = {arXiv},
  primaryclass  = {astro-ph.HE},
  doi           = {10.3847/1538-4357/ad9a85},
  journal       = {Astrophys. J.},
  volume        = {979},
  number        = {2},
  pages         = {192},
  year          = {2025}
}

@article{Shakura:1972te,
  author  = {Shakura, N. I. and Sunyaev, R. A.},
  title   = {{Black holes in binary systems. Observational appearance}},
  journal = {Astron. Astrophys.},
  volume  = {24},
  pages   = {337--355},
  year    = {1973}
}

@article{Jiang:2010ba,
  author        = {Jiang, Yanfei and Goodman, Jeremy},
  title         = {{Star Formation in Quasar Disk}},
  eprint        = {1011.3541},
  archiveprefix = {arXiv},
  primaryclass  = {astro-ph.HE},
  doi           = {10.1088/0004-637X/730/1/45},
  journal       = {Astrophys. J.},
  volume        = {730},
  pages         = {45},
  year          = {2011}
}

@article{Goodman:2002gv,
  author        = {Goodman, Jeremy},
  title         = {{Selfgravity and qso disks}},
  eprint        = {astro-ph/0201001},
  archiveprefix = {arXiv},
  doi           = {10.1046/j.1365-8711.2003.06241.x},
  journal       = {Mon. Not. Roy. Astron. Soc.},
  volume        = {339},
  pages         = {937},
  year          = {2003}
}

@article{Sirko:2002ex,
  author        = {Sirko, Edwin and Goodman, Jeremy},
  title         = {{Spectral energy distributions of selfgravitating QSO discs}},
  eprint        = {astro-ph/0209469},
  archiveprefix = {arXiv},
  doi           = {10.1046/j.1365-8711.2003.06431.x},
  journal       = {Mon. Not. Roy. Astron. Soc.},
  volume        = {341},
  pages         = {501},
  year          = {2003}
}

@article{Ennoggi:2025ijc,
  author        = {Ennoggi, Lorenzo and Campanelli, Manuela and Krolik, Julian and Noble, Scott C. and Zlochower, Yosef and de Simone, Maria Chiara},
  title         = {{Merger of Spinning, Accreting Supermassive Black Hole Binaries}},
  eprint        = {2509.10319},
  archiveprefix = {arXiv},
  primaryclass  = {astro-ph.HE},
  doi           = {10.1103/f74l-3c4y},
  journal       = {Phys. Rev. Lett.},
  volume        = {136},
  number        = {11},
  pages         = {111401},
  year          = {2026}
}

@article{Lodato:2007hf,
  author        = {Lodato, Giuseppe and Pringle, J. E.},
  title         = {{Warp diffusion in accretion discs: A numerical investigation}},
  eprint        = {0708.1124},
  archiveprefix = {arXiv},
  primaryclass  = {astro-ph},
  doi           = {10.1111/j.1365-2966.2007.12332.x},
  journal       = {Mon. Not. Roy. Astron. Soc.},
  volume        = {381},
  pages         = {1287},
  year          = {2007}
}

@article{McCartney2014,
  author  = {McCartney, Gerry and Hacker, Thomas and Yang, Baijian},
  journal = {Educause Review},
  title   = {{Empowering Faculty: A Campus Cyberinfrastructure Strategy for Research Communities}},
  year    = {2014},
  url     = {https://er.educause.edu/articles/2014/7/empowering-faculty-a-campus-cyberinfrastructure-strategy-for-research-communities}
}

@inproceedings{10.1145/3491418.3530766,
  author    = {Song, X. Carol and Smith, Preston and Kalyanam, Rajesh and Zhu, Xiao and Adams, Eric and Colby, Kevin and Finnegan, Patrick and Gough, Erik and Hillery, Elizabett and Irvine, Rick and Maji, Amiya and St. John, Jason},
  title     = {Anvil - System Architecture and Experiences from Deployment and Early User Operations},
  year      = {2022},
  isbn      = {9781450391610},
  publisher = {Association for Computing Machinery},
  address   = {New York, NY, USA},
  url       = {https://doi.org/10.1145/3491418.3530766},
  doi       = {10.1145/3491418.3530766},
  booktitle = {Practice and Experience in Advanced Research Computing 2022: Revolutionary: Computing, Connections, You},
  articleno = {23},
  numpages  = {9},
  location  = {Boston, MA, USA},
  series    = {PEARC '22}
}

@article{Begelman:1980vb,
  author  = {Begelman, M. C. and Blandford, R. D. and Rees, M. J.},
  title   = {{Massive black hole binaries in active galactic nuclei}},
  doi     = {10.1038/287307a0},
  journal = {Nature},
  volume  = {287},
  pages   = {307--309},
  year    = {1980}
}

@article{Milosavljevic:2001vi,
  author        = {Milosavljevic, Milos and Merritt, David},
  title         = {{Formation of galactic nuclei}},
  eprint        = {astro-ph/0103350},
  archiveprefix = {arXiv},
  reportnumber  = {RUTGERS-AP-300},
  doi           = {10.1086/323830},
  journal       = {Astrophys. J.},
  volume        = {563},
  pages         = {34--62},
  year          = {2001}
}

@misc{deSimone:2025cqg,
  archiveprefix = {arXiv},
  journal       = {},
  author        = {de Simone, Maria Chiara and Campanelli, Manuela and Ennoggi, Lorenzo and Lousto, Carlos O. and Zlochower, Yosef},
  eprint        = {2510.05883},
  month         = {10},
  primaryclass  = {astro-ph.GA},
  year          = {2025}
}

@article{McKernan:2019hqs,
  author        = {McKernan, B. and Ford, K. E. S. and Bartos, I. and Graham, M. J. and Lyra, W. and Marka, S. and Marka, Z. and Ross, N. P. and Stern, D. and Yang, Y.},
  title         = {{Ram-pressure stripping of a kicked Hill sphere: Prompt electromagnetic emission from the merger of stellar mass black holes in an AGN accretion disk}},
  eprint        = {1907.03746},
  archiveprefix = {arXiv},
  primaryclass  = {astro-ph.HE},
  doi           = {10.3847/2041-8213/ab4886},
  journal       = {Astrophys. J. Lett.},
  volume        = {884},
  number        = {2},
  pages         = {L50},
  year          = {2019}
}

@article{McKernan:2020lgr,
  author        = {McKernan, B. and Ford, K. E. S. and O'Shaughnessy, R.},
  title         = {{Black hole, neutron star, and white dwarf merger rates in AGN discs}},
  eprint        = {2002.00046},
  archiveprefix = {arXiv},
  primaryclass  = {astro-ph.HE},
  doi           = {10.1093/mnras/staa2681},
  journal       = {Mon. Not. Roy. Astron. Soc.},
  volume        = {498},
  number        = {3},
  pages         = {4088--4094},
  year          = {2020}
}

@article{Tagawa:2023uqa,
  author        = {Tagawa, Hiromichi and Kimura, Shigeo S. and Haiman, Zolt{\'a}n and Perna, Rosalba and Bartos, Imre},
  title         = {{Observable Signature of Merging Stellar-mass Black Holes in Active Galactic Nuclei}},
  eprint        = {2301.07111},
  archiveprefix = {arXiv},
  primaryclass  = {astro-ph.HE},
  doi           = {10.3847/1538-4357/acc4bb},
  journal       = {Astrophys. J.},
  volume        = {950},
  number        = {1},
  pages         = {13},
  year          = {2023}
}

@article{Ford:2025bbk,
  author        = {Ford, K. E. Saavik and McKernan, Barry},
  title         = {{Using gravitational waves {\&} multi-messenger astronomy to reverse-engineer the properties of galactic nuclei}},
  eprint        = {2506.08801},
  archiveprefix = {arXiv},
  primaryclass  = {astro-ph.HE},
  doi           = {10.1088/1361-6382/ae0aad},
  journal       = {Class. Quant. Grav.},
  volume        = {43},
  number        = {3},
  pages         = {033001},
  year          = {2026}
}

@article{Kim:2024zjb,
  archiveprefix = {arXiv},
  author        = {Kim, Yoonsoo and Most, Elias R.},
  doi           = {10.1103/PhysRevD.111.083025},
  eprint        = {2409.12359},
  journal       = {Phys. Rev. D},
  number        = {8},
  pages         = {083025},
  primaryclass  = {astro-ph.HE},
  title         = {{General relativistic magnetized Bondi-Hoyle-Lyttleton accretion with a spin-field misalignment: Jet nutation, polarity reversals, and Magnus drag}},
  volume        = {111},
  year          = {2025}
}

@article{Chen:2023xrm,
  author        = {Chen, Ken and Dai, Zi-Gao},
  title         = {{Electromagnetic Counterparts Powered by Kicked Remnants of Black Hole Binary Mergers in AGN Disks}},
  eprint        = {2311.10518},
  archiveprefix = {arXiv},
  primaryclass  = {astro-ph.HE},
  doi           = {10.3847/1538-4357/ad0dfd},
  journal       = {Astrophys. J.},
  volume        = {961},
  number        = {2},
  pages         = {206},
  year          = {2024}
}

@article{Liska:2017alm,
  author        = {Liska, Matthew and Hesp, Casper and Tchekhovskoy, Alexander and Ingram, Adam and van der Klis, Michiel and Markoff, Sera},
  title         = {{Formation of Precessing Jets by Tilted Black-hole Discs in 3D General Relativistic MHD Simulations}},
  eprint        = {1707.06619},
  archiveprefix = {arXiv},
  primaryclass  = {astro-ph.HE},
  doi           = {10.1093/mnrasl/slx174},
  journal       = {Mon. Not. Roy. Astron. Soc.},
  volume        = {474},
  number        = {1},
  pages         = {L81--L85},
  year          = {2018}
}

@article{Liska:2018ayk,
  author        = {Liska, M. and Tchekhovskoy, A. and Ingram, A. and van der Klis, M.},
  title         = {{Bardeen{\textendash}Petterson alignment, jets, and magnetic truncation in GRMHD simulations of tilted thin accretion discs}},
  eprint        = {1810.00883},
  archiveprefix = {arXiv},
  primaryclass  = {astro-ph.HE},
  doi           = {10.1093/mnras/stz834},
  journal       = {Mon. Not. Roy. Astron. Soc.},
  volume        = {487},
  number        = {1},
  pages         = {550--561},
  year          = {2019}
}

@article{Merritt:2002hc,
  author        = {Merritt, David and Ekers, R. D.},
  title         = {{Tracing black hole mergers through radio lobe morphology}},
  eprint        = {astro-ph/0208001},
  archiveprefix = {arXiv},
  reportnumber  = {RUTGERS-AP-344},
  doi           = {10.1126/science.1074688},
  journal       = {Science},
  volume        = {297},
  pages         = {1310--1313},
  year          = {2002}
}

@article{Dennett-Thorpe:2001pdn,
  author        = {Dennett-Thorpe, J. and Scheuer, P. A. G. and Laing, R. A. and Bridle, A. H. and Pooley, Guy G. and Reich, W.},
  title         = {{Jet reorientation in agn: two winged radio galaxies}},
  eprint        = {astro-ph/0110339},
  archiveprefix = {arXiv},
  doi           = {10.1046/j.1365-8711.2002.05106.x},
  journal       = {Mon. Not. Roy. Astron. Soc.},
  volume        = {330},
  pages         = {609},
  year          = {2002}
}

@article{Bogdanovic:2021aav,
  author        = {Bogdanovic, Tamara and Miller, M. Coleman and Blecha, Laura},
  title         = {{Electromagnetic counterparts to massive black-hole mergers}},
  eprint        = {2109.03262},
  archiveprefix = {arXiv},
  primaryclass  = {astro-ph.HE},
  doi           = {10.1007/s41114-022-00037-8},
  journal       = {Living Rev. Rel.},
  volume        = {25},
  number        = {1},
  pages         = {3},
  year          = {2022}
}

@article{Liu:2003fx,
  author        = {Liu, Fu-Kun},
  title         = {{X-shaped radio galaxies as observational evidence for the interaction of supermassive binary black holes and accretion disk at pc scale}},
  eprint        = {astro-ph/0310691},
  archiveprefix = {arXiv},
  doi           = {10.1111/j.1365-2966.2004.07325.x},
  journal       = {Mon. Not. Roy. Astron. Soc.},
  volume        = {347},
  pages         = {1357},
  year          = {2004}
}

@article{Roberts:2015wja,
  author        = {Roberts, David H. and Saripalli, Lakshmi and Subrahmanyan, Ravi},
  title         = {{The Abundance of X-shaped Radio Sources: Implications for the Gravitational Wave Background}},
  eprint        = {1503.02021},
  archiveprefix = {arXiv},
  primaryclass  = {astro-ph.GA},
  doi           = {10.1088/2041-8205/810/1/L6},
  journal       = {Astrophys. J. Lett.},
  volume        = {810},
  number        = {1},
  pages         = {L6},
  year          = {2015}
}

@article{Kimura:2021xxu,
  author        = {Kimura, Shigeo S. and Murase, Kohta and Bartos, Imre},
  title         = {{Outflow Bubbles from Compact Binary Mergers Embedded in Active Galactic Nuclei: Cavity Formation and the Impact on Electromagnetic Counterparts}},
  eprint        = {2103.02461},
  archiveprefix = {arXiv},
  primaryclass  = {astro-ph.HE},
  doi           = {10.3847/1538-4357/ac0535},
  journal       = {Astrophys. J.},
  volume        = {916},
  number        = {2},
  pages         = {111},
  year          = {2021}
}

@article{OLeary:2008tpn,
  author        = {O'Leary, Ryan M. and Loeb, Abraham},
  title         = {{Star Clusters Around Recoiled Black Holes in the Milky Way Halo}},
  eprint        = {0809.4262},
  archiveprefix = {arXiv},
  primaryclass  = {astro-ph},
  doi           = {10.1111/j.1365-2966.2009.14611.x},
  journal       = {Mon. Not. Roy. Astron. Soc.},
  volume        = {395},
  pages         = {781},
  year          = {2009}
}

@article{OLeary:2011ykg,
  author        = {O'Leary, Ryan M. and Loeb, Abraham},
  title         = {{Recoiled star clusters in the Milky Way halo: N-body simulations and a candidate search through SDSS}},
  eprint        = {1102.3695},
  archiveprefix = {arXiv},
  primaryclass  = {astro-ph.CO},
  doi           = {10.1111/j.1365-2966.2011.20078.x},
  journal       = {Mon. Not. Roy. Astron. Soc.},
  volume        = {421},
  pages         = {2737},
  year          = {2012}
}

@article{Ogilvie:1998js,
  author        = {Ogilvie, G. I.},
  title         = {{The nonlinear fluid dynamics of a warped accretion disc}},
  eprint        = {astro-ph/9812073},
  archiveprefix = {arXiv},
  doi           = {10.1046/j.1365-8711.1999.02340.x},
  journal       = {Mon. Not. Roy. Astron. Soc.},
  volume        = {304},
  pages         = {557},
  year          = {1999}
}

@article{2011ApJS..192....9T,
  author        = {{Turk}, M.~J. and {Smith}, B.~D. and {Oishi}, J.~S. and {Skory}, S. and
                   {Skillman}, S.~W. and {Abel}, T. and {Norman}, M.~L.},
  title         = {{yt: A Multi-code Analysis Toolkit for Astrophysical Simulation Data}},
  journal       = {The Astrophysical Journal Supplement Series},
  archiveprefix = {arXiv},
  eprint        = {1011.3514},
  primaryclass  = {astro-ph.IM},
  year          = 2011,
  month         = jan,
  volume        = 192,
  eid           = {9},
  pages         = {9},
  doi           = {10.1088/0067-0049/192/1/9},
  adsurl        = {https://ui.adsabs.harvard.edu/abs/2011ApJS..192....9T}
}

@article{harris2020array,
  title     = {Array programming with {NumPy}},
  author    = {Charles R. Harris and K. Jarrod Millman and St{\'{e}}fan J.
               van der Walt and Ralf Gommers and Pauli Virtanen and David
               Cournapeau and Eric Wieser and Julian Taylor and Sebastian
               Berg and Nathaniel J. Smith and Robert Kern and Matti Picus
               and Stephan Hoyer and Marten H. van Kerkwijk and Matthew
               Brett and Allan Haldane and Jaime Fern{\'{a}}ndez del
               R{\'{i}}o and Mark Wiebe and Pearu Peterson and Pierre
               G{\'{e}}rard-Marchant and Kevin Sheppard and Tyler Reddy and
               Warren Weckesser and Hameer Abbasi and Christoph Gohlke and
               Travis E. Oliphant},
  year      = {2020},
  month     = sep,
  journal   = {Nature},
  volume    = {585},
  number    = {7825},
  pages     = {357--362},
  doi       = {10.1038/s41586-020-2649-2},
  publisher = {Springer Science and Business Media {LLC}},
  url       = {https://doi.org/10.1038/s41586-020-2649-2}
}

@article{2020SciPy-NMeth,
  author  = {Virtanen, Pauli and Gommers, Ralf and Oliphant, Travis E. and
             Haberland, Matt and Reddy, Tyler and Cournapeau, David and
             Burovski, Evgeni and Peterson, Pearu and Weckesser, Warren and
             Bright, Jonathan and {van der Walt}, St{\'e}fan J. and
             Brett, Matthew and Wilson, Joshua and Millman, K. Jarrod and
             Mayorov, Nikolay and Nelson, Andrew R. J. and Jones, Eric and
             Kern, Robert and Larson, Eric and Carey, C J and
             Polat, {\.I}lhan and Feng, Yu and Moore, Eric W. and
             {VanderPlas}, Jake and Laxalde, Denis and Perktold, Josef and
             Cimrman, Robert and Henriksen, Ian and Quintero, E. A. and
             Harris, Charles R. and Archibald, Anne M. and
             Ribeiro, Ant{\^o}nio H. and Pedregosa, Fabian and
             {van Mulbregt}, Paul and {SciPy 1.0 Contributors}},
  title   = {{{SciPy} 1.0: Fundamental Algorithms for Scientific
             Computing in Python}},
  journal = {Nature Methods},
  year    = {2020},
  volume  = {17},
  pages   = {261--272},
  adsurl  = {https://rdcu.be/b08Wh},
  doi     = {10.1038/s41592-019-0686-2}
}

@book{paraview:2015,
  author    = {Ayachit, Utkarsh},
  title     = {The ParaView Guide: A Parallel Visualization Application},
  year      = {2015},
  isbn      = {1930934300},
  publisher = {Kitware, Inc.},
  address   = {Clifton Park, NY, USA}
}

@incollection{paraview:2005,
  title     = {36 - ParaView: An End-User Tool for Large-Data Visualization},
  editor    = {Charles D. Hansen and Chris R. Johnson},
  booktitle = {Visualization Handbook},
  publisher = {Butterworth-Heinemann},
  address   = {Burlington},
  pages     = {717-731},
  year      = {2005},
  isbn      = {978-0-12-387582-2},
  doi       = {https://doi.org/10.1016/B978-012387582-2/50038-1},
  url       = {https://www.sciencedirect.com/science/article/pii/B9780123875822500381},
  author    = {James Ahrens and Berk Geveci and Charles Law}
}

@article{Hunter:2007,
  author    = {Hunter, J. D.},
  title     = {Matplotlib: A 2D graphics environment},
  journal   = {Computing in Science \& Engineering},
  volume    = {9},
  number    = {3},
  pages     = {90--95},
  publisher = {IEEE COMPUTER SOC},
  doi       = {10.1109/MCSE.2007.55},
  year      = 2007
}

@software{the_matplotlib_development_team_2026_22700776,
  author    = {{The Matplotlib Development Team}},
  title     = {Matplotlib: Visualization with Python},
  month     = sep,
  year      = 2026,
  publisher = {Zenodo},
  version   = {v3.11.2},
  doi       = {10.5281/zenodo.22700776},
  url       = {https://doi.org/10.5281/zenodo.22700776}
}

@article{Healy:2014yta,
  author        = {Healy, James and Lousto, Carlos O. and Zlochower, Yosef},
  title         = {{Remnant mass, spin, and recoil from spin aligned black-hole binaries}},
  eprint        = {1406.7295},
  archiveprefix = {arXiv},
  primaryclass  = {gr-qc},
  doi           = {10.1103/PhysRevD.90.104004},
  journal       = {Phys. Rev. D},
  volume        = {90},
  number        = {10},
  pages         = {104004},
  year          = {2014}
}

@article{Merritt:2008kg,
  author        = {Merritt, David and Schnittman, Jeremy D. and Komossa, Stefanie},
  title         = {{Hypercompact Stellar Systems Around Recoiling Supermassive Black Holes}},
  eprint        = {0809.5046},
  archiveprefix = {arXiv},
  primaryclass  = {astro-ph},
  doi           = {10.1088/0004-637X/699/2/1690},
  journal       = {Astrophys. J.},
  volume        = {699},
  pages         = {1690--1710},
  year          = {2009}
}

@article{Tchekhovskoy:2011zx,
  archiveprefix = {arXiv},
  author        = {Tchekhovskoy, Alexander and Narayan, Ramesh and McKinney, Jonathan C.},
  doi           = {10.1111/j.1745-3933.2011.01147.x},
  eprint        = {1108.0412},
  journal       = {Mon. Not. Roy. Astron. Soc.},
  pages         = {L79--L83},
  primaryclass  = {astro-ph.HE},
  title         = {{Efficient Generation of Jets from Magnetically Arrested Accretion on a Rapidly Spinning Black Hole}},
  volume        = {418},
  year          = {2011}
}

@article{Blandford:1977ds,
  author  = {Blandford, R. D. and Znajek, R. L.},
  title   = {{Electromagnetic extractions of energy from Kerr black holes}},
  doi     = {10.1093/mnras/179.3.433},
  journal = {Mon. Not. Roy. Astron. Soc.},
  volume  = {179},
  pages   = {433--456},
  year    = {1977}
}

@article{McKernan:2023xio,
  author        = {McKernan, B. and Ford, K. E. S.},
  title         = {{Constraining the LVK AGN channel with black hole spins}},
  eprint        = {2309.15213},
  archiveprefix = {arXiv},
  primaryclass  = {astro-ph.HE},
  doi           = {10.1093/mnras/stae1351},
  journal       = {Mon. Not. Roy. Astron. Soc.},
  volume        = {531},
  number        = {3},
  pages         = {3479--3485},
  year          = {2024}
}

@article{Tagawa:2019osr,
  author        = {Tagawa, Hiromichi and Haiman, Zoltan and Kocsis, Bence},
  title         = {{Formation and Evolution of Compact Object Binaries in AGN Disks}},
  eprint        = {1912.08218},
  archiveprefix = {arXiv},
  primaryclass  = {astro-ph.GA},
  doi           = {10.3847/1538-4357/ab9b8c},
  journal       = {Astrophys. J.},
  volume        = {898},
  number        = {1},
  pages         = {25},
  year          = {2020}
}

@article{Chatterjee:2023ber,
  author        = {Chatterjee, Koushik and Kaaz, Nicholas and Liska, Matthew and Tchekhovskoy, Alexander and Markoff, Sera},
  title         = {{Misaligned magnetized accretion flows onto spinning black holes: Magneto-spin alignment, outflow power, and intermittent jets}},
  eprint        = {2311.00432},
  archiveprefix = {arXiv},
  primaryclass  = {astro-ph.HE},
  doi           = {10.1103/hgj9-v4fk},
  journal       = {Phys. Rev. D},
  volume        = {112},
  number        = {6},
  pages         = {063013},
  year          = {2025}
}

@article{Li:2022eup,
  author        = {Li, Rixin and Lai, Dong},
  title         = {{Hydrodynamical evolution of black hole binaries embedded in AGN discs: II. dependence on equation~of state, binary mass, and separation scales}},
  eprint        = {2207.01125},
  archiveprefix = {arXiv},
  primaryclass  = {astro-ph.HE},
  doi           = {10.1093/mnras/stad1117},
  journal       = {Mon. Not. Roy. Astron. Soc.},
  volume        = {522},
  number        = {2},
  pages         = {1881--1894},
  year          = {2023}
}

@article{Li:2022pnc,
  author        = {Li, Rixin and Lai, Dong},
  title         = {{Hydrodynamical evolution of black-hole binaries embedded in AGN discs}},
  eprint        = {2202.07633},
  archiveprefix = {arXiv},
  primaryclass  = {astro-ph.HE},
  doi           = {10.1093/mnras/stac2577},
  journal       = {Mon. Not. Roy. Astron. Soc.},
  volume        = {517},
  number        = {2},
  pages         = {1602--1624},
  year          = {2022}
}

@article{Burrows:2023ffl,
  author        = {Burrows, Adam and Wang, Tianshu and Vartanyan, David and Coleman, Matthew S. B.},
  title         = {{A Theory for Neutron Star and Black Hole Kicks and Induced Spins}},
  eprint        = {2311.12109},
  archiveprefix = {arXiv},
  primaryclass  = {astro-ph.HE},
  doi           = {10.3847/1538-4357/ad2353},
  journal       = {Astrophys. J.},
  volume        = {963},
  number        = {1},
  pages         = {63},
  year          = {2024}
}

@article{Coleman:2022lwr,
  author        = {Coleman, Matthew S. B. and Burrows, Adam},
  title         = {{Kicks and induced spins of neutron stars at birth}},
  eprint        = {2209.02711},
  archiveprefix = {arXiv},
  primaryclass  = {astro-ph.HE},
  doi           = {10.1093/mnras/stac2573},
  journal       = {Mon. Not. Roy. Astron. Soc.},
  volume        = {517},
  number        = {3},
  pages         = {3938--3961},
  year          = {2022}
}

@article{Wongwathanarat:2012zp,
  author        = {Wongwathanarat, A. and Janka, H. -Th. and Mueller, E.},
  title         = {{Three-dimensional neutrino-driven supernovae: Neutron star kicks, spins, and asymmetric ejection of nucleosynthesis products}},
  eprint        = {1210.8148},
  archiveprefix = {arXiv},
  primaryclass  = {astro-ph.HE},
  doi           = {10.1051/0004-6361/201220636},
  journal       = {Astron. Astrophys.},
  volume        = {552},
  pages         = {A126},
  year          = {2013}
}

@article{Popov:2025hva,
  author        = {Popov, Sergei and M{\"u}ller, Bernhard and Mandel, Ilya},
  title         = {{Natal kicks of compact objects}},
  eprint        = {2509.01430},
  archiveprefix = {arXiv},
  primaryclass  = {astro-ph.HE},
  doi           = {10.1016/j.newar.2025.101734},
  journal       = {New Astron. Rev.},
  volume        = {101},
  pages         = {101734},
  year          = {2025}
}

@article{Chan:2020lnd,
  author        = {Chan, Conrad and M{\"u}ller, Bernhard and Heger, Alexander},
  title         = {{The impact of fallback on the compact remnants and chemical yields of core-collapse supernovae}},
  eprint        = {2003.04320},
  archiveprefix = {arXiv},
  primaryclass  = {astro-ph.SR},
  doi           = {10.1093/mnras/staa1431},
  journal       = {Mon. Not. Roy. Astron. Soc.},
  volume        = {495},
  number        = {4},
  pages         = {3751--3762},
  year          = {2020}
}

@article{Janka:2024xbp,
  author        = {Janka, Hans-Thomas and Kresse, Daniel},
  title         = {{Interplay between neutrino kicks and hydrodynamic kicks of neutron stars and black holes}},
  eprint        = {2401.13817},
  archiveprefix = {arXiv},
  primaryclass  = {astro-ph.HE},
  doi           = {10.1007/s10509-024-04343-1},
  journal       = {Astrophys. Space Sci.},
  volume        = {369},
  number        = {8},
  pages         = {80},
  year          = {2024}
}

@misc{Li:2026pxl,
  author        = {Li, Sean E. and Gottlieb, Ore and Metzger, Brian D.},
  eprint        = {2609.10694},
  archiveprefix = {arXiv},
  primaryclass  = {astro-ph.HE},
  month         = {9},
  year          = {2026}
}

@article{Kaaz:2025udg,
  author        = {Kaaz, Nicholas and Liska, Matthew and Ward, Charlotte and Davelaar, Jordy},
  title         = {{Changing-look Active Galactic Nuclei Powered by Disk Tearing}},
  eprint        = {2511.09626},
  archiveprefix = {arXiv},
  primaryclass  = {astro-ph.HE},
  doi           = {10.3847/1538-4357/ae665a},
  journal       = {Astrophys. J.},
  volume        = {1003},
  number        = {2},
  pages         = {161},
  year          = {2026}
}

@article{Manikantan:2025eqa,
  author        = {Manikantan, Vikram and Paschalidis, Vasileios},
  title         = {{Magnetically arrested minidisks around nonspinning binary black holes: The MAD scenario for circumbinary disk accretion}},
  eprint        = {2509.18254},
  archiveprefix = {arXiv},
  primaryclass  = {astro-ph.HE},
  doi           = {10.1103/g98b-m8m1},
  journal       = {Phys. Rev. D},
  volume        = {112},
  number        = {10},
  pages         = {103050},
  year          = {2025}
}

@article{Most:2024qus,
  author        = {Most, Elias R. and Wang, Hai-Yang},
  title         = {{Magnetically Arrested Circumbinary Accretion Flows}},
  eprint        = {2408.00757},
  archiveprefix = {arXiv},
  primaryclass  = {astro-ph.HE},
  doi           = {10.3847/2041-8213/ad7713},
  journal       = {Astrophys. J. Lett.},
  volume        = {973},
  number        = {1},
  pages         = {L19},
  year          = {2024}
}

@misc{He:2026rle,
  author        = {He, Lei and Zhu, Liang-Gui and Chen, Ken and Dai, Zi-Gao and Yuan, Ye-Fei and Wang, Jian-Min and Zhao, Wen},
  eprint        = {2609.39224},
  archiveprefix = {arXiv},
  primaryclass  = {astro-ph.HE},
  month         = {9},
  year          = {2026}
}

@misc{Kalinani:2026mqe,
  author        = {Kalinani, Jay V. and Campanelli, Manuela and Chabanov, Michail and de Simone, Maria Chiara and Lousto, Carlos O. and Witek, Helvi and Zlochower, Yosef},
  eprint        = {2609.24837},
  archiveprefix = {arXiv},
  primaryclass  = {astro-ph.HE},
  month         = {9},
  year          = {2026}
}

@article{LIGOScientific:2020ufj,
  author        = {Abbott, R. and others},
  collaboration = {LIGO Scientific, Virgo},
  title         = {{Properties and Astrophysical Implications of the 150 M$_\odot$ Binary Black Hole Merger GW190521}},
  eprint        = {2009.01190},
  archiveprefix = {arXiv},
  primaryclass  = {astro-ph.HE},
  reportnumber  = {LIGO-P2000021},
  doi           = {10.3847/2041-8213/aba493},
  journal       = {Astrophys. J. Lett.},
  volume        = {900},
  number        = {1},
  pages         = {L13},
  year          = {2020}
}

@article{Yuan:2021jjt,
  author        = {Yuan, Chengchao and Murase, Kohta and Zhang, B. Theodore and Kimura, Shigeo S. and M{\'e}sz{\'a}ros, Peter},
  title         = {{Post-Merger Jets from Supermassive Black Hole Coalescences as Electromagnetic Counterparts of Gravitational Wave Emission}},
  eprint        = {2101.05788},
  archiveprefix = {arXiv},
  primaryclass  = {astro-ph.HE},
  doi           = {10.3847/2041-8213/abee24},
  journal       = {Astrophys. J. Lett.},
  volume        = {911},
  number        = {1},
  pages         = {L15},
  year          = {2021}
}

@article{Abramowicz:2011xu,
  author        = {Abramowicz, Marek A. and Fragile, P. Chris},
  title         = {{Foundations of Black Hole Accretion Disk Theory}},
  eprint        = {1104.5499},
  archiveprefix = {arXiv},
  primaryclass  = {astro-ph.HE},
  reportnumber  = {NSF-KITP-12-055},
  doi           = {10.12942/lrr-2013-1},
  journal       = {Living Rev. Rel.},
  volume        = {16},
  pages         = {1},
  year          = {2013}
}

@article{Fragile:2008sv,
  author        = {Fragile, P. Chris and Blaes, Omer M.},
  title         = {{Epicyclic Motions and Standing Shocks in Numerically Simulated Tilted Black-Hole Accretion Disks}},
  eprint        = {0807.2453},
  archiveprefix = {arXiv},
  primaryclass  = {astro-ph},
  doi           = {10.1086/591936},
  journal       = {Astrophys. J.},
  volume        = {687},
  pages         = {757},
  year          = {2008}
}

@article{Fairbairn:2021,
  author        = {{Fairbairn}, Callum W. and {Ogilvie}, Gordon I.},
  title         = {{Non-linear resonant torus oscillations as a model of Keplerian disc warp dynamics}},
  journal       = {Monthly Notices of the Royal Astronomical Society},
  year          = 2021,
  month         = dec,
  volume        = {508},
  number        = {2},
  pages         = {2426-2446},
  doi           = {10.1093/mnras/stab2717},
  archiveprefix = {arXiv},
  eprint        = {2109.08889},
  primaryclass  = {astro-ph.SR},
  adsurl        = {https://ui.adsabs.harvard.edu/abs/2021MNRAS.508.2426F}
}

@article{McKinney:2012wd,
  author        = {McKinney, Jonathan C. and Tchekhovskoy, Alexander and Blandford, Roger D.},
  title         = {{Alignment of Magnetized Accretion Disks and Relativistic Jets with Spinning Black Holes}},
  eprint        = {1211.3651},
  archiveprefix = {arXiv},
  primaryclass  = {astro-ph.CO},
  doi           = {10.1126/science.1230811},
  journal       = {Science},
  volume        = {339},
  pages         = {49--52},
  year          = {2013}
}

@article{Tagawa:2026lqy,
  author        = {Tagawa, Hiromichi and Haiman, Zolt{\'a}n and Kimura, Shigeo S. and Yesuf, Hassen M. and Guo, Hengxiao},
  title         = {{Electromagnetic Flares from Compact-object Mergers in Active Galactic Nucleus Disks: Signatures and Predictions}},
  eprint        = {2604.05020},
  archiveprefix = {arXiv},
  primaryclass  = {astro-ph.HE},
  doi           = {10.3847/1538-4357/ae7d29},
  journal       = {Astrophys. J.},
  volume        = {1006},
  number        = {2},
  pages         = {121},
  year          = {2026}
}

@article{Dittmann:2023sha,
  author        = {Dittmann, Alexander J. and Dempsey, Adam M. and Li, Hui},
  title         = {{The Evolution of Inclined Binary Black Holes in the Disks of Active Galactic Nuclei}},
  eprint        = {2310.03832},
  archiveprefix = {arXiv},
  primaryclass  = {astro-ph.HE},
  doi           = {10.3847/1538-4357/ad23ce},
  journal       = {Astrophys. J.},
  volume        = {964},
  number        = {1},
  pages         = {61},
  year          = {2024}
}

@article{Dempsey:2022iws,
  author        = {Dempsey, Adam M. and Li, Hui and Mishra, Bhupendra and Li, Shengtai},
  title         = {{Contracting and Expanding Binary Black Holes in 3D Low-mass AGN Disks: The Importance of Separation}},
  eprint        = {2203.06534},
  archiveprefix = {arXiv},
  primaryclass  = {astro-ph.HE},
  doi           = {10.3847/1538-4357/ac9d92},
  journal       = {Astrophys. J.},
  volume        = {940},
  number        = {2},
  pages         = {155},
  year          = {2022}
}

@article{LaMassa:2015,
  author        = {{LaMassa}, Stephanie M. and {Cales}, Sabrina and {Moran}, Edward C. and {Myers}, Adam D. and {Richards}, Gordon T. and {Eracleous}, Michael and {Heckman}, Timothy M. and {Gallo}, Luigi and {Urry}, C. Megan},
  title         = {{The Discovery of the First {\textquotedblleft}Changing Look{\textquotedblright} Quasar: New Insights Into the Physics and Phenomenology of Active Galactic Nucleus}},
  journal       = {\apj},
  year          = 2015,
  month         = feb,
  volume        = {800},
  number        = {2},
  eid           = {144},
  pages         = {144},
  doi           = {10.1088/0004-637X/800/2/144},
  archiveprefix = {arXiv},
  eprint        = {1412.2136},
  primaryclass  = {astro-ph.GA},
  adsurl        = {https://ui.adsabs.harvard.edu/abs/2015ApJ...800..144L}
}

@article{MacLeod:2016,
  author        = {{MacLeod}, Chelsea L. and {Ross}, Nicholas P. and {Lawrence}, Andy and {Goad}, Mike and {Horne}, Keith and {Burgett}, William and {Chambers}, Ken C. and {Flewelling}, Heather and {Hodapp}, Klaus and {Kaiser}, Nick and {Magnier}, Eugene and {Wainscoat}, Richard and {Waters}, Christopher},
  title         = {{A systematic search for changing-look quasars in SDSS}},
  journal       = {\mnras},
  year          = 2016,
  month         = mar,
  volume        = {457},
  number        = {1},
  pages         = {389-404},
  doi           = {10.1093/mnras/stv2997},
  archiveprefix = {arXiv},
  eprint        = {1509.08393},
  primaryclass  = {astro-ph.GA},
  adsurl        = {https://ui.adsabs.harvard.edu/abs/2016MNRAS.457..389M}
}

@article{Shappee:2014,
  author        = {{Shappee}, B.~J. and {Prieto}, J.~L. and {Grupe}, D. and {Kochanek}, C.~S. and {Stanek}, K.~Z. and {De Rosa}, G. and {Mathur}, S. and {Zu}, Y. and {Peterson}, B.~M. and {Pogge}, R.~W. and {Komossa}, S. and {Im}, M. and {Jencson}, J. and {Holoien}, T.~W.-S. and {Basu}, U. and {Beacom}, J.~F. and {Szczygie{\l}}, D.~M. and {Brimacombe}, J. and {Adams}, S. and {Campillay}, A. and {Choi}, C. and {Contreras}, C. and {Dietrich}, M. and {Dubberley}, M. and {Elphick}, M. and {Foale}, S. and {Giustini}, M. and {Gonzalez}, C. and {Hawkins}, E. and {Howell}, D.~A. and {Hsiao}, E.~Y. and {Koss}, M. and {Leighly}, K.~M. and {Morrell}, N. and {Mudd}, D. and {Mullins}, D. and {Nugent}, J.~M. and {Parrent}, J. and {Phillips}, M.~M. and {Pojmanski}, G. and {Rosing}, W. and {Ross}, R. and {Sand}, D. and {Terndrup}, D.~M. and {Valenti}, S. and {Walker}, Z. and {Yoon}, Y.},
  title         = {{The Man behind the Curtain: X-Rays Drive the UV through NIR Variability in the 2013 Active Galactic Nucleus Outburst in NGC 2617}},
  journal       = {\apj},
  year          = 2014,
  month         = jun,
  volume        = {788},
  number        = {1},
  eid           = {48},
  pages         = {48},
  doi           = {10.1088/0004-637X/788/1/48},
  archiveprefix = {arXiv},
  eprint        = {1310.2241},
  primaryclass  = {astro-ph.HE},
  adsurl        = {https://ui.adsabs.harvard.edu/abs/2014ApJ...788...48S}
}

@article{1975ApJ...195L..65B,
  author  = {{Bardeen}, James M. and {Petterson}, Jacobus A.},
  title   = {{The Lense-Thirring Effect and Accretion Disks around Kerr Black Holes}},
  journal = {\apjl},
  year    = 1975,
  month   = jan,
  volume  = {195},
  pages   = {L65},
  doi     = {10.1086/181711},
  adsurl  = {https://ui.adsabs.harvard.edu/abs/1975ApJ...195L..65B}
}

@article{Nixon:2012fm,
  author        = {Nixon, Chris and King, Andrew},
  title         = {{Broken discs: warp propagation in accretion discs}},
  eprint        = {1201.1297},
  archiveprefix = {arXiv},
  primaryclass  = {astro-ph.HE},
  doi           = {10.1111/j.1365-2966.2011.20377.x},
  journal       = {Mon. Not. Roy. Astron. Soc.},
  volume        = {421},
  pages         = {1201},
  year          = {2012}
}

@article{Lodato:2006kv,
  author        = {Lodato, Giuseppe and Pringle, J. E.},
  title         = {{The evolution of misaligned accretion discs and spinning black holes}},
  eprint        = {astro-ph/0602306},
  archiveprefix = {arXiv},
  doi           = {10.1111/j.1365-2966.2006.10194.x},
  journal       = {Mon. Not. Roy. Astron. Soc.},
  volume        = {368},
  pages         = {1196--1208},
  year          = {2006}
}

@article{Nelson:2000mw,
  author        = {Nelson, Richard P. and Papaloizou, John C. B.},
  title         = {{Hydrodynamic simulations of the bardeen-petterson effect}},
  eprint        = {astro-ph/0001439},
  archiveprefix = {arXiv},
  doi           = {10.1046/j.1365-8711.2000.03478.x},
  journal       = {Mon. Not. Roy. Astron. Soc.},
  volume        = {315},
  pages         = {570},
  year          = {2000}
}

@article{Lousto:2009ka,
  author        = {Lousto, Carlos O. and Nakano, Hiroyuki and Zlochower, Yosef and Campanelli, Manuela},
  title         = {{Statistical studies of Spinning Black-Hole Binaries}},
  eprint        = {0910.3197},
  archiveprefix = {arXiv},
  primaryclass  = {gr-qc},
  doi           = {10.1103/PhysRevD.81.084023},
  journal       = {Phys. Rev. D},
  volume        = {81},
  pages         = {084023},
  year          = {2010},
  note          = {[Erratum: Phys.Rev.D 82, 129902 (2010)]}
}

\appendix

\section{Simulation results for non-constant initial surface mass density}
\label{sec:non-const-p-results}

\begin{figure*}
    \centering
    \raisebox{-0.5\height}{\includegraphics[width=0.48\linewidth]{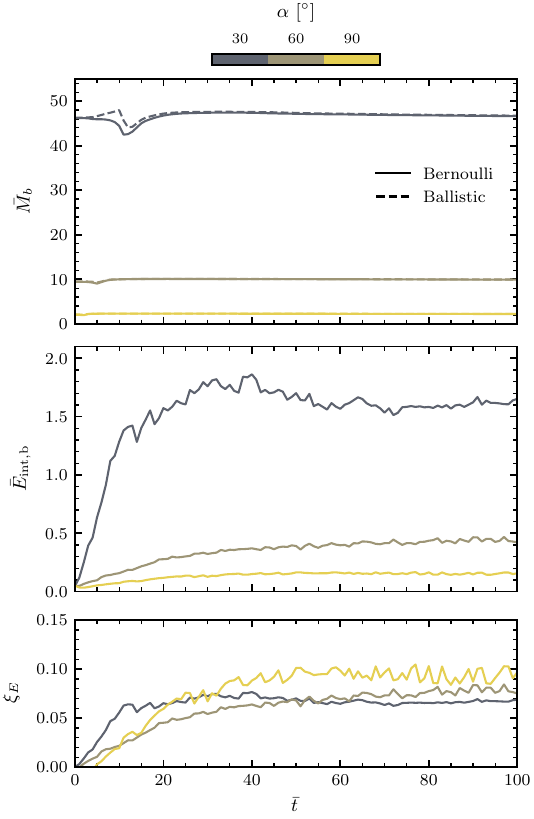}}
    \hfill
    \raisebox{-0.5\height}{\includegraphics[width=0.48\linewidth]{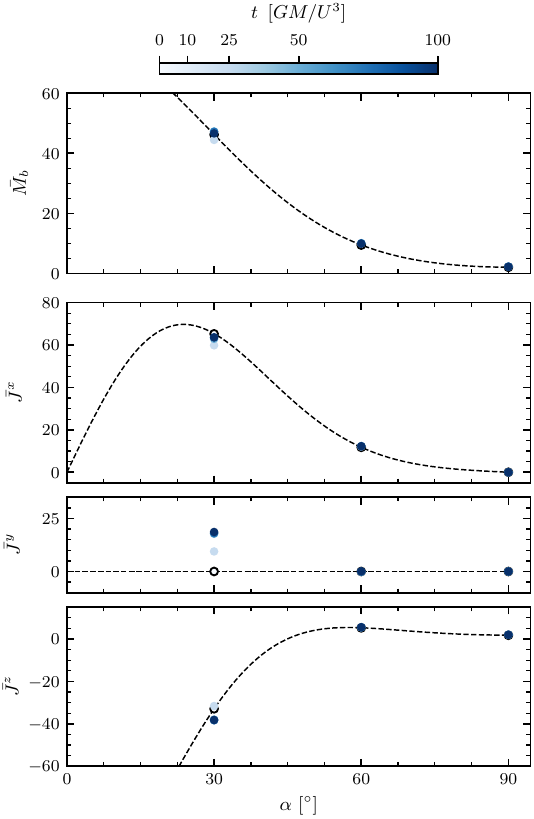}}
    \\[3em]
    \subcaptionbox{$\alpha=90^\circ$ \label{fig:p10-a90}}
    {\includegraphics[width=0.33\linewidth]{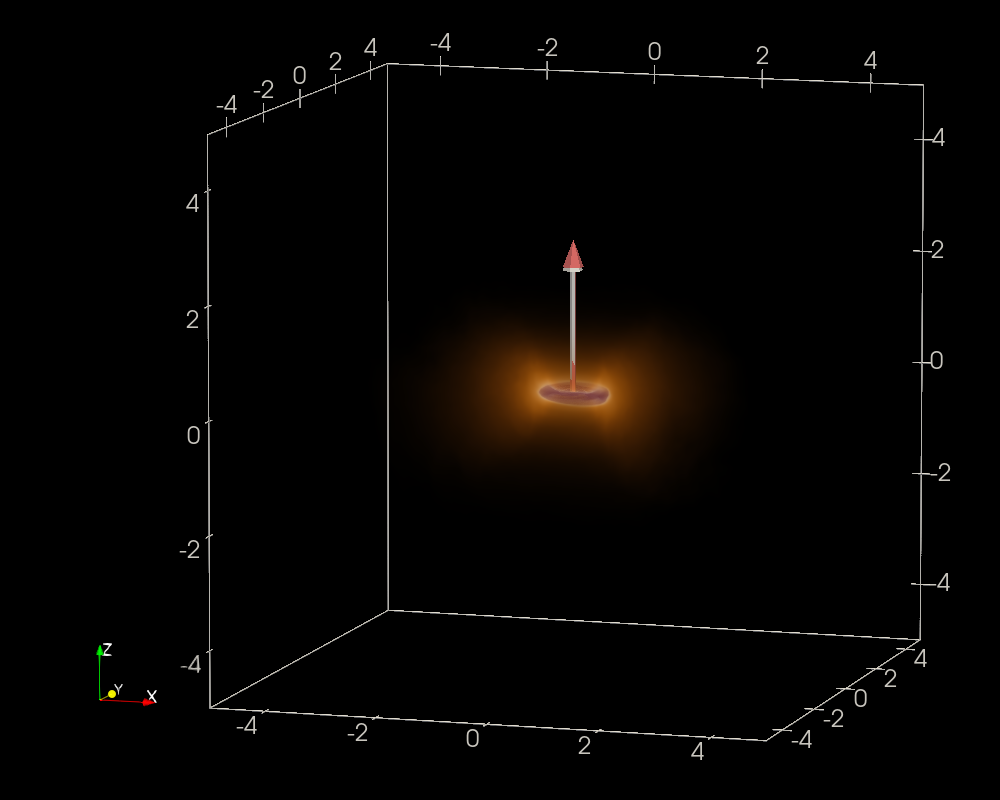}}
    \subcaptionbox{$\alpha=60^\circ$ \label{fig:p10-a60}}
    {\includegraphics[width=0.33\linewidth]{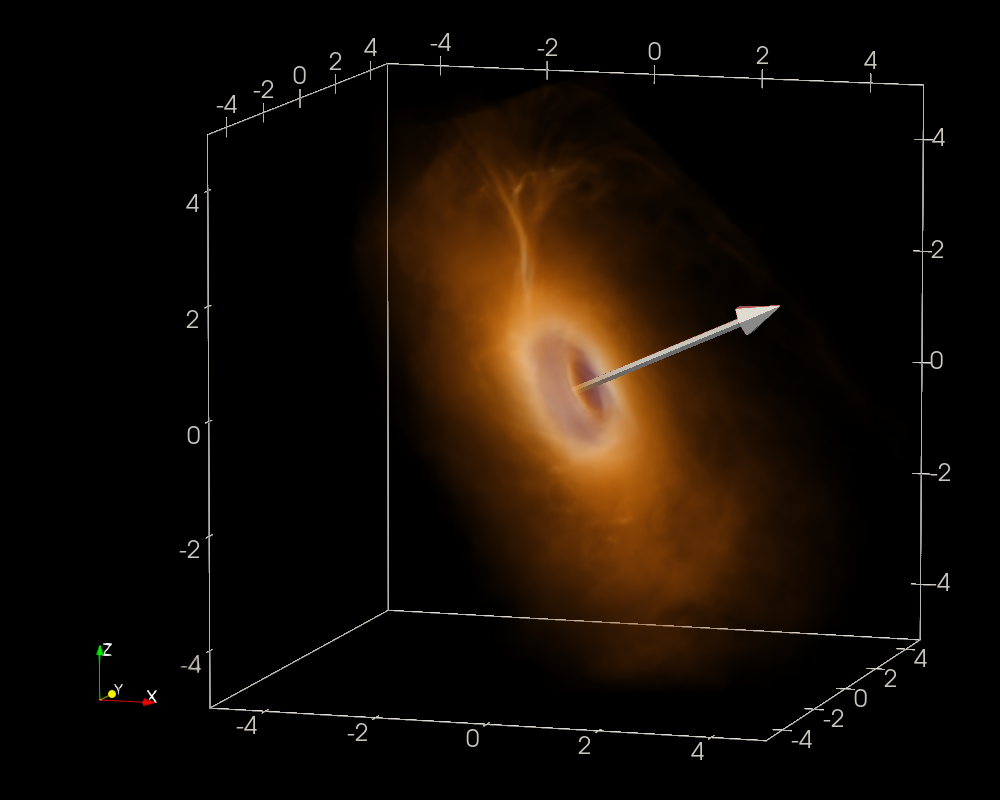}}
    \subcaptionbox{$\alpha=30^\circ$ \label{fig:p10-a30}}
    {\includegraphics[width=0.33\linewidth]{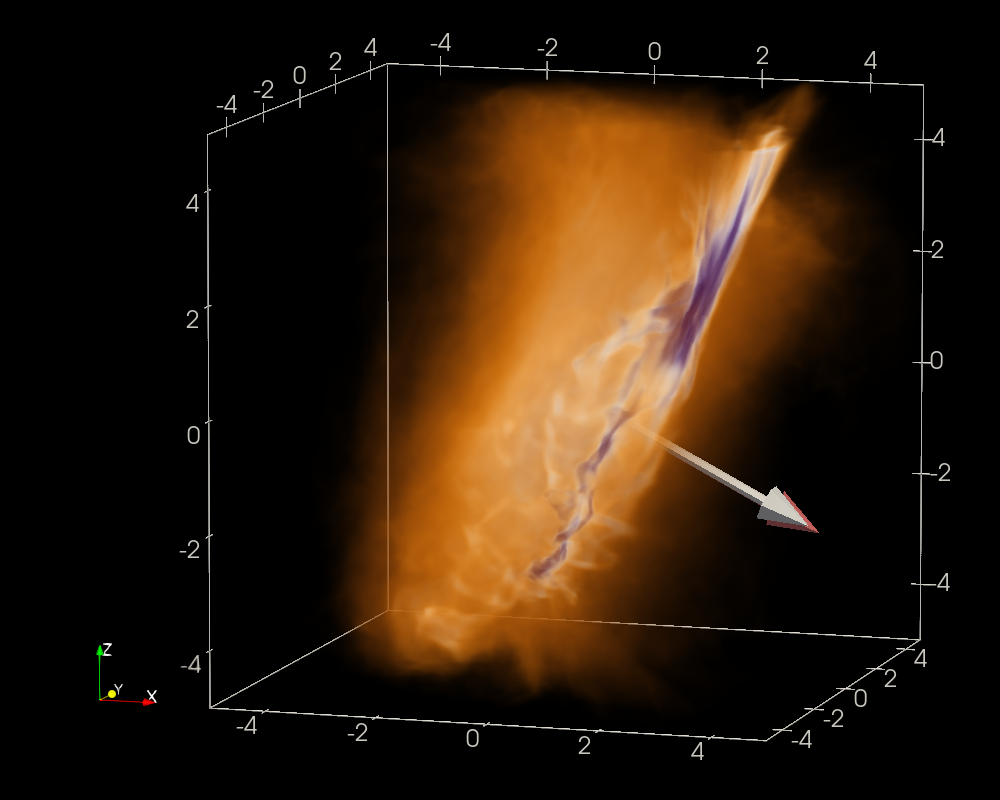}}
    \\[1em]
    \caption{Simulation results from the $p=1$ initial surface mass density
        profile. Each figure corresponds to Fig.~\ref{fig:p00-mass-eint-bound},
        \ref{fig:comparison-p00}, \ref{fig:3d-p00-final-tilt} in the main text.}
    \label{fig:p10-results}
\end{figure*}

\begin{figure*}
    \centering
    \raisebox{-0.5\height}{\includegraphics[width=0.48\linewidth]{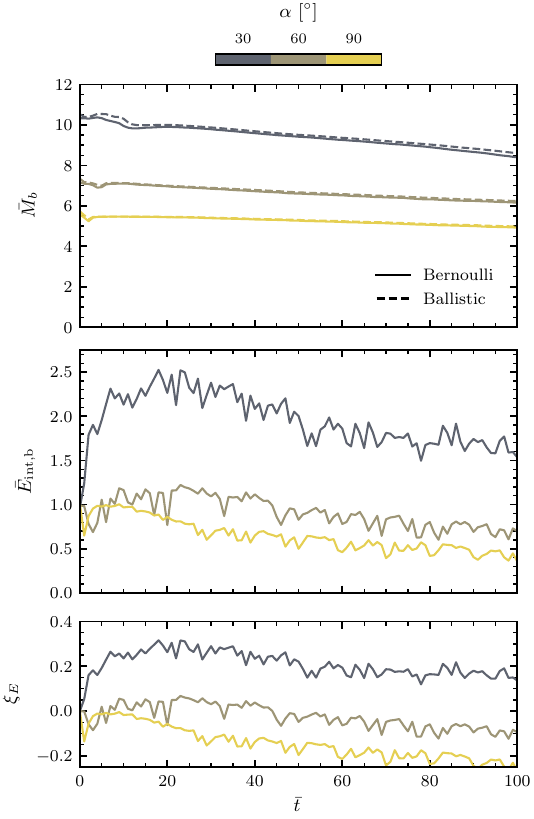}}
    \hfill
    \raisebox{-0.5\height}{\includegraphics[width=0.48\linewidth]{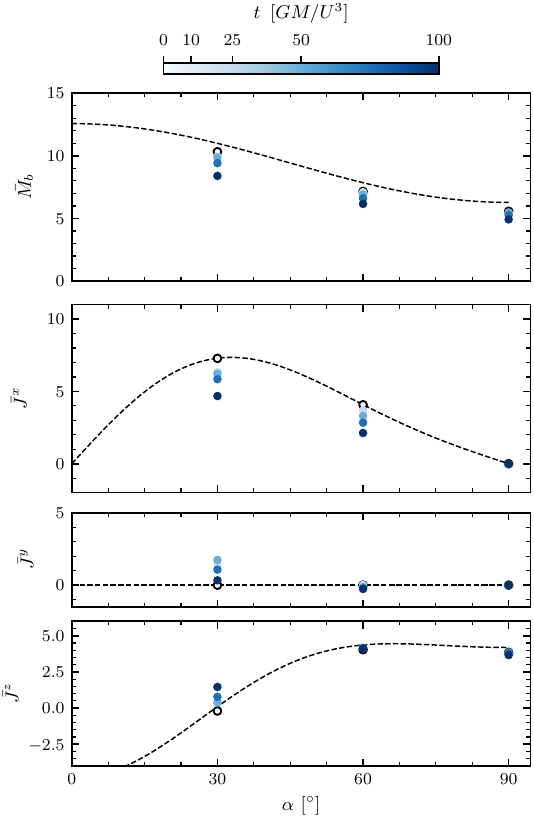}}
    \\[3em]
    \subcaptionbox{$\alpha=90^\circ$ \label{fig:pm10-a90}}
    {\includegraphics[width=0.33\linewidth]{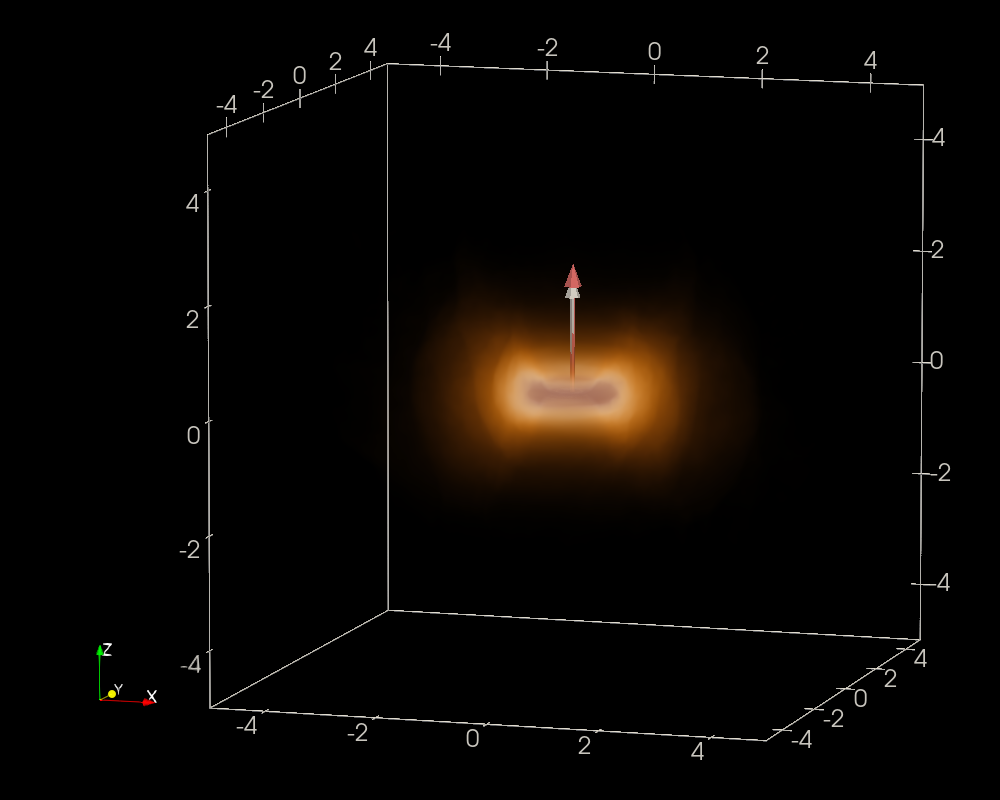}}
    \subcaptionbox{$\alpha=60^\circ$ \label{fig:pm10-a60}}
    {\includegraphics[width=0.33\linewidth]{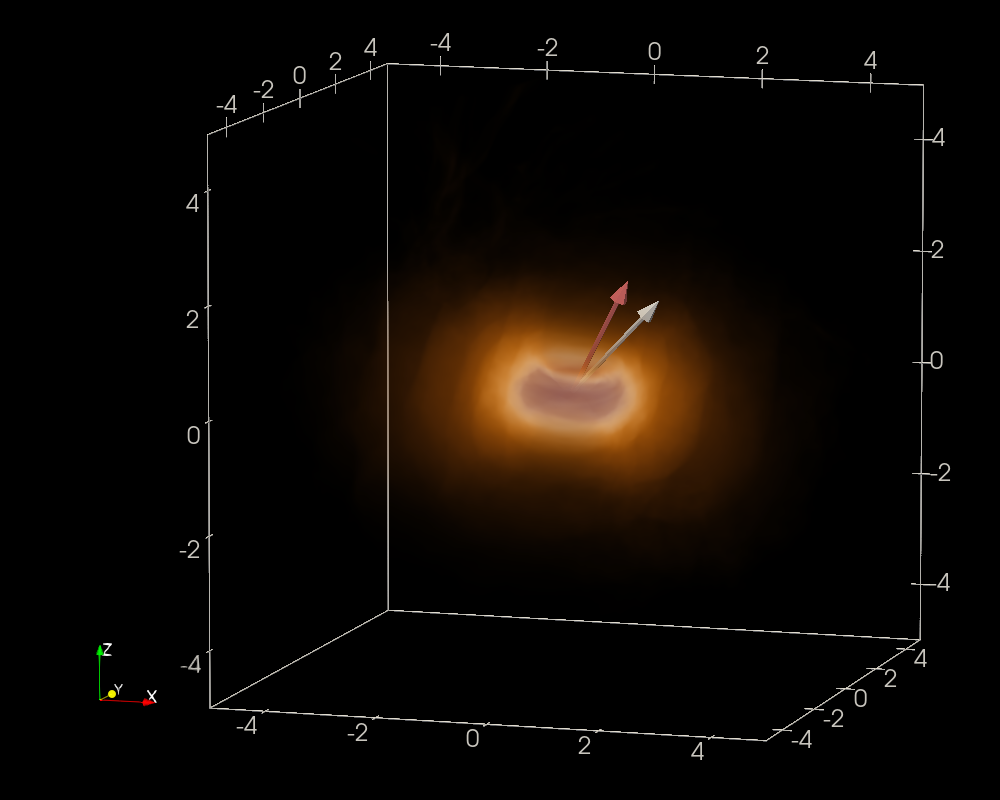}}
    \subcaptionbox{$\alpha=30^\circ$ \label{fig:pm10-a30}}
    {\includegraphics[width=0.33\linewidth]{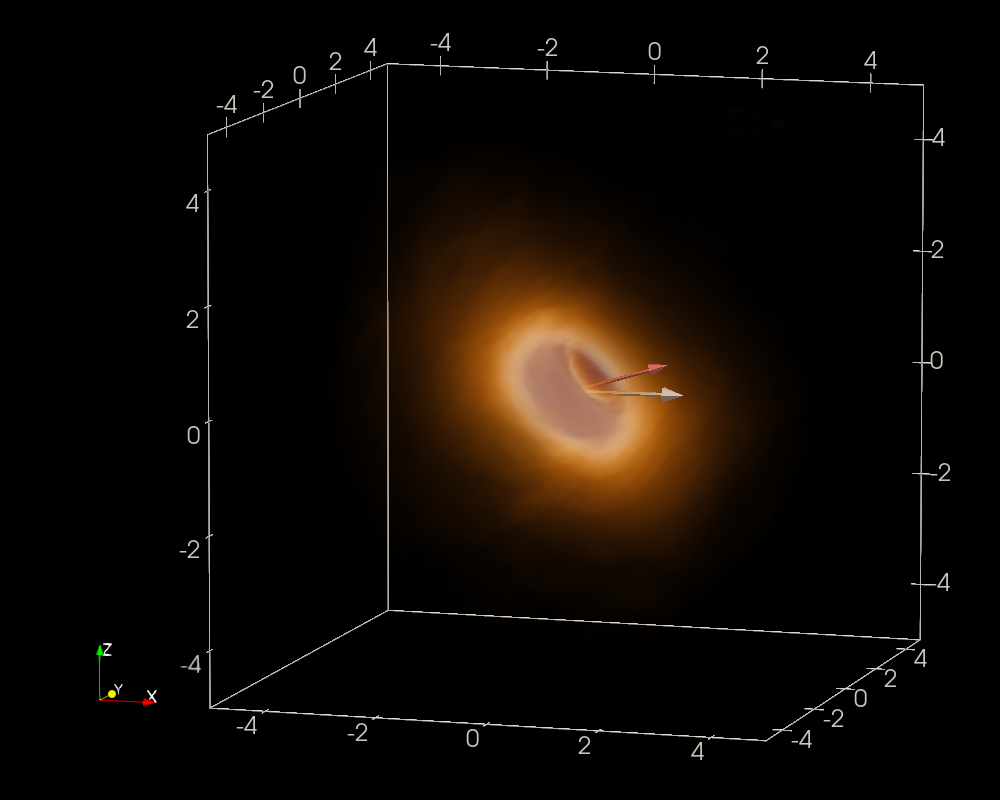}}
    \\[1em]
    \caption{Same as Fig.~\ref{fig:p10-results}, for $p=-1$ simulations.}
    \label{fig:pm10-results}
\end{figure*}

\begin{figure*}
    \centering
    \includegraphics[width=0.98\linewidth]{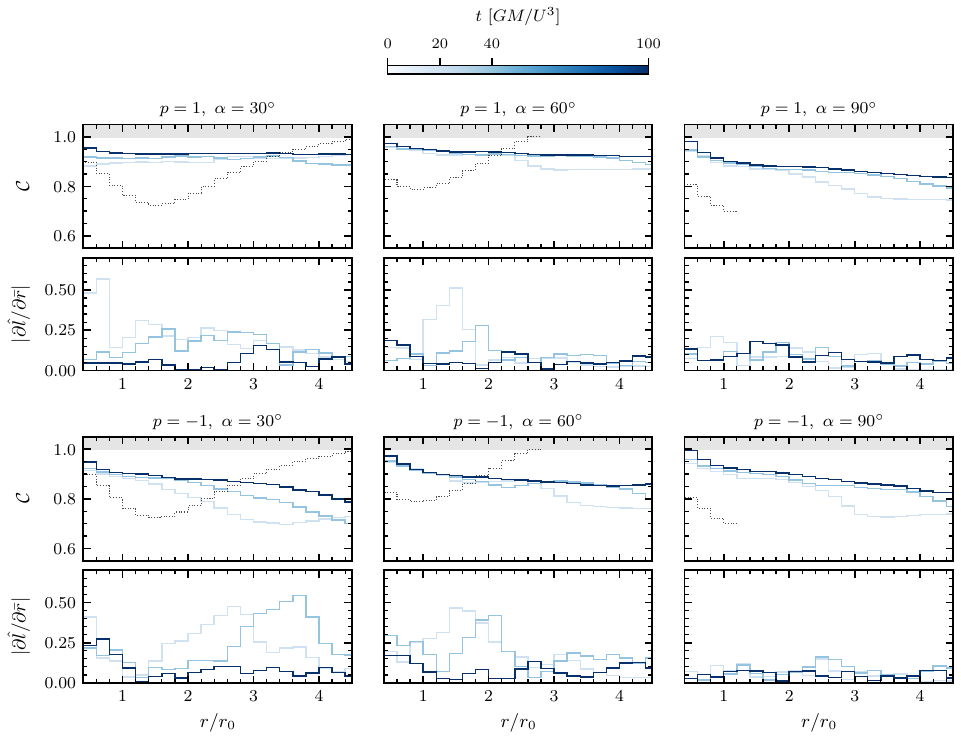}
    \caption{Same as Fig.~\ref{fig:coherence-and-warp-p00}, for $p=\pm 1$
        simulations.}
    \label{fig:coherence-and-warp-ppm1}
\end{figure*}

Fig.~\ref{fig:p10-results}, Fig.~\ref{fig:pm10-results}, and
Fig.~\ref{fig:coherence-and-warp-ppm1} summarize the results from the
simulations with $p=1$ and $p=-1$ initial surface mass density profiles. For the
$p=-1$ case, the $\approx$10\% deficit of the total bound mass at $\bar{t}=0$ is
due to the inner radial truncation applied to the initial mass density profile
(Eq.~\ref{eq:disk-truncation-inner}).

\bsp	%
\label{lastpage}
\end{document}